\documentclass[a4paper,12pt]{article}

\usepackage{xcolor}
\definecolor{nosaka}{rgb}{0.0, 0.5, 0.0}
\definecolor{nosaka2}{rgb}{0.6, 0.0, 0.0}
\usepackage{listings}
\usepackage{longtable}

\usepackage[utf8]{inputenc}

\usepackage{url}
\usepackage{tikz}
\usetikzlibrary{decorations.pathmorphing}

\usepackage{amsmath}

\makeatletter
\newcommand{\subscripts}[3]{%
  \@mathmeasure\z@\displaystyle{#2}%
  \global\setbox\@ne\vbox to\ht\z@{}\dp\@ne\dp\z@
  \setbox\tw@\box\@ne
  \@mathmeasure4\displaystyle{\copy\tw@_{#1}}%
  \@mathmeasure6\displaystyle{{#2}_{#3}}%
  \dimen@-\wd6 \advance\dimen@\wd4 \advance\dimen@\wd\z@
  \hbox to\dimen@{}\mathop{\kern-\dimen@\box4\box6}%
}
\makeatother

\usepackage{graphicx}

\usepackage{mathrsfs}

\usepackage{tikz}
\usepackage{bm}
\usepackage{arydshln}
\usepackage{color}
\usepackage{ulem}
\usepackage{amssymb}

\usepackage{framed}
\usepackage{here}
\usepackage{comment}

\newcommand{\be}{\begin{equation}}
\newcommand{\ee}{\end{equation}}
\def\ba#1\ea{\begin{align}#1\end{align}}
\newcommand{\f}{\frac}

\usepackage{geometry}
\numberwithin{equation}{section}
\allowdisplaybreaks

\begin{document}

\newcommand{\hiduke}[1]{\hspace{\fill}{\small [{#1}]}}
\newcommand{\aff}[1]{${}^{#1}$}
\renewcommand{\thefootnote}{\fnsymbol{footnote}}

\begin{titlepage}
\begin{flushright}
{\footnotesize preprint SISSA 23/2020/FISI,\, MIT-CTP/5242}
\end{flushright}
\begin{center}
{\Large\bf
Chaos exponents of SYK traversable wormholes
}\\
\bigskip\bigskip
\bigskip\bigskip
{\large Tomoki Nosaka\footnote{\tt nosaka@yukawa.kyoto-u.ac.jp}}\aff{1,2}
{\large and Tokiro Numasawa\footnote{\tt tokiro.numasawa@mail.mcgill.ca}}\aff{3,4}\\
\bigskip\bigskip
\aff{1}: {\small
\it INFN Sezione di Trieste, Via Valerio 2, 34127 Trieste, Italy
}\\
\aff{2}: {\small
\it International School for Advanced Studies (SISSA), Via Bonomea 265, 34136 Trieste, Italy
}\\
\bigskip
\aff{3}: {\small
\it Center for Theoretical Physics, Massachusetts Institute of Technology\\
 Cambridge, MA 02139, USA
}\\
\aff{4}: {\small
\it Department of Physics, McGill University\\
3600 Rue University, Montreal, Quebec H3A 2T8, Canada 
}\\
\end{center}
\bigskip
\bigskip
\begin{abstract}
In this paper we study the chaos exponent, the exponential growth rate of the out-of-time-ordered four point functions, in a two coupled SYK models which exhibits a first order phase transition between the high temperature black hole phase and the low temperature gapped phase interpreted as a traversable wormhole.
% In this paper we study the chaos exponent, the exponential growth rate of the out-of-time-ordered four point functions, in a two coupled SYK models which exhibits a first order phase transition between the high temperature black hole phase and the low temperature gapped phase interpreted as the traversable wormhole.
We see that as the temperature decreases the chaos exponent exhibits a discontinuous fall-off from the value of order the universal bound $2\pi/\beta$ at the critical temperature of the phase transition, which is consistent with the expected relation between black holes and strong chaos.
Interestingly, the chaos exponent is small but non-zero even in the wormhole phase.
This is surprising but consistent with the observation on the decay rate of the two point function \cite{Qi:2020ian}, and we found the chaos exponent and the decay rate indeed obey the same temperature dependence in this regime.
% This is consistent with the observation on the decay rate of the two point function \cite{Qi:2020ian}, and we found the chaos exponent and the decay rate indeed obey the same temperature dependence in this regime.
We also studied the chaos exponent of a closely related model with single SYK term, and found that the chaos exponent of this model is always greater than that of the two coupled model in the entire parameter space.
% We also studied the chaos exponent of closely related model with a single SYK term, and found that the chaos exponent of this model is always greater than that of the two coupled model in the entire parameter space.
%
\end{abstract}

\bigskip\bigskip\bigskip

\end{titlepage}

\renewcommand{\thefootnote}{\arabic{footnote}}
\setcounter{footnote}{0}

\tableofcontents

% \newpage

\section{Introduction and Summary}

The SYK model \cite{PhysRevLett.70.3339,KitaevTalk} and its variants are useful toy models to study various aspects of quantum chaos and its gravity dual related to the black hole dynamics \cite{Hayden:2007cs,Sekino:2008he}.
The SYK model is a disordered quantum mechanical model where $N$ Majorana fermions are coupled by $q$-body interactions with random couplings $J_{i_1i_2\cdots i_q}$.
This model is simple enough to study directly at finite parameter regime.
The perturbative expansion of the correlation functions simplifies in the large $N$ limit, where only the melonic diagrams survives.
% The perturbative expansion of the correlation functions simplifies in the large $N$ limit, where only the melonic diagrams survives \cite{}.
As a result one can resum the perturbation series and write down the Schwinger-Dyson equation explicitly, with which one can study the thermalization property (decay of autocorrelation function) and the chaos exponent (out-of-time-ordered four point function) \cite{Larkin:1969aaa,Maldacena:2016hyu} directly at finite coupling.
We can also study the fluctuation properties of the spectrum and the eigenvectors \cite{1977RSPSA.356..375B,Bohigas:1983er} for finite $N$ by the exact diagonalization of the Hamiltonian as a $2^{N/2}\times 2^{N/2}$ matrix for each realization of $J_{i_1i_2\cdots i_q}$.
Despite these simplicities the dynamics of the SYK model is highly chaotic.
By solving the Schwinger-Dyson equation at the strong coupling limit we find that the chaos exponent saturates the universal upper bound \cite{Maldacena:2015waa} for $q\ge 4$.
The SYK model for $q\ge 4$ also enjoys the random matrix theory like level statistics \cite{Cotler:2016fpe,Gharibyan:2018jrp,Kobrin:2020xms,You:2016ldz,Garcia-Garcia:2016mno,Jia:2019orl}, which are distinctive criteria for the quantum chaos.
% The SYK model for $q\ge 4$ also enjoys the random matrix theory like level statistics \cite{Cotler:2016fpe,Gharibyan:2018jrp,Kobrin:2020xms,You:2016ldz,Garcia-Garcia:2016mno,Jia:2019orl}, which are distinctive criteria for the quantum chaos \cite{}.
For $q=2$ the SYK model is not chaotic.

Although there are no direct argument on the gravity dual of the SYK model, the SYK model has in common with $\text{AdS}_2$ spacetime at finite distance from the boundary (which is called nearly $\text{AdS}_2$ or N$\text{AdS}_2$) in the following sence \cite{Maldacena:2016upp}.
In the low energy limit the large $N$ SYK model enjoys an emergent symmetry corresponding to the reparametrization of the time variable.
This symmetry is spontaneously broken to $\text{SL}(2,\mathbb{R})$ by choosing a single solution to the Schwinger-Dyson equation (or equivalently, a single reparametrization), and is broken explicitly once we take into account the term of time derivative in the Schwinger-Dyson equation.
The low energy effective theory of the reparametrization modes is given by the Schwarzian action.
Whole these structures are same as what we encounter for the dynamics of the shape of the cutoff boundary of N$\text{AdS}_2$.
% The whole these structures are same as what we encounter for the dynamics of the shape of the cutoff boundary of the N$\text{AdS}_2$.

We can construct various models by using the SYK models as building blocks, often keeping the aformentioned tractabilities of the original SYK model and play.
Such models would play the role of experiments to understand various phenomena related to the quantum chaos.
For example, we can study the thermalization process under various quantum quench caused by SYK-like deformations \cite{Bhattacharya:2018fkq}, can introduce spatial directions \cite{Garcia-Garcia:2018pwt,Gu:2016oyy}, can realize a model with tunable chaoticity by coupling $\text{SYK}_{q\ge 4}$ with $\text{SYK}_2$ to compare different characterizations of the quantum chaos \cite{Garcia-Garcia:2017bkg,Nosaka:2018iat}, and so on.

In this paper we consider the model of two SYK systems (which we call L system and R system) coupled by a uniform quadratic interaction, where the random coupling of the two SYK systems are completely correlated.
This model was proposed \cite{Maldacena:2018lmt} to be dual to the two sided $\text{AdS}_2$ black hole or the global $\text{AdS}_2$ spacetime depending on the strength of the LR coupling, where in the latter situation can be interpreted as a traversable wormhole created by negative null energy due to the direct coupling between the two boundaries \cite{Gao:2016bin,Maldacena:2017axo}.
% This model was proposed \cite{Maldacena:2018lmt} to be dual to the two sided $\text{AdS}_2$ black hole or the global $\text{AdS}_2$ spacetime depending on the strength of the LR coupling, where in the latter situation can be interpreted as a traversable wormhole created by negative null energy due to the direct coupling between the two boundaries \cite{Gao:2016bin,Maldacena:2017axo}.
Indeed from the analysis of the large $N$ free energy this model was found to exhibit a first order phase transition between the low temperature gapped phase and the high temperature (or small LR coupling) large entropy phase, which correspond respectively to the traversable wormhole and the two-sided black hole \cite{Maldacena:2018lmt}.
% Indeed from the analysis of the large $N$ free energy this model was found to a first order phase transition between the low temperature gapped phase and the high temperature (or small LR coupling) large entropy phase, which corresponds respectively to the traversable wormhole and the two-sided black hole \cite{Maldacena:2018lmt}.

Note that the thermodynamic quantities which characterize the black hole phase and the Hawking-Page like phase transition mentioned above are not by themselves direct criteria for the quantum chaos.
% Note that the thermodynamic quantities which characterize the black hole phase and the Hawking-Page like phase transition mentioned above are not by themselves any direct criteria for the quantum chaos.
However, since various holographic arguments suggests that the system dual to a black hole spacetime is highly chaotic \cite{Festuccia:2006sa,Roberts:2014isa,Shenker:2013pqa,Shenker:2013yza,Shenker:2014cwa}, it would be natural to expect that the Hawking-Page like transition is indeed related to the quantum chaos \cite{Garcia-Garcia:2017bkg}.
% However, since the various holographic argumentes suggests that the system dual to a black hole spacetime is highly chaotic \cite{Festuccia:2006sa,Roberts:2014isa,Shenker:2013pqa,Shenker:2013yza,Shenker:2014cwa}, it would be natural to expect that the Hawking-Page like transition is indeed related to the quantum chaos \cite{Garcia-Garcia:2017bkg}.
This motivate us to study in detail how the chaotic property of a system varies around the phase transition.
As the phase transition takes place only in the large $N$ limit, in this paper we focus on the chaos exponent which we can study directly in the large $N$ limit by solving the real time Schwinger-Dyson equation, rather than the level statistics which would require a non-trivial extrapolation to address the large $N$ limit \cite{Garcia-Garcia:2019poj}.

Here we briefly summarize our results.
First of all, at high temperature far from the phase transition regime the chaos exponent of the two coupled model agrees with that for the single SYK model.
% First of all, at high temperature far from the phase transition regime the chaos exponent of the two coupled model agrees with that for the singe SYK model.
This is because the LR coupling is essentially a mass term and hence irrelevant in the high energy limit.
As the temperature is decreased the two results start to deviate; while the chaos exponent for the SYK model monotonically approaches the upper bound $\frac{2\pi}{\beta}$, for the two coupled model $\lambda_L/(2\pi/\beta)$ starts to decrease at some temperature above the phase transition temperature $T_c$.
% As the temperature is decreased the two results start to deviate; while the chaos exponent for the SYK model monotonically approaches the upper bound $\frac{2\pi}{\beta}$, for the two coupled model $\frac{\lambda_L}{(2\pi/\beta)}$ at some temperature above the phase transition temperature $T_c$.
This is in contrast to the behavior of the free energy whose temperature dependence in the black hole phase is almost same as that for the uncoupled case even near $T=T_c$.
% This is in contrast to the free energy whose temperature dependence in the black hole phase is almost same as that for the uncoupled case even near $T=T_c$.
At $T=T_c$ the chaos exponent jumps due to the interchange of the dominant configuration among the two distinctive solutions to the Schwinger-Dyson equation.

We have also studied the chaos exponent in the low-temperature wormhole phase in detail.
At first thought one may expect that the system is not chaotic at all in the wormhole phase.
% At first thought one may expect the system is not chaotic at all in the wormhole phase.
For example if we consider the decay rate of a large $N$ two point function, the decaying behavior in the black hole phase can be understood as the fact that the infalling mode does not come out from the black hole again \cite{Maldacena:2001kr}.
In the wormhole geometry, on the other hand, the signal from the right boundary reaches the left boundary and then reflects back to reach the right boundary again, which seems to suggest that the two point function continues to oscillate and the system never thermalizes.
This is indeed the case for example for the confining phase of the 4d $\text{U}(N)$ Yang-Mills theory on $S^3$ \cite{Festuccia:2006sa,Amado:2017kgr,Engelsoy:2020tsp}.
However, it was found \cite{Qi:2020ian} that the two point functions exhibit exponential decay even in the wormhole phase, although the decay rate is small so that the signal can traverse between the two boundaries many times before it disappears \cite{Plugge:2020wgc}.
We have found that the chaos exponent in the wormhole phase is also small but non-zero, which is consistent with the results in \cite{Qi:2020ian}.
We have further discovered a simple relation between the chaos exponent $\lambda_L$ and the energy gap $E_\text{gap}$ holds in the low temperature regime:
\begin{align}
\lambda_L\sim e^{-\frac{\frac{q}{2}-2}{2}\beta E_\text{gap}}
,\quad\quad (q=4).
\label{lowtemperatureformula}
\end{align}
We found this is true also when the LR coupling is sufficiently large so that the phase transition does not exist any more \cite{Maldacena:2018lmt,Garcia-Garcia:2019poj}, as long as the temperature is sufficiently low.
This formula is reminiscent of the low temperature limit of the chaos exponent for the weakly coupled matrix field theory $\lambda_L\sim \lambda^2 e^{-m\beta}$ \cite{Stanford:2015owe} where $m$ is the mass of the matrix scalar field and $\lambda$ is the 't Hooft coupling.

As a comparison, we have also studied the chaos exponent of the single SYK model with the same quadratic deformation \cite{Kourkoulou:2017zaj}.
Although this single sided model is similar to the two coupled model when the quadratic coupling is zero or large enough, it was found \cite{Nosaka:2019tcx} that this model does not exhibit phase transition in any parameter regime.
% Although this single sided model is similar to the two coupled model when the quadratic coupling is zero or large enough, this model does not exhibht phase transition in any parameter regime \cite{Nosaka:2019tcx}.
We have found that as we decrease the temperature the chaos exponent of the single sided model behaves qualitatively similarly to that of the two coupled model in the black hole phase, while the temperature where $\lambda_L/(2\pi/\beta)$ starts to decrease is slightly lower than that in the two coupled model.
At low temperature, the chaos exponent is significantly large compared with the two coupled model due to the absence of the phase transition.
We have also found that the chaos exponent of the single sided model also obeys the same formula \eqref{lowtemperatureformula} when the energy gap is significant compared with the thermal fluctuations (i.e.~when the spectral function shows well separated peaks).
% At low temparature, while the chaos exponent is significantly large compared with the two coupled model due to the absense of the phase transition, we found that the chaos exponent of the single sided model also obeys the same formula \eqref{lowtemperatureformula} when the energy gap is significant compared with the thermal fluctuations (i.e.~when the spectral function shows well separated peaks).

This paper is organized as follows.
In section \ref{sec_models}, we introduce the models we will study: the two coupled model \cite{Maldacena:2018lmt} and the single sided model \cite{Kourkoulou:2017zaj}, and review their large $N$ effective descriptions by the bilocal fields ($G\Sigma$ formalism).
% We also briefly comment on the phase structures of these models which were obtained respectively in \cite{Maldacena:2018lmt} and \cite{Nosaka:2019tcx}.
In section \ref{sec_chaosexponent} we continue the $G\Sigma$ formalism to the Lorentzian real time to study the OTOC and the chaos exponent of the two models.
In section \ref{sec_results}, after reviewing the phase structures of the two models, we display the results of the real time numerical analysis.
% In section \ref{sec_results} we display the results of the numerical analysis.
In particular, we display the chaos exponent of the two models in the whole parameter regime including the vicinity of the phase transition point in the case of the two coupled model.
We observe an interesting similarity between the critical behavior of the chaos exponent and that of the specific heat.
We also argue an analytic derivation of the chaos exponent in the low temperature regime.
In section \ref{sec_Discussion} we discuss implications of our results and propose future directions.

% We also try to explain the difference between the quantum chaotic property of the two models by proposing a generalization of the two coupled model which unify the two models.

Although in section \ref{sec_results} we focus on the cases where the two models are built from the SYK model with $q=4$, in appendix \ref{sec_q6q8} we also display some results for the two models built from $\text{SYK}_{q=6}$ or $\text{SYK}_{q=8}$.
% Although in section \ref{sec_results} we focus on the cases where the two models are built from the SYK model with $q=4$, in appendix \ref{sec_q6q8} we display some results for the two models built from $\text{SYK}_{q=6}$ or $\text{SYK}_{q=8}$.

%%%%%%%%%%%%%%%%%%%%
%%%%%%%%%%%%%%%%%%%%

\section{Models}
\label{sec_models}
In this paper we consider the following two models.
The first model consists of the two SYK systems with $N/2$ fermions per each side,\footnote{
Here we put $N/2$, not $N$, fermions per each side, which is a different notation from \cite{Maldacena:2018lmt}.}
coupled with a simple quadratic interaction:
% The first model is the two SYK system with $N/2$ fermions per each side,
coupled with a simple quadratic interaction:
\begin{align}
H_\text{two}&=i^{\frac{q}{2}}\sum_{i_1<i_2<\cdots<i_q}^{\frac{N}{2}}J_{i_1i_2\cdots i_q}(\psi_{i_1}^L\psi_{i_2}^L\cdots\psi_{i_q}^L+(-1)^{\frac{q}{2}}\psi_{i_1}^R\psi_{i_2}^R\cdots\psi_{i_q}^R)+i\mu\sum_{i=1}^{\frac{N}{2}}\psi_i^L\psi_i^R,
\label{Htwocoupled}
\end{align}
where $\{\psi_i^a,\psi_j^b\}=\delta^{ab}\delta_{ij}$ and
\begin{align}
\langle J_{i_1i_2\cdots i_q}\rangle=0,\quad
\langle (J_{i_1i_2\cdots i_q})^2\rangle=\frac{{\cal J}^2\cdot 2^{q-1}(q-1)!}{q(N/2)^{q-1}}.\quad (\text{no sum over }i_1,i_2,\cdots,i_q)
\end{align}
% with
% \begin{align}
% \xi=\frac{{\cal J}^2\cdot 2^{q-1}(q-1)!}{q}.
% \label{xi}
% \end{align}
The second model is the single SYK system with $N$ fermions with the same mass deformation:
\begin{align}
H_\text{single}&=i^{\frac{q}{2}}\sum_{i_1<i_2<\cdots<i_q}^N J'_{i_1i_2\cdots i_q}\chi_{i_1}\chi_{i_2}\cdots\chi_{i_q}+i\mu\sum_{i=1}^{\frac{N}{2}}\chi_{2i-1}\chi_{2i},
\label{Hsinglesided}
\end{align}
where $\{\chi_i,\chi_j\}=\delta_{ij}$ and
\begin{align}
\langle J'_{i_1i_2\cdots i_q}\rangle=0,\quad
\langle (J'_{i_1i_2\cdots i_q})^2\rangle=\frac{{\cal J}^2\cdot 2^{q-1}(q-1)!}{q N^{q-1}}.\quad (\text{no sum over }i_1,i_2,\cdots,i_q)
\end{align}

In the following sections we shall call these models respectively as ``two coupled model'' and ``single sided model''.
These models show interesting thermodynamical properties \cite{Maldacena:2018lmt,Nosaka:2019tcx}.
We will review some of these properties in section \ref{sec_results} which are particularly relevant to the study of the chaos exponent.

The two coupled model \eqref{Htwocoupled} has a $\mathbb{Z}_4$ symmetry \cite{Garcia-Garcia:2019poj} that is generated by 
\be
\psi_i^L \to \psi_i^R, \qquad \psi_i^R \to -\psi_i^L. \label{Z4symmetry}
\ee

%%%%%%%%%%%%%%%%%%%%
%%%%%%%%%%%%%%%%%%%%

\subsection{$G\Sigma$ formalism}
In these models we can rewrite the partition function into an expression without disorder by introducing new variables of bi-local fields.
% In these models we can rewrite the partition function without disorder by introducing new variables of bi-local fields.
This formalism turns out to be useful for analyzing the system in the large $N$ limit.

\subsubsection{Two coupled model}
First let us consider the two coupled model \eqref{Htwocoupled}, whose partition function is defined as
\begin{align}
Z_\text{two}(\beta)=\biggl\langle
\int{\cal D}\psi_i^a(u)
\exp\biggl[-\int du\Bigl(\frac{1}{2}\sum_{a=L,R}\sum_{i=1}^{N/2}\psi_i^a\partial_u\psi_i^a+H_\text{two}\Bigr)
\biggr]
\biggr\rangle_{J_{i_1i_2\cdots i_q}}.
\label{<ZMQ>_J}
\end{align}
We can perform the disorder average first by writing it explicitly as the Gaussian integration over $J_{i_1i_2\cdots i_q}$, to obtain
% Here we can perform the disorder average first by writingn it explicitly as the Gaussian integration over $J_{i_1i_2\cdots i_q}$, to obtain
\begin{align}
Z_\text{two}(\beta)&=
\Bigl(\frac{\pi \mathcal{J}^2 2^{q-1} (q-1)!}{q (N/2)^{q-1}}\Bigr)^{
-\frac{1}{2}
{{N/2}\choose{q}}
}
\int dJ_{i_1i_2\cdots i_q}\exp\biggl[-\frac{(N/2)^{q-1} q}{2\mathcal{J}^2 2^{q-1}(q-1)!}\sum_{i_1<i_2<\cdots<i_q}J_{i_1i_2\cdots i_q}^2\biggr]\nonumber \\
&\quad \int {\cal D}\psi_i^a
\exp\biggl[-\int du\Bigl(\frac{1}{2}\sum_{a,i}\psi_i^a\partial_u\psi_i^a+H_\text{two}\Bigr)\biggr]\nonumber \\
&=\int {\cal D}\psi_i^a
\exp\biggl[\frac{i^q\mathcal{J}^2 2^{q-1} (q-1)!}{2q (N/2)^{q-1}}\sum_{i_1<i_2<\cdots<i_q}\Bigl
(\int du(\psi_{i_1}^L\psi_{i_2}^L\cdots\psi_{i_q}^L+(-1)^{\frac{q}{2}}\psi_{i_1}^R\psi_{i_2}^R\cdots\psi_{i_q}^R)\Bigr)^2\nonumber \\
&\quad -\int du\Bigl(\frac{1}{2}\sum_{a,i}\psi_i^a\partial_u\psi_i^a+i\mu\sum_i\psi_i^L\psi_i^R\Bigr)
\biggr].
\end{align}
If we define the bi-local fields
\begin{align}
G_{ab}(u,u')=\frac{1}{(N/2)}\sum_{i=1}^{N/2}\psi_i^a(u)\psi_i^b(u'), \label{introduceG_inMQmodel}
\end{align}
the last expression can be written as
\begin{align}
Z_\text{two}(\beta)
&=\int {\cal D}\psi_i^a
\exp\biggl[
-\frac{1}{2}\sum_{a,i}\int du\psi_i^a\partial_u\psi_i^a\nonumber \\
&\quad+\sum_{a,b}\int dudu'\Bigl[\frac{N \mathcal{J}^2 2^{q-1} }{4 q^2 }s_{ab}G_{ab}(u,u')-\frac{i\mu}{4}\epsilon_{ab}G_{ab}(u,u')\delta(u-u')\Bigr]
\biggr],
\label{ZMQwithjustDpsi}
\end{align}
where we have defined $G_{ab}(u,u')=\frac{1}{(N/2)}\sum_{i=1}^{N/2}\psi_i^a(u)\psi_i^b(u')$ and also the following constant matrices
\begin{align}
s_{ab}=\begin{pmatrix}
1&(-1)^{\frac{q}{2}}\\
(-1)^{\frac{q}{2}}&1
\end{pmatrix},\quad
\epsilon_{ab}=\begin{pmatrix}
0&1\\
-1&0
\end{pmatrix}.
\end{align}

If we further introduce Lagrange multiplier bilocal field $\Sigma_{ab}(u,u')$
\begin{align}
1&=\int {\cal D}G_{ab}(u,u')\prod_{u,u'}\delta\Bigl(G_{ab}(u,u')-\frac{1}{(N/2)}\sum_i\psi_i^a(u)\psi_i^b(u')\Bigr)\nonumber \\
&=\int {\cal D}G_{ab}{\cal D}\Sigma_{ab}e^{-\frac{N}{4}\int du du'(\Sigma_{ab}(u,u')-i\mu\epsilon_{ab}\delta(u-u'))(G_{ab}(u,u')-\frac{1}{(N/2)}\sum_i\psi_i^a(u)\psi_i^b(u'))},
\end{align}
to regard $G_{ab}(u,u')$ as an independent set of the integration variables from $\psi_i^a(u)$, we can perform the inntegration over $\psi_i^a$ in \eqref{ZMQwithjustDpsi} explicitly as\footnote{
% to regard $G_{ab}(u,u')$ as an independent set of the integration variables from $\psi_i^a(u)$, we can perform the inntegration over $\psi_i^a$ explicitly as\footnote{
In this paper we do not impose anti-symmetry property on $G_{ab}(u,u')$ in the $G\Sigma$ formalism, and treat $G_{ab}(u,u')$ as four independent bilocal fields without any restriction on the $u,u'$-dependence.
This approach allows, when we discuss variational problems, us to treat all of $\delta G_{ab}(u,u')$ and $\delta\Sigma_{ab}(u,u')$ as independent variational modes.
Also note that here we have introduced the auxiliary field $\Sigma_{ab}(u,u')$ with a shift by a fixed configuration $-i\mu\epsilon_{ab}\delta(u-u')$ for later convenience in section \ref{sec_realtimeSDeqMQ}.
}
\begin{align}
&\int {\cal D}\psi_i^a(u)\exp\Biggl[
\frac{1}{2}\int dudu'
\begin{pmatrix}
\psi_i^L(u)&
\psi_i^R(u)
\end{pmatrix}\nonumber \\
&\quad
\begin{pmatrix}
-\delta(u-u')\partial_{u'}+\frac{\Sigma_{LL}(u,u')-\Sigma_{LL}(u',u)}{2}&
\frac{\Sigma_{LR}(u,u')-\Sigma_{RL}(u',u)}{2}-i\mu\delta(u-u')\\
\frac{\Sigma_{RL}(u,u')-\Sigma_{LR}(u',u)}{2}+i\mu\delta(u-u')&
-\delta(u-u')\partial_{u'}+\frac{\Sigma_{RR}(u,u')-\Sigma_{RR}(u',u)}{2}&
\end{pmatrix}
\begin{pmatrix}
\psi_i^L(u')\\
\psi_i^R(u')
\end{pmatrix}
\Biggr]\nonumber \\
&=\text{Pf}
\Biggl[
\begin{pmatrix}
-\delta(u-u')\partial_{u'}+\frac{\Sigma_{LL}(u,u')-\Sigma_{LL}(u',u)}{2}&
\frac{\Sigma_{LR}(u,u')-\Sigma_{RL}(u',u)}{2}-i\mu\delta(u-u')\\
\frac{\Sigma_{RL}(u,u')-\Sigma_{LR}(u',u)}{2}+i\mu\delta(u-u')&
-\delta(u-u')\partial_{u'}+\frac{\Sigma_{RR}(u,u')-\Sigma_{RR}(u',u)}{2}&
\end{pmatrix}
\Biggr]^{\frac{N}{2}}.
\end{align}
As a result we obtain \cite{Maldacena:2018lmt}
\begin{align}
Z_\text{two}(\beta)=\int {\cal D}G_{ab}{\cal D}\Sigma_{ab}e^{-NS_\text{two}}
\label{MQGSigmafinal}
\end{align}
with
\begin{align}
S_\text{two}&=-\frac{1}{4}\log\det
\begin{pmatrix}
-\delta(u-u')\partial_{u'}+\frac{\Sigma_{LL}(u,u')-\Sigma_{LL}(u',u)}{2}&
\frac{\Sigma_{LR}(u,u')-\Sigma_{RL}(u',u)}{2}-i\mu\delta(u-u')\\
\frac{\Sigma_{RL}(u,u')-\Sigma_{LR}(u',u)}{2}+i\mu\delta(u-u')&
-\delta(u-u')\partial_{u'}+\frac{\Sigma_{RR}(u,u')-\Sigma_{RR}(u',u)}{2}&
\end{pmatrix}
\nonumber \\
&\quad +\sum_{a,b}\frac{1}{4}\int du du'\Bigl(\Sigma_{ab}(u,u')G_{ab}(u,u')
-\frac{\mathcal{J}^2}{2 q^2}s_{ab}[2G_{ab}(u,u')]^q
\Bigr).
\end{align}
In the large $N$ limit, the partition function is dominated by the contribution from the saddle point configurations, which are the solutions of the following Schwinger-Dyson equations
\begin{align}
&\frac{\delta S_{\text{two}}}{\delta\Sigma_{ab}(u,u')}=0\quad \leftrightarrow \nonumber \\
&G_{ab}(u,u')\nonumber \\
&=
-\begin{pmatrix}
-\delta(u-u'')\partial_{u''}+\frac{\Sigma_{LL}(u,u'')-\Sigma_{LL}(u'',u)}{2}&\frac{\Sigma_{LR}(u,u'')-\Sigma_{RL}(u'',u)}{2}-i\mu\delta(u-u'')\\
\frac{\Sigma_{RL}(u,u'')-\Sigma_{LR}(u'',u)}{2}+i\mu\delta(u-u'')           &-\delta(u-u'')\partial_{u''}+\frac{\Sigma_{RR}(u,u'')-\Sigma_{RR}(u'',u)}{2}
\end{pmatrix}^{-1}_{ab}(u,u'),\label{EuclideanEoMoftwocoupledmodel1} \\
&\frac{\delta S_{\text{two}}}{\delta G_{ab}(u,u')}=0\quad \leftrightarrow\quad
\Sigma_{ab}(u,u')=\f{\mathcal{J}^2}{q} s_{ab}[2G_{ab}(u,u')]^{q-1}.
\label{EuclideanEoMoftwocoupledmodel2}
\end{align}
Note that in the $G\Sigma$ formalism we do not impose the symmetry property $G_{ab}(u,u')=-G_{ba}(u',u)$ which follows from the original way we have introduced them \eqref{introduceG_inMQmodel}, and treat each of $G_{ab}(u,u')$, $\Sigma_{ab}(u,u')$ as independent bilocal fields.
This symmetry property, however, must be recovered once we integrate out the auxiliary bilocal fields $\Sigma_{ab}(u,u')$.
Indeed, in the first line of the equation of motion \eqref{EuclideanEoMoftwocoupledmodel1} since the right-hand side is anti-symmetric under $(u,a)\leftrightarrow (u',b)$ it follows that a saddle solution satisfies
\begin{align}
G_{ab}(u,u')=-G_{ba}(u',u),\quad
\Sigma_{ab}(u,u')=-\Sigma_{ba}(u',u).
\label{antisymmetryfollowsfromEoM_MQ}
\end{align}
Here we have also recalled the second line of \eqref{EuclideanEoMoftwocoupledmodel2} to obtain the latter result.
Taking into account these relations we can rewrite the first line of the equations of motion $\delta S_\text{two}/\delta \Sigma_{ab}(u,u')=0$ simpliy as
% Taking into account these relations we can rewrite the first line of the equations of motion $\frac{\delta S_\text{two}}{\delta \Sigma_{ab}(u,u')}=0$ simpliy as
\begin{align}
\partial_uG_{ab}(u,u')-\sum_c\Bigl(-i\mu\epsilon_{ac}G_{cb}(u,u')+\int du''\Sigma_{ac}(u,u'')G_{cb}(u'',u')\Bigr)=\delta_{ab}\delta(u-u'),
% \partial_uG_{ab}(u,u')-\sum_c\int du''\Sigma_{ac}(u,u'')G_{cb}(u'',u')=\delta_{ab}\delta(u-u'),
\label{MQEoM1usedforrealtimeSD}
\end{align}
which we will use to derive the real time continuation in the next section.

Lastly, note that the solution we are interested in is the one which we can indeed interpret as the two point function of $\psi_i^a(u)$ at finite temperature\footnote{
In this paper we adopt the annealed average
\begin{align}
\langle{\cal O}\rangle \equiv \frac{\langle \int {\cal D}\psi_i^a {\cal O}e^{-\int du(\frac{1}{2}\psi_i^a\partial_u\psi_i^a+H)}\rangle_{J_{i_1i_2\cdots i_q}}}
{\langle \int {\cal D}\psi_i^a e^{-\int du(\frac{1}{2}\psi_i^a\partial_u\psi_i^a+H)}\rangle_{J_{i_1i_2\cdots i_q}}}
\end{align}
so that we can treat the random coupling in the same way as a constant field and integrate them in the partition function \eqref{<ZMQ>_J}.
% so that we can treat the random coupling in the same way as a constant field and integrate for the partition function \eqref{<ZMQ>_J}.
Although this is different from the quenched average
\begin{align}
\langle{\cal O}\rangle_{\text{quenched}} \equiv 
\biggl\langle
\frac{\int {\cal D}\psi_i^a {\cal O}e^{-\int du(\frac{1}{2}\psi_i^a\partial_u\psi_i^a+H)}}
{\int {\cal D}\psi_i^a e^{-\int du(\frac{1}{2}\psi_i^a\partial_u\psi_i^a+H)}}
\biggr\rangle_{J_{i_1i_2\cdots i_q}}
\end{align}
which was originally adopted for finite $N$, the two results agrees in the large $N$ limit.
}
\begin{align}
G_{ab}(u,u')&=\frac{1}{(N/2)}\sum_{i=1}^{\frac{N}{2}}\langle{\cal T}\psi_i^a(u)\psi_i^b(u')\rangle\nonumber \\
&=\begin{cases}
\frac{1}{(N/2)}\sum_{i=1}^{\frac{N}{2}}\langle \text{tr}e^{{\widehat H}u}\psi_i^a(0)e^{-{\widehat H}(u-u')}\psi_i^b(0)e^{-{\widehat H}u'}e^{-\beta{\widehat H}}\rangle_{J_{i_1i_2\cdots i_q}}\quad (\text{Re}[u]>\text{Re}[u'])\\
-\frac{1}{(N/2)}\sum_{i=1}^{\frac{N}{2}}\langle \text{tr}e^{{\widehat H}u'}\psi_i^b(0)e^{-{\widehat H}(u'-u)}\psi_i^a(0)e^{-{\widehat H}u}e^{-\beta{\widehat H}}\rangle_{J_{i_1i_2\cdots i_q}}\quad (\text{Re}[u]<\text{Re}[u'])
\end{cases},
\end{align}
which obeys the following properties:
\begin{align}
G_{ab}(u,u')^*&=-G_{ab}(-u^*,-{u'}^*),\label{MQ_physicalansatz2} \\
G_{ab}(u+\beta,u')&=-G_{ab}(u,u')\quad \text{(if }\text{Re}[u]<\text{Re}[u']<\text{Re}[u+\beta]\text{)}.\label{MQ_physicalansatz3}
\end{align}
Hence when we solve the equations of motions we should further impose these properties as ansatze, although they are neither imposed on the integration measure ${\cal D}G_{ab}(u,u')$ in \eqref{MQGSigmafinal} nor consequences of the equations of motion \eqref{MQEoM1usedforrealtimeSD},\eqref{EuclideanEoMoftwocoupledmodel2}.

\subsubsection{Single sided model}
\label{sec_GSigmaformalism_singlesided}
One can do the same rewriting for the single sided model \eqref{Hsinglesided} by introducing\footnote{
Note that \eqref{introduceGinKM} is redundant; one may also proceed by introducing only two bi-local fields $G(u,u')=\frac{1}{N}\sum_{i=1}^N\chi_i(u)\chi_i(u')$ and $G_\text{off}(u,u')=\frac{1}{N/2}\sum_{i=1}^{N/2}\chi_{2i-1}(u)\chi_{2i}(u')$, as in \cite{Nosaka:2019tcx}.
Nevertheless we found it more convenient to introduce the four bi-local fields $G_{ab}(u,u')$ and the subsequent four auxiliary bilocal fields $\Sigma_{ab}(u,u')$ as they allow a completely parallel treatment of the one-loop determinant contribution and the $\Sigma G$ bilinear term when we discuss the variations of the action $S_\text{single}$ to derive the equations of motion \eqref{EoMinusinglesided1},\eqref{EoMinusinglesided2} and the ladder kernel for the four point functions \eqref{KMladder1}.
% Nevertheless we found it more convenient to introduce the four bi-local fields $G_{ab}(u,u')$ and the subsequent four auxiliary bilocal fields $\Sigma_{ab}(u,u')$ as they allow a completely parallell treatment of the one-loop determinant contribution and the $\Sigma G$ bilinear term when we discuss the variations of the action $S_\text{single}$.
}
\begin{align}
G_{LL}(u,u')&=\frac{1}{N/2}\sum_{i=1}^{\frac{N}{2}}\chi_{2i-1}(u)\chi_{2i-1}(u'),\quad
G_{LR}(u,u')=\frac{1}{N/2}\sum_{i=1}^{\frac{N}{2}}\chi_{2i-1}(u)\chi_{2i}(u'),\nonumber \\
G_{RL}(u,u')&=\frac{1}{N/2}\sum_{i=1}^{\frac{N}{2}}\chi_{2i}(u)\chi_{2i-1}(u'),\quad
G_{RR}(u,u')=\frac{1}{N/2}\sum_{i=1}^{\frac{N}{2}}\chi_{2i}(u)\chi_{2i}(u'),
\label{introduceGinKM}
\end{align}
as
\begin{align}
Z_\text{single}&=
\biggl\langle
\int{\cal D}\chi_i(u)\exp\biggl[-\int du\Bigl(
\frac{1}{2}\sum_{i=1}^N\chi_i\partial_u\chi_i\text{+}H_\text{single}
\Bigr)\biggr]
\biggr\rangle_{J'_{i_1i_2\cdots i_q}}\nonumber \\
&=\int {\cal D}G_{ab}{\cal D}\Sigma_{ab}e^{-NS_\text{single}},
\label{redundantGSigmaformalismKM1}
\end{align}
with
\begin{align}
S_\text{single}&=-\frac{1}{4}\log\det\begin{pmatrix}
-\delta(u-u')\partial_{u'}+\frac{\Sigma_{LL}(u,u')-\Sigma_{LL}(u',u)}{2}&\frac{\Sigma_{LR}(u,u')-\Sigma_{RL}(u',u)}{2}-i\mu\delta(u-u')\\
\frac{\Sigma_{RL}(u,u')-\Sigma_{LR}(u',u)}{2}+i\mu\delta(u-u')&-\delta(u-u')\partial_{u'}+\frac{\Sigma_{RR}(u,u')-\Sigma_{RR}(u',u)}{2}
\end{pmatrix}\nonumber \\
&\quad +\frac{1}{4}\int dudu'\Bigl(\sum_{a,b}\Sigma_{ab}(u,u')G_{ab}(u,u')
-\frac{\mathcal{J}^2 2^q}{q^2}\Bigl(\frac{G_{LL}(u,u')+G_{RR}(u,u')}{2}\Bigr)^q
\Bigr).
\label{redundantGSigmaformalismKM2}
\end{align}
The equations of motion are
\begin{align}
&\frac{\delta S_\text{single}}{\delta \Sigma_{ab}}=0\quad\leftrightarrow\nonumber \\
&G_{ab}(u,u')\nonumber \\
&=-\begin{pmatrix}
-\delta(u-u')\partial_{u'}+\frac{\Sigma_{LL}(u,u')-\Sigma_{LL}(u',u)}{2}&\frac{\Sigma_{LR}(u,u')-\Sigma_{RL}(u',u)}{2}-i\mu\delta(u-u')\\
\frac{\Sigma_{RL}(u,u')-\Sigma_{LR}(u',u)}{2}+i\mu\delta(u-u')&-\delta(u-u')\partial_{u'}+\frac{\Sigma_{RR}(u,u')-\Sigma_{RR}(u',u)}{2}
\end{pmatrix}^{-1}_{ab}(u,u'),\label{EoMinusinglesided1} \\
&\frac{\delta S_\text{single}}{\delta G_{ab}}=0\quad\leftrightarrow\nonumber \\
&\Sigma_{LL}(u,u')=
\Sigma_{RR}(u,u')=
\frac{\mathcal{J}^2}{q}\Bigl(G_{LL}(u,u')+G_{RR}(u,u')\Bigr)^{q-1},\quad
\Sigma_{LR}(u,u')=\Sigma_{RL}(u,u')=0.
\label{EoMinusinglesided2}
\end{align}
Similarly to the case of the two coupled model, from the equations of motion we immediately find
% Similarly to the case of two coupled model, from the equations of motion we immediately find
\begin{align}
G_{ab}(u,u')&=-G_{ba}(u',u),\quad
\Sigma_{ab}=-\Sigma_{ba}(u',u),\nonumber \\
\Sigma_{LL}(u,u')&=\Sigma_{RR}(u,u'),\quad
\Sigma_{LR}(u,u')=\Sigma_{RL}(u,u')=0.
\label{symmetrypropertyfromEoM_KM}
\end{align}
Using the last three equations of \eqref{symmetrypropertyfromEoM_KM}, we can simplify the first line of the equaitons of motion \eqref{EoMinusinglesided1} as
\begin{align}
\partial_uG_{LL}(u,u')-\int du''\Sigma_{LL}(u,u'')G_{LL}(u'',u')+i\mu G_{RL}(u,u')&=\delta(u-u'),\nonumber \\
\partial_uG_{LR}(u,u')-\int du''\Sigma_{LL}(u,u'')G_{LR}(u'',u')+i\mu G_{RR}(u,u')&=0,\nonumber \\
\partial_uG_{RL}(u,u')-\int du''\Sigma_{LL}(u,u'')G_{RL}(u'',u')-i\mu G_{LL}(u,u')&=0,\nonumber \\
\partial_uG_{RR}(u,u')-\int du''\Sigma_{LL}(u,u'')G_{RR}(u'',u')-i\mu G_{LR}(u,u')&=\delta(u-u').
\end{align}

Remarkably, in the single sided model we can solve these equations of motion explicitly with respect to $G_{LR}$, $G_{RL}$
\begin{align}
G_{LR}(u,u')&=-\frac{i}{\mu}(\partial_uG_{RR}(u,u')-\int du''\Sigma_{LL}(u,u'')G_{RR}(u'',u')-\delta(u-u')),\nonumber \\
G_{RL}(u,u')&=\frac{i}{\mu}(\partial_uG_{LL}(u,u')-\int du''\Sigma_{LL}(u,u'')G_{LL}(u'',u')-\delta(u-u')),
\end{align}
with which we obtain a closed set of equations only for $G_{LL},G_{RR},\Sigma_{LL}$:
\begin{align}
&\frac{i}{\mu}\int du''(\delta(u-u'')\partial_{u''}-\Sigma_{LL}(u,u''))\Bigl[\int du'''(\delta(u''-u''')\partial_{u'''}-\Sigma_{LL}(u'',u'''))G_{LL}(u''',u')\nonumber \\
&\quad\quad\quad\quad\quad\quad\quad\quad\quad\quad\quad\quad\quad\quad\quad\quad -\delta(u''-u')\Bigr]-i\mu G_{LL}(u,u')=0,\nonumber \\
&\frac{i}{\mu}\int du''(\delta(u-u'')\partial_{u''}-\Sigma_{LL}(u,u''))\Bigl[\int du'''(\delta(u''-u''')\partial_{u'''}-\Sigma_{LL}(u'',u'''))G_{RR}(u''',u')\nonumber \\
&\quad\quad\quad\quad\quad\quad\quad\quad\quad\quad\quad\quad\quad\quad\quad\quad -\delta(u''-u')\Bigr]-i\mu G_{RR}(u,u')=0,
\end{align}
together with the first equation of \eqref{EoMinusinglesided2}.

Lastly, as in the case of the two coupled model, our interest is restricted to the solutions which satisfies the following additional properties
\begin{align}
G_{ab}(u,u')^*&=-G_{ab}(-u^*,-{u'}^*),\label{KM_physicalansatz2} \\
G_{ab}(u+\beta,u')&=-G_{ab}(u,u')\quad \text{(if }\text{Re}[u]<\text{Re}[u']<\text{Re}[u+\beta]\text{)},\label{KM_physicalansatz3}
\end{align}
such that we can interpret $G_{ab}(u,u')$ as
\begin{align}
G_{LL}(u,u')&=\frac{1}{N/2}\sum_{i=1}^{\frac{N}{2}}\langle{\cal T}\chi_{2i-1}(u)\chi_{2i-1}(u')\rangle,\quad
G_{LR}(u,u')=\frac{1}{N/2}\sum_{i=1}^{\frac{N}{2}}\langle{\cal T}\chi_{2i-1}(u)\chi_{2i}(u')\rangle,\nonumber \\
G_{RL}(u,u')&=\frac{1}{N/2}\sum_{i=1}^{\frac{N}{2}}\langle{\cal T}\chi_{2i}(u)\chi_{2i-1}(u')\rangle,\quad
G_{RR}(u,u')=\frac{1}{N/2}\sum_{i=1}^{\frac{N}{2}}\langle{\cal T}\chi_{2i}(u)\chi_{2i}(u')\rangle.
\end{align}

%%%%%%%%%%%%%%%%%%%%
%%%%%%%%%%%%%%%%%%%%

\section{Chaos exponent}
\label{sec_chaosexponent}

\subsection{Two coupled model}

The quantum chaoticity of the two coupled model can be characterized by the following four point functions called the out-of-time-ordered correlators (OTOC)
\begin{align}
&\frac{1}{(N/2)^2}\sum_{i,j}\Bigl\langle
\psi_i^a\Bigl(\frac{3\beta}{4}+it_1\Bigr)
\psi_i^b\Bigl(\frac{\beta}{4}+it_2\Bigr)
\psi_j^c\Bigl(\frac{\beta}{2}\Bigr)
\psi_j^d(0)
\Bigr\rangle_{J_{i_1i_2\cdots i_1}}
\nonumber \\
&=
\frac{1}{(N/2)^2}\sum_{i,j}\Bigl\langle
\psi_i^a\Bigl(\frac{3\beta}{4}+it_1\Bigr)
\psi_i^b\Bigl(\frac{\beta}{4}+it_2\Bigr)
\Bigr\rangle
\Bigl\langle
\psi_j^c\Bigl(\frac{\beta}{2}\Bigr)
\psi_j^d(0)
\Bigr\rangle_{J_{i_1i_2\cdots i_1}}
+\frac{1}{N/2}{\cal F}_{abcd}(t_1,t_2).
\label{4ptfcn}
\end{align}
When the system is chaotic, the connected part ${\cal F}_{abcd}(t_1,t_2)$ of an OTOC behaves at late time as
\begin{align}
{\cal F}_{abcd}(t_1,t_2)\sim e^{\frac{\lambda_L(t_1+t_2)}{2}}
\end{align}
where $\lambda_L$ is the chaos exponent which quantify the chaoticity of the system.
Since the left-hand side of \eqref{4ptfcn} inside $\langle \cdots\rangle_{J_{i_1i_2\cdots i_q}}$ is written in terms of the bi-local field \eqref{introduceG_inMQmodel} as $G_{ab}(\beta/2+i(t_1-t_2))G_{cd}(\beta/2)$, we can calculate this four point function as well as the connected part in the large $N$ limit within the $G\Sigma$ formalism, with the Euclidean time variables continued appropriately.
Below we first demonstrate the analytic continuation and derive the real time Schwinger-Dyson equations \eqref{MQrealtimeSDfinal1},\eqref{MQrealtimeSDfinal2}, and then explain how to obtain the chaos exponent from the real time two point functions.

\subsubsection{Real time Schwinger-Dyson equation}
\label{sec_realtimeSDeqMQ}

\begin{figure}
\begin{center}
\includegraphics[width=4cm]{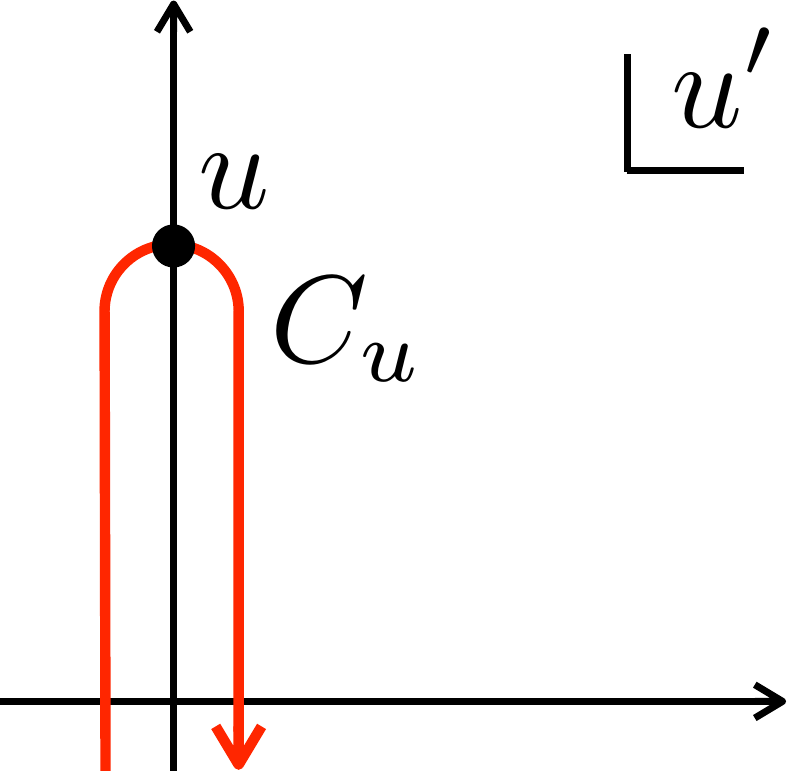}\quad\quad\quad\quad\quad\quad\quad\quad
\includegraphics[width=6cm]{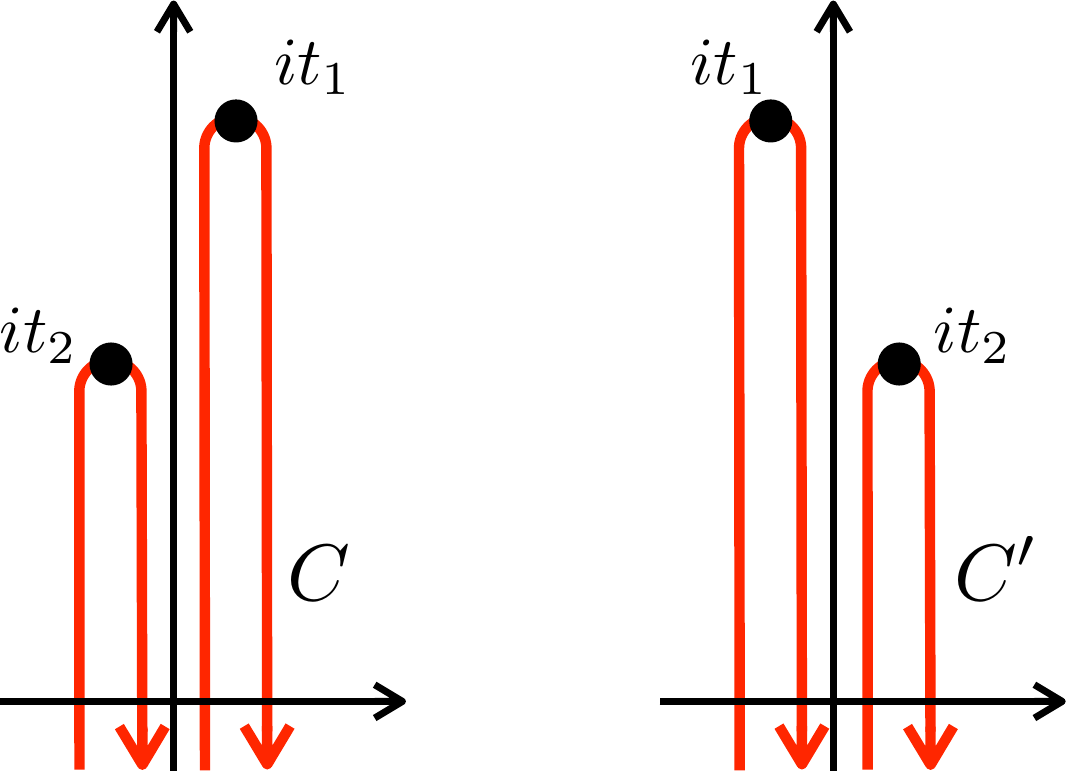}
\end{center}
\caption{
Left: Keldysh contour for the insertion of single operator.
Right: Contours $C,C'$ used in \eqref{continuation1},\eqref{continuation2}.
}
\label{fig_200921Keldysh}
\end{figure}

Our starting point is the Schwinger-Dyson equations \eqref{MQEoM1usedforrealtimeSD},\eqref{EuclideanEoMoftwocoupledmodel2} together with the symmetry properties \eqref{antisymmetryfollowsfromEoM_MQ} and ansatz \eqref{MQ_physicalansatz2},\eqref{MQ_physicalansatz3}.
To obtain the Schwinger-Dyson equation in Lorentzian time $t$, we continue $u$ to $u=it$.
There are two different ways to continue $G_{ab}(u_1,u_2)$ when $\text{Re}[u_1]=\text{Re}[u_2]$ corresponding to the ordering in the operator formalism, which define the following two independent components
\begin{align}
G_{ab}^>(t_1,t_2)&=-iG_{ab}(it_1^-,it_2^+)=-i\lim_{\epsilon\rightarrow +0}G_{ab}(\epsilon+it_1,-\epsilon+it_2),\nonumber \\
G_{ab}^<(t_1,t_2)&=-iG_{ab}(it_1^+,it_2^-)=-i\lim_{\epsilon\rightarrow +0}G_{ab}(-\epsilon+it_1,\epsilon+it_2).  \label{definitionof><}
\end{align}
When an operator is inserted at some $u$ the forward/backward time evolution around $u$ does not cancel, which result in the Keldysh contour (see figure \ref{fig_200921Keldysh}) in the path integral formalism.
As a result we obtain the following two real time equations from the continuation of \eqref{MQEoM1usedforrealtimeSD}
\begin{align}
-i\partial_{t_1}G_{ab}(it_1^-,it_2^+)-\sum_c\Bigl(-i\mu\epsilon_{ac}G_{cb}(it_1^-,it_2^+)+\int_{C} du'\Sigma_{ac}(it_1^-,u')G_{cb}(u',it_2^+)\Bigr)&=0,
\label{continuation1} \\
-i\partial_{t_1}G_{ab}(it_1^+,it_2^-)-\sum_c\Bigl(-i\mu\epsilon_{ac}G_{cb}(it_1^+,it_2^-)+\int_{C'} du'\Sigma_{ac}(it_1^+,u')G_{cb}(u',it_2^-)\Bigr)&=0,
\label{continuation2}
\end{align}
where the integrations are over the contours depicted in Fig.~\ref{fig_200921Keldysh} and can be rewritten as
\begin{align}
\int_{C} du'\Sigma_{ac}(it_1^-,u')G_{cb}(u',it_2^+)&=
-i\int_{-\infty}^\infty dt_3(\Sigma_{ac}^R(t_1,t_3)G^>_{cb}(t_3,t_2)+\Sigma_{ac}^>(t_1,t_3)G^A_{cb}(t_3,t_2))\nonumber \\
\int_{C'} du'\Sigma_{ac}(it_1^+,u')G_{cb}(u',it_2^-)&=
-i\int_{-\infty}^\infty dt_3(\Sigma_{ac}^R(t_1,t_3)G^<_{cb}(t_3,t_2)+\Sigma_{ac}^<(t_1,t_3)G^A_{cb}(t_3,t_2)).\label{Langreth}
\end{align}
Here we have defined the retarded/advanced component of the two point funcitons
\begin{align}
G_{ab}^R(t_1,t_2)&=\theta(t_1-t_2)(G^>(t_1,t_2)-G^<(t_1,t_2)), \\
G_{ab}^A(t_1,t_2)&=\theta(t_2-t_1)(G^<(t_1,t_2)-G^>(t_1,t_2)),  \label{definitionofRA}
\end{align}
and $\Sigma^R_{ab},\Sigma^A_{ab}$ in the same way.
Taking the difference between \eqref{continuation1} and \eqref{continuation2}, and using the formulas \eqref{Langreth}, we obtain
\begin{align}
&-i\partial_{t_1}G_{ab}^R(t_1,t_2)-\sum_c\Bigl(-i\mu \rho_{ac}G_{cb}^R(t_1,t_2)-\int dt_3(\Sigma_{ac}^{R}(t_1,t_3)G_{cb}^R(t_3,t_2)-\Sigma_{ac}^{A}(t_1,t_3)G_{cb}^A(t_3,t_2))\Bigr)\nonumber \\
%&\quad =\delta_{ab}(\delta^>_C(t_1,t_2)-\delta^<_C(t_1,t_2))
&\quad =  - i \delta(t_1 - t_2) \cdot  2G^>_{ab}(t_1,t_1)
% &\quad \textcolor{cyan}{ =  - i \delta(t_1 - t_2) \cdot  2G^>_{ab}(t_1,t_1)}
=-\delta_{ab}\delta(t_1-t_2).
\label{manipulation2}
\end{align}
Here the second term in the integrand $\Sigma_{ac}^{A}(t_1,t_3)G_{cb}^A(t_3,t_2)$ vanishes for $t_1>t_2$, hence we end up with the following set of equations:
\begin{align}
&-i\partial_{t_1}G_{ab}^R(t_1,t_2)-\sum_c\Bigl(-i\mu \rho_{ac}G_{cb}^R(t_1,t_2)-\int dt_3\Sigma_{ac}^{R}(t_1,t_3)G_{cb}^R(t_3,t_2)\Bigr)=-\delta_{ab}\delta(t_1-t_2), \label{manipulation3} \\
&\Sigma_{ab}^{>}(t_1,t_2)=-\frac{i^q\mathcal{J}^2}{q}s_{ab}G^>_{ab}(t_1,t_2)^{q-1},\label{MQrealtimeSDeqsecondline}\\
&\Sigma_{ab}^{R}(t_1,t_2)=\theta(t_1-t_2)(\Sigma_{ab}^{>}(t_1,t_2)+\Sigma_{ba}^{>}(t_2,t_1))\label{MQSigmaRinSigma>}.
\end{align}
Here we have also written the continuation of the second line of the equations of motion \eqref{EuclideanEoMoftwocoupledmodel2} and the definition of the retarded component \eqref{definitionofRA} with $\Sigma_{ab}^{<}$ eliminated with the help of the anti-symmetric property $\Sigma_{ab}(u_1,u_2)=-\Sigma_{ba}(u_2,u_1)$ \eqref{antisymmetryfollowsfromEoM_MQ}.
If we assume $G_{ab}^>(t_1,t_2)$ and $G_{ab}^R(t_1,t_2)$ depends only on $t_1-t_2$, we can write the two point functions also in the Fourier modes
\begin{align}
{\widetilde f}^X(\omega)=\int_{-\infty}^{\infty}dt e^{i\omega t}f^X(t),\quad
f^X(t)=\int_{-\infty}^{\infty}\frac{d\omega}{2\pi} e^{-i\omega t}{\widetilde f}^X(t),\quad (f=G_{ab},\Sigma_{ab},\quad X=>,<,R,A).
\end{align}
The first equation \eqref{manipulation3} can be written in the Fourier modes as
\begin{align}
{\widetilde G}_{LL}^R(\omega)=\frac{-(-\omega+{\widetilde\Sigma}_{RR}^{R}(\omega))}{(-\omega+{\widetilde\Sigma}_{LL}^{R}(\omega))(-\omega+{\widetilde\Sigma}_{RR}^{R}(\omega))-({\widetilde\Sigma}_{LR}^{R}+i\mu)({\widetilde\Sigma}_{RL}^{R}-i\mu)},\nonumber \\
{\widetilde G}_{LR}^R(\omega)=\frac{{\widetilde\Sigma}_{LR}^{R}(\omega)+i\mu}{(-\omega+{\widetilde\Sigma}_{LL}^{R}(\omega))(-\omega+{\widetilde\Sigma}_{RR}^{R}(\omega))-({\widetilde\Sigma}_{LR}^{R}+i\mu)({\widetilde\Sigma}_{RL}^{R}-i\mu)},\nonumber \\
{\widetilde G}_{RL}^R(\omega)=\frac{{\widetilde\Sigma}_{RL}^{R}(\omega)-i\mu}{(-\omega+{\widetilde\Sigma}_{LL}^{R}(\omega))(-\omega+{\widetilde\Sigma}_{RR}^{R}(\omega))-({\widetilde\Sigma}_{LR}^{R}+i\mu)({\widetilde\Sigma}_{RL}^{R}-i\mu)},\nonumber \\
{\widetilde G}_{RR}^R(\omega)=\frac{-(-\omega+{\widetilde\Sigma}_{LL}^{R}(\omega))}{(-\omega+{\widetilde\Sigma}_{LL}^{R}(\omega))(-\omega+{\widetilde\Sigma}_{RR}^{R}(\omega))-({\widetilde\Sigma}_{LR}^{R}+i\mu)({\widetilde\Sigma}_{RL}^{R}-i\mu)}.
\label{MQrealtimeSDeqfirstlineinomega}
\end{align}

Apparently the equations \eqref{manipulation3}-\eqref{MQSigmaRinSigma>}, \eqref{MQrealtimeSDeqfirstlineinomega}, and \eqref{definitionofRA} are not closed by themselves as $G_{ab}^R$ does not completely determine $G_{ab}^>$ throught \eqref{definitionofRA}.
% Apparently the equations \eqref{manipulation3} (or \eqref{MQrealtimeSDeqfirstlineinomega}), \eqref{MQrealtimeSDeqsecondline}, \eqref{MQSigmaRinSigma>} and \eqref{definitionof><RA} are not closed by themselves as $G_{ab}^R$ does not completely determine $G_{ab}^>$ throught \eqref{definitionof><RA}.
This problem is fixed by taking into account the KMS condition \eqref{MQ_physicalansatz3} in the following way.
First using the KMS relation in Lorentzian signature, we obtain
\begin{align}
G_{ab}^R(t_1,t_2)&=\theta(t_1-t_2)(G_{ab}^>(t_1,t_2)-G^<_{ab}(t_1,t_2))\nonumber \\
&=\theta(t_1-t_2)(G_{ab}^>(t_1,t_2)+G_{ab}^>(t_1-i\beta,t_2)).
\label{write>inR1}
\end{align}
Next we consider $(G_{ba}^R(t_2,t_1))^*$, use \eqref{MQ_physicalansatz2} to rewrite $(G_{ba}^{>,<})^*$ in terms of $G_{ab}^>$, and then do the same rewriting as above: 
\begin{align}
(G_{ba}^R(t_2,t_1))^* &= \theta(t_2-t_1)(G^>_{ba}(t_2,t_1)^*-G^<_{ba}(t_2,t_1)^*) \nonumber \\
&=\theta(t_2-t_1)(G^<_{ba}(t_2,t_1)-G^>_{ba}(t_2,t_1))\nonumber \\
&= \theta(t_2-t_1)(-G^>_{ab}(t_1,t_2)+G^<_{ab}(t_1,t_2)) \nonumber \\
&= \theta(t_2-t_1)(-G^>_{ab}(t_1,t_2)-G_{ab}^>(t_1-i\beta,t_2)).
\end{align}
Combining these relations, we obtain
\begin{align}
G_{ab}^>(t_1,t_2)+G_{ab}^>(t_1-i\beta,t_2)=G_{ab}^R(t_1,t_2)-(G_{ba}^R(t_2,t_1))^*.
\end{align}
In the Fourier modes this is written as
\begin{align}
{\widetilde G}_{ab}^>(\omega)=\frac{{\widetilde G}_{ab}^R(\omega)-({\widetilde G}_{ba}^R(\omega))^*}{1+e^{-\beta \omega}}.
\label{MQ_relationbetween>andR_byKMS}
\end{align}
Hence \eqref{manipulation3}-\eqref{MQSigmaRinSigma>}, \eqref{MQrealtimeSDeqfirstlineinomega} and \eqref{MQ_relationbetween>andR_byKMS} together form a closed system of the equations for $G_{ab}^R(t)$ which we can solve numerically.

Note that once we obtain the retarded component $G^R_{ab}(t)$, we can compute $G_{ab}(u)$ for general $u\in\mathbb{C}$ with $0<\text{Re}[u]<\beta$ as\footnote{
We can also compute $G_{ab}(u)$ with $-\beta<\text{Re}[u]<0$ by using the anti-symmetry property $G_{ab}(u)=-G_{ba}(-u)$ \eqref{antisymmetryfollowsfromEoM_MQ}.
}
\begin{align}
G_{ab}(u)=iG_{ab}^>(t=-iu)=i\int \frac{d\omega}{2\pi}e^{-\omega u}\frac{{\widetilde G}_{ab}^R(\omega)-({\widetilde G}_{ba}^R(\omega))^*}{1+e^{-\beta\omega}},
\label{fixed200806}
\end{align}
which we use to compute the chaos exponent in section \ref{sec_chaosexponent_MQ}.
Also note that by setting $u=\tau$ $(0<\tau<\beta)$ the formula \eqref{fixed200806} reproduces the Euclidean propagator which we can obtain relatively easily by solving the Schwinger-Dyson equations \eqref{MQEoM1usedforrealtimeSD},\eqref{EuclideanEoMoftwocoupledmodel2} on the real contour, hence \eqref{fixed200806} can be also used as a trivial check for the validity of the real time computation.

\subsubsection{Further symmetry ansatz}
We can further impose the following symmetry properties consistently with the Schwinger-Dyson equations \eqref{manipulation3}-\eqref{MQSigmaRinSigma>}, \eqref{MQrealtimeSDeqfirstlineinomega} and the physical ansatz \eqref{MQ_physicalansatz2}-\eqref{MQ_physicalansatz3}, \eqref{MQ_relationbetween>andR_byKMS}
% We can further impose the following symmetry properties consistently with the Schwinger-Dyson equations \eqref{manipulation3} (or \eqref{MQrealtimeSDeqfirstlineinomega})-\eqref{MQSigmaRinSigma>} and the physical ansatz \eqref{MQ_physicalansatz2}-\eqref{MQ_physicalansatz3} (or \eqref{MQ_relationbetween>andR_byKMS})
\begin{align}
G_{RR}^>(t)=G_{LL}^>(t),\quad G_{RL}^>(t)=-G_{LR}^>(t),\quad
\Sigma_{RR}^>(t)=\Sigma_{LL}^>(t),\quad \Sigma_{RL}^>(t)=-\Sigma_{LR}^>(t).
\label{furthersymmetryMQ}
\end{align}
Note that this corresponds to imposing the $\mathbb{Z}_4$ symmetry \eqref{Z4symmetry}.
Under these additional constraints, the Schwinger-Dyson equations \eqref{manipulation3}-\eqref{MQSigmaRinSigma>}, \eqref{MQrealtimeSDeqfirstlineinomega} reduce to the following set of equations:
% Under these additional constraints, the Schwinger-Dyson equations \eqref{manipulation3} (or \eqref{MQrealtimeSDeqfirstlineinomega})-\eqref{MQSigmaRinSigma>} reduce to the following set of equations:
\begin{align}
{\widetilde G}_{LL}^R(\omega)&=\frac{-(-\omega+{\widetilde \Sigma}_{LL}^R(\omega))}{(-\omega+{\widetilde\Sigma}_{LL}^R(\omega))^2+({\widetilde\Sigma}_{LR}^R(\omega)+i\mu)^2},\quad
{\widetilde G}_{LR}^R(\omega)=\frac{{\widetilde \Sigma}_{LR}^R(\omega)+i\mu}{(-\omega+{\widetilde\Sigma}_{LL}^R(\omega))^2+({\widetilde\Sigma}_{LR}^R(\omega)+i\mu)^2},\nonumber \\
% {\widetilde G}_{LL}^R(\omega)&=\frac{-(-\omega+{\widetilde \Sigma}_{LL}^R(\omega))}{(-\omega+{\widetilde\Sigma}_{LL}^R)^2+({\widetilde\Sigma}_{LR}^R(\omega)+i\mu)^2},\quad
% {\widetilde G}_{LR}^R(\omega)=\frac{{\widetilde \Sigma}_{LR}^R(\omega)+i\mu}{(-\omega+{\widetilde\Sigma}_{LL}^R)^2+({\widetilde\Sigma}_{LR}^R(\omega)+i\mu)^2},\nonumber \\
\Sigma_{LL}^>(t)&=-\frac{i^q\mathcal{J}^2}{q}[2G_{LL}^>(t)]^{q-1},\quad
\Sigma_{LR}^>(t)=-\frac{\mathcal{J}^2}{q}[2G_{LR}^>(t)]^{q-1},\nonumber \\
\Sigma_{LL}^R(t)&=\theta(t)(\Sigma_{LL}^>(t)+\Sigma_{LL}^>(-t)),\quad
\Sigma_{LR}^R(t)=\theta(t)(\Sigma_{LR}^>(t)-\Sigma_{LR}^>(-t)),
\label{MQrealtimeSDfinal1}
\end{align}
while the constraints of the physical ansatz are now written as
\begin{align}
G_{LL}^>(t)^*&=-G_{LL}^>(-t),\quad
G_{LR}^>(t)^*=G_{LR}^>(-t),\nonumber \\
{\widetilde G}_{LL}^>(\omega)
&=\frac{2i\text{Im}[{\widetilde G}_{LL}^R(\omega)]}{1+e^{-\beta\omega}}
=-\frac{i\rho_{LL}(\omega)}{1+e^{-\beta\omega}},\quad
{\widetilde G}_{LR}^>(\omega)
=\frac{2\text{Re}[{\widetilde G}_{LR}^R(\omega)]}{1+e^{-\beta\omega}}
=-\frac{\rho_{LR}(\omega)}{1+e^{-\beta\omega}}.
\label{MQrealtimeSDfinal2}
\end{align}
Here we have defined the spectral functions
\begin{align}
\rho_{LL}(\omega)=-2\text{Im}[{\widetilde G}_{LL}^R(\omega)],\quad
\rho_{LR}(\omega)=-2\text{Re}[{\widetilde G}_{LR}^R(\omega)].
\label{eq_spectralfunction}
\end{align}
As we see in section \ref{sec_results}, these quantities are useful to characterize the gapped regime.
% which are useful quantities to characterize the gapped regime.

%%%%%%%%%%%%%%%%%%%%
%%%%%%%%%%%%%%%%%%%%

\subsubsection{Four point function}
\label{sec_4ptfcnMQ}
We consider the following four point function which is written as the two point function in the $G\Sigma$ formalism:
% We consider the following four point function which is written as the two point function in the bilocal field formalism:
\begin{align}
\frac{1}{(N/2)^2}\sum_{i,j}^{N/2}\langle
\psi_i^a(u_1)
\psi_i^b(u_2)
\psi_j^c(u_3)
\psi_j^d(u_4)
\rangle=
\frac{1}{Z_\text{two}}\int {\cal D}G_{ab}{\cal D}\Sigma_{ab}G_{ab}(u_1,u_2)G_{cd}(u_1,u_2)e^{-NS_\text{two}}.
\end{align}
In the large $N$ limit we can evaluate this correlation function by expanding $S_{\text{two}}$ around a solution of the Schwinger-Dyson equations \eqref{EuclideanEoMoftwocoupledmodel1},\eqref{EuclideanEoMoftwocoupledmodel2}, $G_{ab}=G_{ab}^{(0)}+N^{-\frac{1}{2}}\delta G_{ab}$, $\Sigma_{ab}=\Sigma_{ab}^{(0)}+N^{-\frac{1}{2}}\delta\Sigma_{ab}$ as
\begin{align}
S_\text{two}&=S_{\text{two}}^{(0)}
+\sum_{a,b,c,d}
\frac{1}{8N}
\int du_1du_2du_3du_4
\nonumber \\
&\quad \begin{pmatrix}
-\delta(u-u')\partial_{u'}+\frac{\Sigma_{LL}^{(0)}(u,u')-\Sigma_{LL}^{(0)}(u',u)}{2}
&\frac{\Sigma_{LR}^{(0)}(u,u')-\Sigma_{RL}^{(0)}(u',u)}{2}-i\mu\delta(u-u')\\
\frac{\Sigma_{RL}^{(0)}(u,u')-\Sigma_{LR}^{(0)}(u',u)}{2}+i\mu\delta(u-u')&
-\delta(u-u')\partial_{u'}+\frac{\Sigma_{RR}^{(0)}(u,u')-\Sigma_{RR}^{(0)}(u',u)}{2}
\end{pmatrix}^{-1}_{da}(u_4,u_1)\nonumber \\
&\quad \frac{
\delta\Sigma_{ab}(u_1,u_2)
-\delta\Sigma_{ba}(u_2,u_1)
}{2}
\nonumber \\
&\quad \begin{pmatrix}
-\delta(u-u')\partial_{u'}+\frac{\Sigma_{LL}^{(0)}(u,u')-\Sigma_{LL}^{(0)}(u',u)}{2}
&\frac{\Sigma_{LR}^{(0)}(u,u')-\Sigma_{RL}^{(0)}(u',u)}{2}-i\mu\delta(u-u')\\
\frac{\Sigma_{RL}^{(0)}(u,u')-\Sigma_{LR}^{(0)}(u',u)}{2}+i\mu\delta(u-u')&
-\delta(u-u')\partial_{u'}+\frac{\Sigma_{RR}^{(0)}(u,u')-\Sigma_{RR}^{(0)}(u',u)}{2}
\end{pmatrix}^{-1}_{bc}(u_2,u_3)\nonumber \\
&\quad
\frac{
\delta\Sigma_{cd}(u_3,u_4)
-\delta\Sigma_{dc}(u_4,u_3)
}{2}
\nonumber \\
&\quad +\sum_{a,b}\frac{1}{4N}\int du_1du_2\nonumber \\
&\quad \Bigl(\delta\Sigma_{ab}(u_1,u_2)\delta G_{ab}(u_1,u_2)-\frac{\mathcal{J}^2 2^{q-1}(q-1)}{2q}s_{ab}G^{(0)}_{ab}(u_1,u_2)^{q-2}\delta G_{ab}(u_1,u_2)^2\Bigr)
\label{Stwoaroundsaddletodelta2}
\end{align}
where the terms of ${\cal O}(\delta G_{ab},\delta \Sigma_{ab})$ trivially vanish since we are expanding $G_{ab},\Sigma_{ab}$ around a solution of the equations of motion.
% where the terms of ${\cal O}(\delta G_{ab},\delta \Sigma_{ab})$ trivially vanish since we are expanding $G_{ab},\Sigma_{ab}$ around a solution of the Schwinger-Dyson eqautions.
Note that the matrix elements in the first term can be replaced with $-G_{da}(u_4,u_1)$ and $-G_{bc}(u_2,u_3)$ with the help of \eqref{EuclideanEoMoftwocoupledmodel1}.
% Note that the matrix elements in the first term can be replaced with $-G_{da}(u_4,u_1)$ and $-G_{bc}(u_2,u_3)$ with the help of the equations of motion \eqref{EuclideanEoMoftwocoupledmodel1}.
Also noticing that $G_{ab}^{(0)}(u,u')=-G_{ba}^{(0)}(u',u)$, we obtain
\begin{align}
S_\text{two}&=S^{(0)}_\text{two}+\sum_{a,b,c,d}\frac{1}{8N}\int du_1du_2du_3du_4\frac{G_{ac}^{(0)}(u_1,u_3)G_{bd}^{(0)}(u_2,u_4)-G_{ad}^{(0)}(u_1,u_4)G_{bc}^{(0)}(u_2,u_3)}{2}\nonumber \\
&\quad \delta \Sigma_{ab}(u_1,u_2)\delta\Sigma_{cd}(u_3,u_4)
+\sum_{a,b}\frac{1}{4N}\int du_1du_2\nonumber \\
&\quad \Bigl(\delta\Sigma_{ab}(u_1,u_2)\delta G_{ab}(u_1,u_2)-\frac{\mathcal{J}^2 2^{q-1}(q-1)}{2q}s_{ab}G^{(0)}_{ab}(u_1,u_2)^{q-2}\delta G_{ab}(u_1,u_2)^2\Bigr)\nonumber \\
&=S_{\text{two}}^{(0)}+\sum_{A,B}\frac{1}{8N}\int dUdV{\cal G}_{AB}(U,V)\delta\Sigma_A(U)\delta\Sigma_B(V)\nonumber \\
&\quad +\sum_A\frac{1}{4N}\Bigl(\delta\Sigma_A(U)\delta G_A(U)-\frac{\mathcal{J}^2 2^{q-1} (q-1)}{2q}s_AG_A^{(0)}(U)^{q-2}\delta G_A(U)^2\Bigr),
\end{align}
where in the second line we have abbreviated the pair of indices/coordinates as $A=(a,b)$, $U=(u_1,u_2)$, and denoted the kernel of $\delta\Sigma$ as ${\cal G}_A(U,V)$:
\begin{align}
{\cal G}_{AB}(U,V)=\frac{
G^{(0)}_{A_1B_1}(U_1,V_1)G^{(0)}_{A_2B_2}(U_2,V_2)
-G^{(0)}_{A_1B_2}(U_1,V_2)G^{(0)}_{A_2B_1}(U_2,V_1)
}{2}.
\end{align}
Since the inserted operator $G_{ab}(u_1,u_2)G_{cd}(u_3,u_4)$ does not depends on $\Sigma_{ab}$ we can integrate $\delta\Sigma_{ab}$ first, which is under the current approximation simply a Gaussian integration:
\begin{align}
&\int {\cal D}\Sigma_{ab}e^{-NS_\text{two}}\nonumber \\
&=e^{-S_\text{two}^{(0)}}
\int {\cal D}\delta\Sigma_A\exp\biggl[
-\sum_{A,B}\frac{1}{8}\int dUdV{\cal G}_{AB}(U,V)\delta\Sigma_A(U)\delta\Sigma_B(V)\nonumber \\
&\quad -\frac{1}{4}\sum_A\int dU\Bigl(\delta G_A(U)\delta\Sigma_A(U)-\frac{\mathcal{J}^2 2^{q-1}(q-1)}{2q}G_A^{(0)}(U)^{q-2}\delta G_A(U)^2\Bigr)
\biggr]\nonumber \\
&=e^{-S_\text{two}^{(0)}}\int{\cal D}\delta\Sigma_A\exp\biggl[
-\frac{1}{8}
\sum_{A,B}\int dUdV
\Bigl(\delta\Sigma_A(U)-\sum_{C}\int dW\delta G_C(W)({\cal G}^{-1})_{CA}(W,U)\Bigr)\nonumber \\
&\quad {\cal G}_{AB}(U,V)
\Bigl(\delta\Sigma_B(V)-\sum_{D}\int dX ({\cal G}^{-1})_{BD}(V,X) \delta G_D(X)\Bigr)\nonumber \\
&\quad +\sum_{A,B}\int dUdV\delta G_A(U)\Bigl(\frac{1}{8}({\cal G}^{-1})_{AB}(U,V)\nonumber \\
&\quad +\frac{\mathcal{J}^2 2^{q-1}(q-1)}{8q}s_AG_A^{(0)}(U)^{q-2}\delta_{AB}\delta(U-V)
\Bigr)
\delta G_B(V)
\biggr]\nonumber \\
&=e^{-S_\text{two}^{(0)}}\exp\biggl[-\frac{1}{8}\sum_{A,B}\int dUdV\delta G_A(U)\Bigl(({\cal G}^{-1})_{AB}(U,V)\nonumber \\
&\quad +\frac{\mathcal{J}^2 2^{q-1}(q-1)}{q}s_AG_A^{(0)}(U)^{q-2}\delta_{AB}\delta(U-V)\Bigr)\delta G_B(V)\biggr].
\end{align}
Expanding the inserted $G_{ab}(u_1,u_2)G_{cd}(u_3,u_4)$ also around the saddle configuration, now we are left with the Gaussian integration in $\delta G_A(U)$ which we can perform as
\begin{align}
\frac{1}{(N/2)^2}\sum_{i,j}^{N/2}\langle
\psi_i^a(u_1)
\psi_i^b(u_2)
\psi_j^c(u_3)
\psi_j^d(u_4)
\rangle
=
G^{(0)}_{ab}(u_1,u_2)G^{(0)}_{cd}(u_3,u_4)+
\frac{1}{(N/2)}{\cal F}_{abcd}(u_1,u_2,u_3,u_4)
\end{align}
Swiching the notation back to $A\rightarrow (a,b)$, $U\rightarrow (u_1,u_2)$, the connected part ${\cal F}_{abcd}(u_1,u_2,u_3,u_4)$ of the four point function is written as
% Swiching the notation back $A\rightarrow (a,b)$, $U\rightarrow (u_1,u_2)$, the connected part ${\cal F}_{abcd}(u_1,u_2,u_3,u_4)$ of the four point function is written as
\begin{align}
&{\cal F}_{abcd}(u_1,u_2,u_3,u_4)\nonumber \\
&=-2\Bigl[({\cal G}^{-1})_{AB}(U,V)\nonumber \\
&\quad +\frac{\mathcal{J}^2 2^{q-1}(q-1)}{q}s_AG_A^{(0)}(U)^{q-2}\delta_{AB}\delta(U-V)\Bigr]^{-1}_{A=(a,b),B=(c,d)}(U=(u_1,u_2),V=(u_3,u_4))\nonumber \\
&=\sum_{n=0}^\infty{\cal F}_{n,abcd}(u_1,u_2,u_3,u_4).
\end{align}
with
\begin{align}
&{\cal F}_{0,abcd}(u_1,u_2,u_3,u_4)
=-2{\cal G}_{(a,b),(c,d)}((u_1,u_2),(v_1,v_2))\nonumber \\
&\quad\quad\quad\quad\quad\quad\quad\quad\quad =-G^{(0)}_{ac}(u_1,u_3)G^{(0)}_{bd}(u_2,u_4)+G_{ad}^{(0)}(u_1,u_4)G_{bc}^{(0)}(u_2,u_3),\nonumber \\
&{\cal F}_{n,abcd}(u_1,u_2,u_3,u_4)\nonumber \\
&=\sum_B\int dV\biggl[-{\cal G}_{(a,b),B}((u_1,u_2),V)\cdot\frac{\mathcal{J}^2 2^{q-1}(q-1)}{q}s_BG_B^{(0)}(V)^{q-1}\biggr]_{(a,b),B}((u_1,u_2),V)\nonumber \\
&\quad\quad\quad\quad {\cal F}_{n,B_1B_2cd}(V_1,V_2,u_3,u_4)\nonumber \\
&=\sum_{e,f}\int dvdv'{\cal K}_{abef}(u_1,u_2,v,v'){\cal F}_{n-1,efcd}(v,v',u_3,u_4),\nonumber \\
&{\cal K}_{abcd}(u_1,u_2,u_3,u_4)
=-\frac{\mathcal{J}^2 2^{q-1}(q-1)}{q}G^{(0)}_{ac}(u_1,u_3)G^{(0)}_{bd}(u_2,u_4)s_{cd}G^{(0)}_{cd}(u_3,u_4)^{q-2}.
\end{align}
Here to write down the ladder kernel ${\cal K}_{abcd}(u_1,u_2,u_3,u_4)$ we have used the fact that the matrix on which ${\cal K}$ acts, ${\cal F}_{n,abcd}(u_1,u_2,u_3,u_4)$, is always anti-symmetric under $(a,u_1)\leftrightarrow (b,u_2)$, which holds inductively.
Note that ${\cal F}_{abcd}$ obeys the following self consistency equation
% Note that ${\cal F}_{abcd}$ obeys the following equation
\begin{align}
{\cal F}_{abcd}(u_1,u_2,u_3,u_4)={\cal F}_{0,abcd}(u_1,u_2,u_3,u_4)+\sum_{e,f}\int dvdv'{\cal K}_{abef}(u_1,u_2,v,v'){\cal F}_{efcd}(v,v',u_3,u_4),
\label{MQEuclideanladdereq}
\end{align}
which we use in the next section.

\subsubsection{Chaos exponent from OTOC at late time}
\label{sec_chaosexponent_MQ}
Now we continue $u_1,u_2,u_3,u_4$ in \eqref{MQEuclideanladdereq} to
% Now we continue the $u_1,u_2,u_3,u_4$ to
\begin{align}
u_1=\frac{3\beta}{4}+it_1,\quad
u_2=\frac{\beta}{4}+it_2,\quad
u_3=\frac{\beta}{2},\quad
u_4=0,
\end{align}
and take the integration contour of $v,v'$ as the following Keldysh contour.
\begin{align*}
\includegraphics[width=5.5cm]{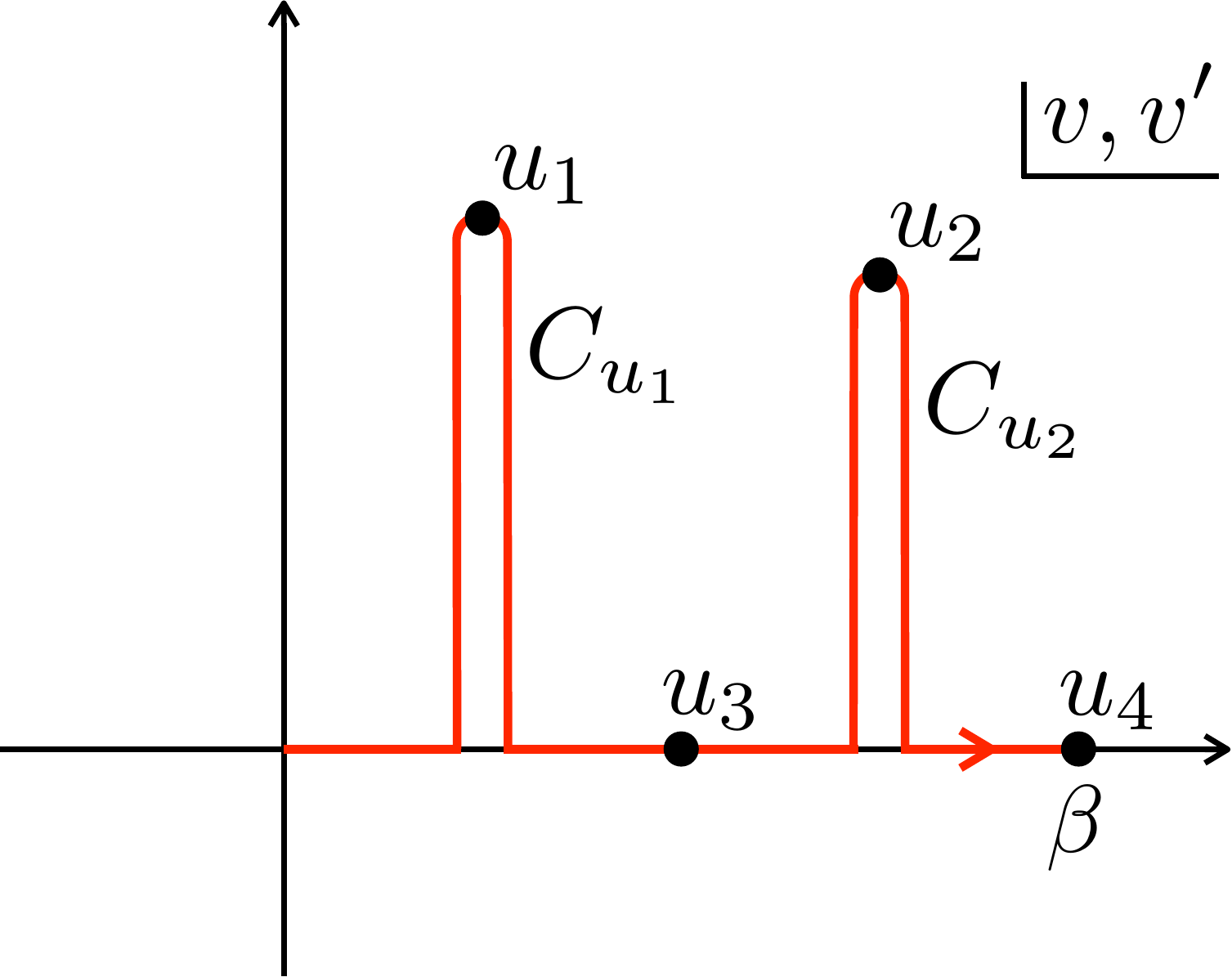}
\end{align*}
% and take the integration contour of $v,v'$ as the Keldysh contour $C=C_{u_1}+C_{u_2}+C_{u_3}+C_{u_4}$ \eqref{Keldyshcontour}.
We are interested in the growing behavior of ${\cal F}(3\beta/4+it_1,\beta/4+it_2,\beta/2,0)\equiv {\cal F}(t_1,t_2)$ at late time $t_1,t_2\gg 1$, where the only relevant contributions in the right-hand side of \eqref{MQEuclideanladdereq} are the second terms with $v\in C_{u_1},v'\in C_{u_2}$; the integrations with $v\in C_{u_2}$ or $v'\in C_{u_1}$ cancel by themselves due to the regularity of the integrand, and all the other terms including the first term ${\cal F}_{0,abcd}$ in \eqref{MQEuclideanladdereq} are suppressed as they contain the two point functions evaluated at $u$ with $\text{Im}[u]\sim t_1,t_2$ being large.
For the same reason, since the integration over $v\in C_{u_1}$, $v'\in C_{u_2}$ is dominated only by the contributions from $\text{Im}[v_1]\sim t_1,\text{Im}[v_2]\sim t_2$, we can freely add to $C_{u_1}$ and $C_{u_2}$ the infinite intervals $\text{Im}[v],\text{Im}[v']\in (-\infty,0)$.
% We are interested in the growing behavior of ${\cal F}(3\beta/4+it_1,\beta/4+it_2,\beta/2,0)\equiv {\cal F}(t_1,t_2)$ at late time $t_1,t_2\gg 1$, where the only relevant contributions in the right-hand side of \eqref{MQEuclideanladdereq} are the second terms with $v\in C_{u_1},v'\in C_{u_2}$; all the other terms trivially cancel out due to the regularity of the integrand or contain the two point function $G^{(0)}(u)$ with $u\sim it_1$ or $u\sim it_2$ which results in the exponential decal at late time.
% We are interested in the growing behavior of ${\cal F}(3\beta/4+it_1,\beta/4+it_2,\beta/2,0)\equiv {\cal F}(t_1,t_2)$ at late time $t_1,t_2\gg 1$, where the only relevant contributions in the right-hand side of \eqref{MQEuclideanladdereq} are the second terms with $v\in C_{u_1},v'\in C_{u_2}$ and $v\in C_{u_2},v'\in C_{u_1}$; all the other terms contains the two point function $G^{(0)}(u)$ with $u\sim it_1$ or $u\sim it_2$ which results in the exponential decal at late time.
Hence we can approximate the ladder relation \eqref{MQEuclideanladdereq} as
\begin{align}
{\cal F}_{abcd}(t_1,t_2)&\approx \sum_{ef}\int dt dt' {\cal K}^R_{abef}(t_1,t_2,t,t'){\cal F}_{efcd}(t,t'),\nonumber \\
{\cal K}^R_{abcd}(t_1,t_2,t_3,t_4)&=
-\frac{\mathcal{J}^2 2^{q-1}(q-1)}{q}G_{ac}^{(0)R}(t_1-t_3)G_{bd}^{(0)R}(t_2-t_4)s_{cd}G_{cd}^{(0)}\Bigl(\frac{\beta}{2}+i(t_3-t_4)\Bigr)^{q-2}.
% \frac{\mathcal{J}^2 2^{q-1}(q-1)}{q}G_{ac}^{(0)R}(t_1-t_3)G_{bd}^{(0)R}(t_2-t_4)s_{cd}G_{cd}^{(0)}\Bigl(\frac{\beta}{2}+i(t_3-t_4)\Bigr)^{q-2}.
\label{MQladderFKF}
\end{align}
If we further pose the following ansatz
\begin{align}
{\cal F}_{abcd}(t_1,t_2)=e^{\frac{\lambda_L(t_1+t_2)}{2}}f_{abcd}(t_{12}),
\label{calFlambdaLfansatzMQ}
\end{align}
we finally obtain, after a little change of the integration variables,
\begin{align}
f_{abcd}(t_{12})&\approx 
-\frac{\mathcal{J}^2 2^{q-1}(q-1)}{q}\sum_{ef}\int dt_-e^{-\frac{\lambda_L(t_{12}-t_-)}{2}}\biggl[\int dt'' G_{ae}^{(0)R}(t_{12}-t_--t'')G_{bf}^{(0)R}(-t'')e^{\lambda_Lt''}\biggr]
\nonumber \\
% \frac{\mathcal{J}^2 2^{q-1}(q-1)}{q}\sum_{ef}\int dt_-e^{-\frac{\lambda_L(t_{12}-t_-)}{2}}\biggl[\int dt'' G_{ae}^{(0)R}(t_{12}-t_--t'')G_{bf}^{(0)R}(-t'')e^{\lambda_Lt''}\biggr]\nonumber \\
&\quad s_{ef}G_{ef}^{(0)}\Bigl(\frac{\beta}{2}+it_-\Bigr)^{q-2}f_{efcd}(t_-),
\label{ladderfinalMQ}
\end{align}
The consequences of this equation are the followings.
First suppose that $\lambda_L$ is less than or equal to the actual value of the largest chaos exponent of the system.
% First suppose $\lambda_L$ is less than or equal to the actual value of the largest chaos exponent of the system.
Then the mode \eqref{calFlambdaLfansatzMQ} indeed exists and hence \eqref{ladderfinalMQ} has a non-trivial solution $f(t)$ corresponding to that mode.
On the other hand, if $\lambda_L$ is larger than the largest chaos exponent, such mode does not exist and hence \eqref{ladderfinalMQ} can only have a trivial solution $f(t)=0$.
Regarding the operation in the right-hand side of \eqref{ladderfinalMQ} as a matrix, the former case is possible only if the largest eigenvalue of the matrix is greater than or equal to $1$.
Therefore we can obtain the chaos exponent by varying the test value $\lambda_L$ and finding the point where the largest eigenvalue crosses $1$.
This procedure can be implemented numerically by the power iteration method.

We cau further simplify the ladder equation \eqref{ladderfinalMQ} as follows.
First of all, notice that the indices $cd$ of $f_{2,abcd}$ are not mixed through the operation of $M_1,M_2$.
This implies that we have only to consider a single choice of $cd$, say $cd=LL$, which hereafter we do not write: $f_{ab}\equiv f_{abLL}$.
Next, using the symmetry properties $G^>_{LL}(t)=G^>_{RR}(t)$, $G^>_{LR}(t)=-G_{RL}^>(t)$ we have imposed by hand \eqref{furthersymmetryMQ}, we can show that the kernel ${\cal K}_{abcd}^R(t_1,t_2,t_3,t_4)$ \eqref{MQladderFKF} is invariant under the simultaneous replacement $L\leftrightarrow R$ in the four indices $abcd$ together with a sign multiplication
\begin{align}
{\cal K}^R_{IJ}\rightarrow
\begin{pmatrix}
0&0&0&1\\
0&0&-1&0\\
0&-1&0&0\\
1&0&0&0
\end{pmatrix}_{IK}
{\cal K}^R_{KL}
\begin{pmatrix}
0&0&0&1\\
0&0&-1&0\\
0&-1&0&0\\
1&0&0&0
\end{pmatrix}_{LJ},\quad\quad
(I=LL,LR,RL,LL)
\end{align}
which is equivalent to the following change of basis of ${\cal F}_{ab}$:
\begin{align}
\begin{pmatrix}
{\cal F}_{LL}\\
{\cal F}_{LR}\\
{\cal F}_{RL}\\
{\cal F}_{RR}
\end{pmatrix}
\rightarrow
\begin{pmatrix}
0&0&0&1\\
0&0&-1&0\\
0&-1&0&0\\
1&0&0&0
\end{pmatrix}
\begin{pmatrix}
{\cal F}_{LL}\\
{\cal F}_{LR}\\
{\cal F}_{RL}\\
{\cal F}_{RR}
\end{pmatrix}.
\label{flipsymmetryofMQladdereq}
\end{align}
Hence the ladder equation splits to the one for the ${\cal F}_{ab}$ which is symmetric under the flip: $({\cal F}_{LL},{\cal F}_{LR},{\cal F}_{RL},{\cal F}_{RR})=({\cal F}_{LL},{\cal F}_{LR},-{\cal F}_{LR},{\cal F}_{LL})$ and the one for ${\cal F}_{ab}$ being antisymmetric: $({\cal F}_{LL},{\cal F}_{LR},{\cal F}_{RL},{\cal F}_{RR})=({\cal F}_{LL},{\cal F}_{LR},{\cal F}_{LR},-{\cal F}_{LL})$.
We can finally write the ladder equation for the symmetric/anti-symmetric sector, which we denote by $\sigma=\pm 1$, as
\begin{align}
&\begin{pmatrix}
f_{2,LL}+\sigma f_{2,RR}\\
f_{2,LR}-\sigma f_{2,RL}
\end{pmatrix}\nonumber \\
&=
\begin{pmatrix}
M_{1,LLLL}+\sigma M_{1,LRLR}&M_{1,LLLR}-\sigma M_{1,LRLL}\\
-(M_{1,LLLR}-\sigma M_{1,LRLL})&M_{1,LLLL}+\sigma M_{1,LRLR}
\end{pmatrix}
\circ
\begin{pmatrix}
M_{2,LL}(f_{2,LL}+\sigma f_{2,RR})\\
M_{2,LR}(f_{2,LR}-\sigma f_{2,RL})
\end{pmatrix}
\label{ladderMQfinalsimplified}
\end{align}
where $\circ$ is the convolution $(f\circ g)(t)=\int dt' f(t-t')g(t')$ and
\begin{align}
f_{2,ab}(t_{12})&=e^{\frac{\lambda_Lt_{12}}{2}}f_{ab}(t_{12}),\nonumber \\
M_{1,abcd}(t)&=\int dt'G_{ab}^{(0)R}(t-t')G^{(0)R}_{cd}(-t')e^{\lambda_Lt'},\nonumber \\
M_{2,ab}(t)&=
-\frac{\mathcal{J}^2 2^{q-1}(q-1)}{q}s_{ab}G_{ab}^{(0)}\Bigl(\frac{\beta}{2}+it\Bigr)^{q-2}.
\end{align}
Note that $M_{1,abcd}$ itself can also be written as a convolution: $M_{1,abcd}=G^{(0)R}_{ab}\circ ({\widehat G}^{(0)R}_{cd}e^{\lambda_Lt})$ with ${\widehat G}^{(0)R}_{ab}(t)=G^{(0)}_{ab}(-t)$.

\subsection{Single sided model}
\subsubsection{Real time Schwinger-Dyson equation}
Thanks to the redundant $G\Sigma$ formalism \eqref{redundantGSigmaformalismKM1},\eqref{redundantGSigmaformalismKM2}, the calculation for the single sided model is completely parallel to those for the two coupled model; the only difference is in the form of the potential term of $G_{ab}(u,u')$ (right-hand side of \eqref{EoMinusinglesided2}), which do not disturb the argument on the analytic continuation in sectoin \ref{sec_realtimeSDeqMQ},\ref{sec_chaosexponent_MQ} and the derivation of the ladder equation in section \ref{sec_4ptfcnMQ}.
Hence we obtain the following set of real time Schwinger-Dyson equations
\begin{align}
&-i\partial_{t_1}G_{LL}^R(t_1,t_2)+i\mu G_{RL}^R(t_1,t_2)+\int dt_3\Sigma_{LL}^{R}(t_1,t_3)G_{LL}^R(t_3,t_2)=-\delta(t_1-t_2),\nonumber \\
&-i\partial_{t_1}G_{LR}^R(t_1,t_2)+i\mu G_{RR}^R(t_1,t_2)+\int dt_3\Sigma_{LL}^{R}(t_1,t_3)G_{LR}^R(t_3,t_2)=0,\nonumber \\
&-i\partial_{t_1}G_{RL}^R(t_1,t_2)-i\mu G_{LL}^R(t_1,t_2)+\int dt_3\Sigma_{LL}^{R}(t_1,t_3)G_{RL}^R(t_3,t_2)=0,\nonumber \\
&-i\partial_{t_1}G_{RR}^R(t_1,t_2)-i\mu G_{LR}^R(t_1,t_2)+\int dt_3\Sigma_{LL}^{R}(t_1,t_3)G_{RR}^R(t_3,t_2)=-\delta(t_1-t_2),\label{realtimeSDeqKM1} \\
&\Sigma_{LL}^{>}(t_1,t_2)=-\frac{i^q\mathcal{J}^2 }{q}\Bigl(G^>_{LL}(t_1,t_2)+G^>_{RR}(t_1,t_2)\Bigr)^{q-1},\label{realtimeSDeqKM2} \\
&\Sigma_{LL}^{R}(t_1,t_2)=\theta(t_1-t_2)(\Sigma_{LL}^{>}(t_1,t_2)+\Sigma_{LL}^{>}(t_2,t_1)), \label{realtimeSDeqKM3}
\end{align}
where we have also used the fact $\Sigma_{RR}(u,u')=\Sigma_{LL}(u,u')$ and $\Sigma_{LR}(u,u')=\Sigma_{RL}(u',u)=0$ \eqref{symmetrypropertyfromEoM_KM}.
If we assume that $G_{ab}^>(t_1,t_2)$ and $G_{ab}^R(t_1,t_2)$ depend only on $t_1-t_2$, we obtain from the first four equations \eqref{realtimeSDeqKM1}
% If we assume $G_{ab}^>(t_1,t_2)$ and $G_{ab}^R(t_1,t_2)$ depends only on $t_1-t_2$, we obtain from the first four equations \eqref{realtimeSDeqKM1}
\begin{align}
{\widetilde G}_{LL}^R(\omega)={\widetilde G}_{RR}^R(\omega)=\frac{-(-\omega+{\widetilde\Sigma}_{LL}^{R}(\omega))}{(-\omega+{\widetilde\Sigma}_{LL}^{R}(\omega))^2-\mu^2},\label{realtimeSDKMfinal1} \\
{\widetilde G}_{LR}^R(\omega)=-{\widetilde G}_{RL}^R(\omega)=\frac{i\mu}{(-\omega+{\widetilde\Sigma}_{LL}^{R}(\omega))^2-\mu^2},\label{realtimeSDKMfinal2}
\end{align}
The greater components $G^{>}_{ab}(t)$ are related to the retarded components as
\begin{align}
{\widetilde G}_{ab}^>(\omega)=\frac{{\widetilde G}_{ab}^R(\omega)-({\widetilde G}_{ba}^R(\omega))^*}{1+e^{-\beta \omega}},
\end{align}
or explicitly
\begin{align}
{\widetilde G}_{LL}^>(\omega)=\frac{2i\text{Im}[{\widetilde G}_{LL}^R(\omega)]}{1+e^{-\beta \omega}}
=-\frac{i\rho_{LL}(\omega)}{1+e^{-\beta \omega}},\quad
{\widetilde G}_{LR}^>(\omega)=\frac{2\text{Re}[{\widetilde G}_{LR}^R(\omega)]}{1+e^{-\beta \omega}}
=-\frac{\rho_{LR}(\omega)}{1+e^{-\beta \omega}},
\label{KMSforKM}
\end{align}
where we have defined the spectral functions $\rho_{LL}(\omega)=-2\text{Im}[{\widetilde G}_{LL}(\omega)]$, $\rho_{LR}(\omega)=-2\text{Re}[{\widetilde G}_{LR}(\omega)]$ in the same was as in the two coupled model \eqref{eq_spectralfunction}.
As we have already mentioned at the end of section \ref{sec_GSigmaformalism_singlesided}, the Schwinger-Dyson equations are decomposed into a closed set of equations only for $G_{LL}^R$, $G_{LL}^>$, $\Sigma_{LL}^R$, $\Sigma_{LL}^>$ \eqref{realtimeSDeqKM2},\eqref{realtimeSDeqKM3},\eqref{realtimeSDKMfinal1},\eqref{KMSforKM} and the rest which gives $G_{LR}^R$, $G_{LR}^>$ explicitly in terms of $\Sigma_{LL}^R$ \eqref{realtimeSDKMfinal2},\eqref{KMSforKM}.

Once we obtain the retarded component $G^R_{ab}(t)$, we can compute $G_{ab}(u)$ for general $u\in\mathbb{C}$ with $0<\text{Re}[u]<\beta$ as
\begin{align}
G_{ab}(u)=iG_{ab}^>(t=-iu)= i\int \frac{d\omega}{2\pi}e^{-\omega u}\frac{{\widetilde G}_{ab}^R(\omega)-({\widetilde G}_{ba}^R(\omega))^*}{1+e^{-\beta\omega}},
\end{align}
and $G_{ab}(u)$ with $-\beta<\text{Re}[u]<0$ by using the anti-symmetry property $G_{ab}(u)=-G_{ba}(-u)$ \eqref{symmetrypropertyfromEoM_KM}.

\subsubsection{Four point function}
The caculation for the four point functions is also in parallel.
% The caculation for the four point functions is also the same.
If we denote $\chi_{2i-1}(u)$ as $\chi_i^L(u)$ and $\chi_{2i}(u)$ as $\chi_i^R(u)$, the four point functions are expressed in the large $N$ limit as follows
\begin{align}
\frac{1}{(N/2)^2}\sum_{i,j}^{\frac{N}{2}}
\langle
\chi_i^a(u_1)\chi_i^b(u_2)
\chi_j^c(u_3)\chi_j^d(u_4)
\rangle
&=\frac{1}{Z_\text{single}}\int {\cal D}G_{ab}{\cal D}\Sigma_{ab}G_{ab}(u_1,u_2)G_{cd}(u_3,u_4)e^{-NS_\text{single}}\nonumber \\
&=G_{ab}^{(0)}(u_1,u_2)G_{cd}^{(0)}(u_3,u_4)+\frac{1}{(N/2)}{\cal F}_{abcd}(u_1,u_2,u_3,u_4)
\end{align}
where $G_{ab}^{(0)}$ are a solutions to the Schwinger-Dyson equations.
The connected part ${\cal F}_{abcd}$ is
\begin{align}
{\cal F}_{abcd}(u_1,u_2,u_3,u_4)=\sum_{n=0}^\infty{\cal F}_{n,abcd}(u_1,u_2,u_3,u_4)
\end{align}
where
\begin{align}
{\cal F}_{0,abcd}(u_1,u_2,u_3,u_4)=-G_{ac}^{(0)}(u_1,u_3)G_{bd}^{(0)}(u_2,u_4)
+G_{ad}^{(0)}(u_1,u_4)G_{bc}^{(0)}(u_2,u_3),\nonumber \\
{\cal F}_{n,abcd}(u_1,u_2,u_3,u_4)=
\sum_{e,f}\int dvdv'{\cal K}_{abef}^{\text{(single)}}(u_1,u_2,v,v')
{\cal F}_{n-1,abcd}(v,v',u_3,u_4),
\end{align}
with
\begin{align}
{\cal K}_{abcd}^{\text{(single)}}(u_1,u_2,u_3,u_4)&=-\frac{\mathcal{J}^2 2^{q-1}(q-1)}{2q}\Bigl(\sum_e G_{ae}^{(0)}(u_1,u_3)G_{be}^{(0)}(u_2,u_4)\Bigr)\nonumber \\
&\quad \Bigl(\frac{G_{LL}^{(0)}(u_3,u_4)+G_{RR}^{(0)}(u_3,u_4)}{2}\Bigr)^{q-2}\delta_{cd}.
\label{KMladder1}
\end{align}

Here we observe an additional simplification which did not occur in the two coupled model: due to the structure of the $cd$ index in ${\cal K}_{abcd}^{\text{(single)}}$ it follows that the recursive relation \eqref{KMladder1} decomposes to the following recursive relation which closes only within ${\cal F}_{ab}(u_1,u_2,u_3,u_4)\equiv {\cal F}_{LLab}(u_1,u_2,u_3,u_4) +{\cal F}_{RRab}(u_1,u_2,u_3,u_4)$
% Here we observe a special simplification which did not occur in the two coupled model: due to the structure of the $cd$ index in ${\cal K}_{abcd}^{\text{(single)}}$ it follows that the recursive relation \eqref{KMladder1} decomposes to the following recursive relation which closes only within ${\cal F}_{ab}(u_1,u_2,u_3,u_4)\equiv {\cal F}_{LLab}(u_1,u_2,u_3,u_4) +{\cal F}_{RRab}(u_1,u_2,u_3,u_4)$
\begin{align}
{\cal F}_{n,ab}(u_1,u_2,u_3,u_4)&=
\int dvdv'{\cal K}^{\text{(single)}}(u_1,u_2,v,v')
{\cal F}_{n-1,ab}(v,v',u_3,u_4),\nonumber \\
{\cal K}^{\text{(single)}}(u_1,u_2,u_3,u_4)&=-\frac{\mathcal{J}^2 2^{q-1} (q-1)}{2q}\Bigl(\sum_{a,b}G_{ab}^{(0)}(u_1,u_3)G_{ab}^{(0)}(u_2,u_4)\Bigr)\nonumber \\
&\quad \Bigl(\frac{G_{LL}^{(0)}(u_3,u_4)+G_{RR}^{(0)}(u_3,u_4)}{2}\Bigr)^{q-2},
\end{align}
and the rest which explicitly determines the other components of ${\cal F}_{abcd}(u_1,u_2,u_3,u_4)$ in terms of ${\cal F}_{cd}(u_1,u_2,u_3,u_4)$:
\begin{align}
{\cal F}_{n,abcd}(u_1,u_2,u_3,u_4)&=
-\frac{\mathcal{J}^2 2^{q-1}(q-1)}{2q}
\int dvdv'
\Bigl(\sum_{e}G_{ae}^{(0)}(u_1,u_3)G_{be}^{(0)}(u_2,u_4)\Bigr)\nonumber \\
&\quad \Bigl(\frac{G_{LL}^{(0)}(u_3,u_4)+G_{RR}^{(0)}(u_3,u_4)}{2}\Bigr)^{q-2}{\cal F}_{n-1,cd}(v,v',u_3,u_4).
\end{align}
Therefore, for the purpose of determining the chaos exponent of the single sided model, it is enough to proceed with only the recursive relation for ${\cal F}_{ab}(u_1,u_2,u_3,u_4)$ written in the form of a self-consistency equation
\begin{align}
{\cal F}_{ab}(u_1,u_2,u_3,u_4)=
{\cal F}_{0,ab}(u_1,u_2,u_3,u_4)
+\int dvdv'{\cal K}^{\text{(single)}}(u_1,u_2,v,v'){\cal F}_{ab}(v,v',u_3,u_4).
\label{KMladder2}
\end{align}
where ${\cal F}_{0,ab}(u_1,u_2,u_3,u_4)={\cal F}_{0,LLab}(u_1,u_2,u_3,u_4)+{\cal F}_{0,RRab}(u_1,u_2,u_3,u_4)$.

\subsubsection{Chaos exponent}
By continuing the ladder equation \eqref{KMladder2} to real time with $u_1=3\beta/4+it_1$, $u_2=\beta/4+it_2$, $u_3=\beta/2$, $u_4=0$ and assuming a growing behavior of ${\cal F}_{ab}(t_1,t_2)\equiv {\cal F}_{ab}(3\beta/4+it_1,\beta/4+it_2,\beta/2,0)$ at late time $t_1,t_2\gg 1$, we obtain the following real time ladder equation
% By continuing the ladder equation \eqref{KMladder2} to real time with $u_1=\frac{3\beta}{4}+it_1$, $u_2=\frac{\beta}{4}+it_2$, $u_3=\frac{\beta}{2}$, $u_4=0$ and assuming a growing behavior of ${\cal F}_{ab}(t_1,t_2)\equiv {\cal F}_{ab}(\frac{3\beta}{4}+it_1,\frac{\beta}{4}+it_2,\frac{\beta}{2},0)$ at late time $t_1,t_2\gg 1$, we obtain the following real time ladder equation
\begin{align}
{\cal F}_{ab}(t_1,t_2)&\approx \int dt dt' {\cal K}^{\text{(single)}R}(t_1,t_2,t,t'){\cal F}_{ab}(t,t'),
\end{align}
with the retarded kernel given as
\begin{align}
&{\cal K}^{\text{(single)}R}(t_1,t_2,t_3,t_4)\nonumber \\
&
=-\frac{\mathcal{J}^2 2^{q-1}(q-1)}{2q}\Bigl(\sum_{a,b} G_{ab}^{(0)R}(t_1-t_3)G_{ab}^{(0)R}(t_2-t_4)\Bigr)
% =\frac{\mathcal{J}^2 2^{q-1}(q-1)}{2q}\Bigl(\sum_{a,b} G_{ab}^{(0)R}(t_1-t_3)G_{ab}^{(0)R}(t_2-t_4)\Bigr)
\nonumber \\
&\quad \Bigl(\frac{G_{LL}^{(0)}(\frac{\beta}{2}+i(t_3-t_4))+G_{RR}^{(0)}(\frac{\beta}{2}+i(t_3-t_4))}{2}\Bigr)^{q-2}\nonumber \\
&=
-\frac{\mathcal{J}^2 2^{q-1}(q-1)}{q}
% \frac{\mathcal{J}^2 2^{q-1}(q-1)}{q}
(G_{LL}^{(0)R}(t_1-t_3)G_{LL}^{(0)R}(t_2-t_4)+G_{LR}^{(0)R}(t_1-t_3)G_{LR}^{(0)R}(t_2-t_4))G_{LL}^{(0)}\nonumber \\
&\quad \Bigl(\frac{\beta}{2}+i(t_3-t_4)\Bigr)^{q-2}, \label{KMretardedkernel}
\end{align}
where in the third line we have used the fact that $G^{(0)R}_{RL}(t)=-G^{(0)R}_{LR}(t)$ and $G^{(0)}_{RR}(u)=G^{(0)}_{LL}(u)$ \eqref{realtimeSDKMfinal1},\eqref{realtimeSDKMfinal2}.
If we further pose the exponentially growing ansatz
\begin{align}
{\cal F}_{ab}(t_1,t_2)=e^{\frac{\lambda_L(t_1+t_2)}{2}}f_{ab}(t_{12}),
\end{align}
the real time ladder equation reduces to
\begin{align}
f_{ab}(t_{12})&\approx 
-\frac{\mathcal{J}^2 2^{q-1}(q-1)}{q}\int dt_-e^{-\frac{\lambda_L(t_{12}-t_-)}{2}}\biggl[\int dt'' (
% \frac{\mathcal{J}^2 2^{q-1}(q-1)}{q}\int dt_-e^{-\frac{\lambda_L(t_{12}-t_-)}{2}}\biggl[\int dt'' (
G_{LL}^{(0)R}(t_{12}-t_--t'')G_{LL}^{(0)R}(-t'')\nonumber \\
&\quad +G_{LR}^{(0)R}(t_{12}-t_--t'')G_{LR}^{(0)R}(-t'')
)
e^{\lambda_Lt''}\biggr]G_{LL}^{(0)}\Bigl(\frac{\beta}{2}+it_-\Bigr)^{q-2}f_{ab}(t_-).
\end{align}

\begin{figure}
\begin{center}
\includegraphics[width=9cm]{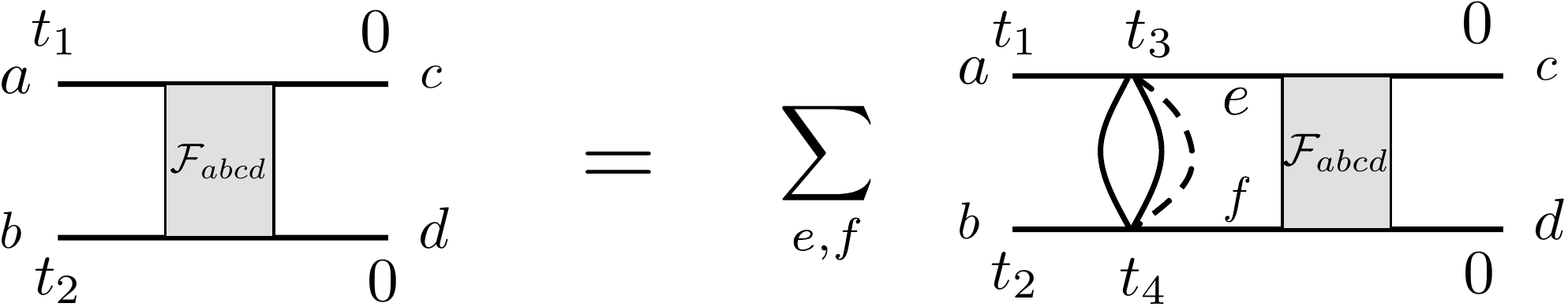} \\
\includegraphics[width=8cm]{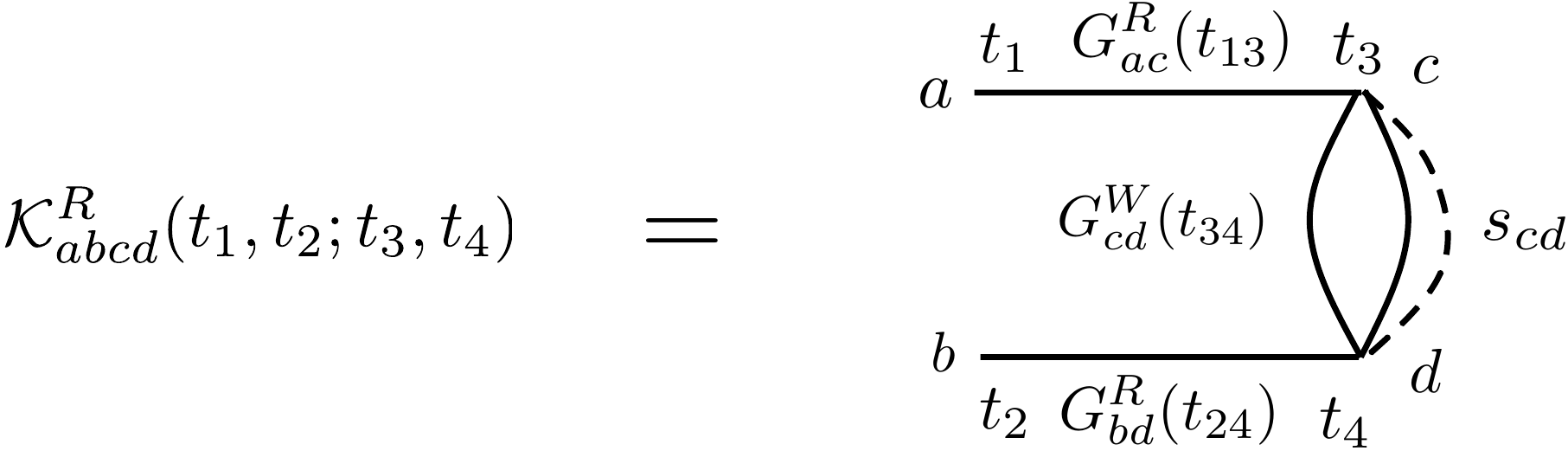}\\
\includegraphics[width=10cm]{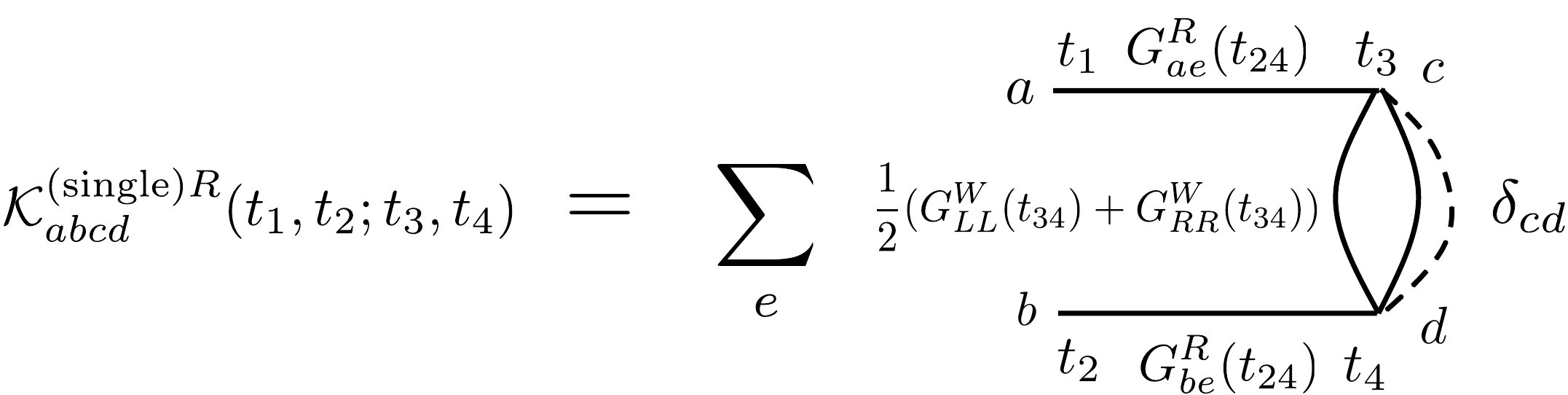}
\caption{
Top/Middle: The diagrammatic representation of the ladder equation/ retarded kernel \eqref{MQladderFKF}. 
Bottom: The diagrammatic representation of the retarded kernel \eqref{KMretardedkernel}. 
}
\label{kerneldiagram}
\end{center}
\end{figure}

We can also understand the structure of the retarded kernels and ladder equations using diagrams (figure \ref{kerneldiagram}).

\section{Results}
\label{sec_results}

In this section we display the numerical results for the real time two point functions and the chaos exponent of the two coupled model and the single sided model.
In all of the following analyses we have chosen $q=4$ and ${\cal J}=1$ for both of the two models.
% In all of the following analysis we have chosen $q=4$ and ${\cal J}=1$ both for the two coupled model and the single sided model.
Some results for different values of $q$ are displayed in appendix \ref{sec_q6q8}.

%%%%%%%%%%%%%%%%%%%%
%%%%%%%%%%%%%%%%%%%%

\subsection{Two coupled model}

\subsubsection{Euclidean propagator $G_{ab}(\tau)$, phase diagram and $E_\text{gap}$}

When the contour of $u$ in \eqref{<ZMQ>_J} is taken as the Euclidean slice $u=\tau\in(0,\beta)$ with $\tau\sim \tau+\beta$, the partition function gives the thermal free energy
\begin{align}
\frac{F(T)}{N}=-\frac{1}{\beta N}\log Z\approx -\frac{1}{\beta N}\sum_{\text{saddles}}e^{-NS_\text{two}[G_{ab}^{\text{(saddle)}},\Sigma_{ab}^{\text{(saddle)}}]},\quad\quad T=\beta^{-1},
\end{align}
where $(G_{ab}^{\text{(saddle)}},\Sigma_{ab}^{\text{(saddle)}})$ are the solutions of the Schwinger-Dyson equations \eqref{MQEoM1usedforrealtimeSD},\eqref{EuclideanEoMoftwocoupledmodel2}.
% where $(G_{ab}^{\text{(saddle)}},\Sigma_{ab}^{\text{(saddle)}})$ are the solutions to the Schwinger-Dyson equations \eqref{MQEoM1usedforrealtimeSD},\eqref{EuclideanEoMoftwocoupledmodel2}.

\begin{figure}
\begin{center}
\includegraphics[width=8cm]{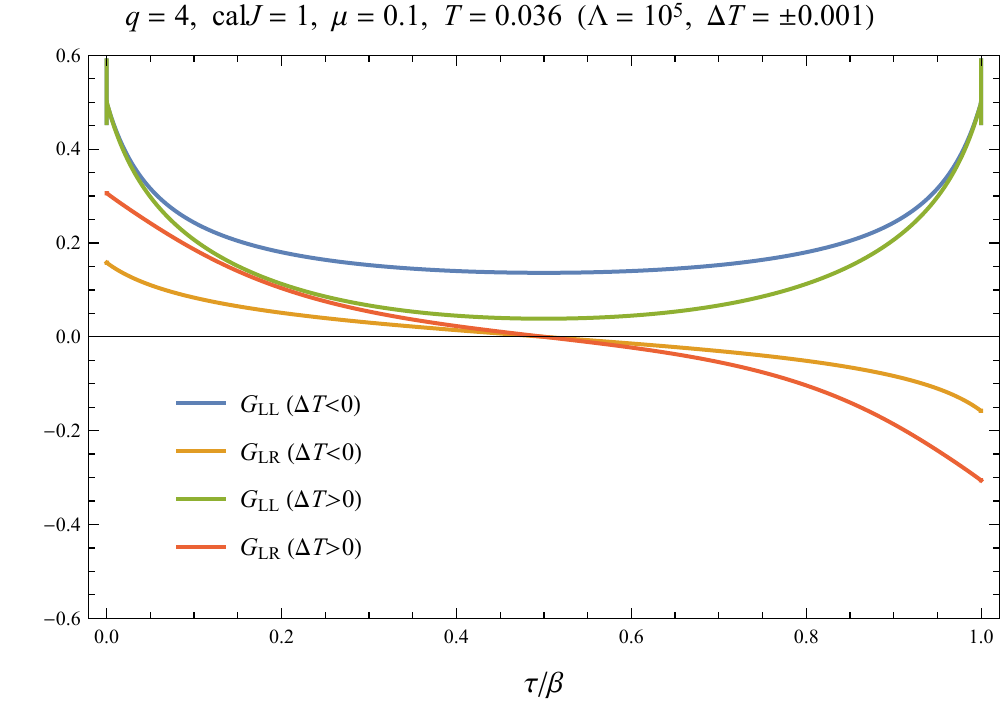}
\caption{
Euclidean propagators for the black hole phase ($\Delta T<0$) and those for the wormhole phase ($\Delta T>0$).
}
\label{MQ_GabEuclidean}
\end{center}
\end{figure}

\begin{figure}
\includegraphics[width=8cm]{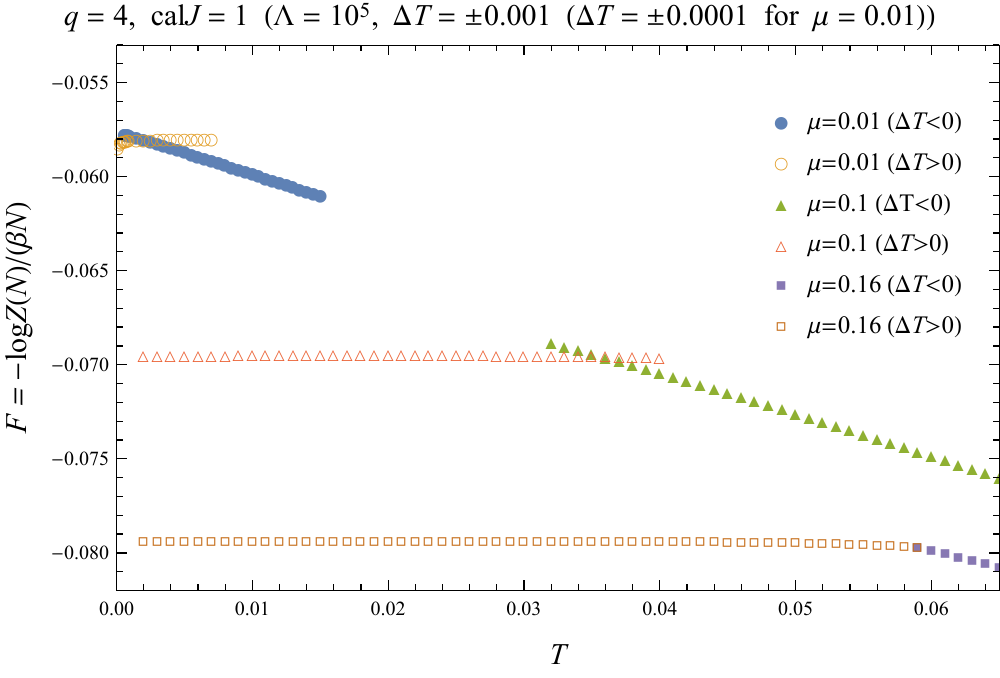}\quad\quad
\includegraphics[width=8cm]{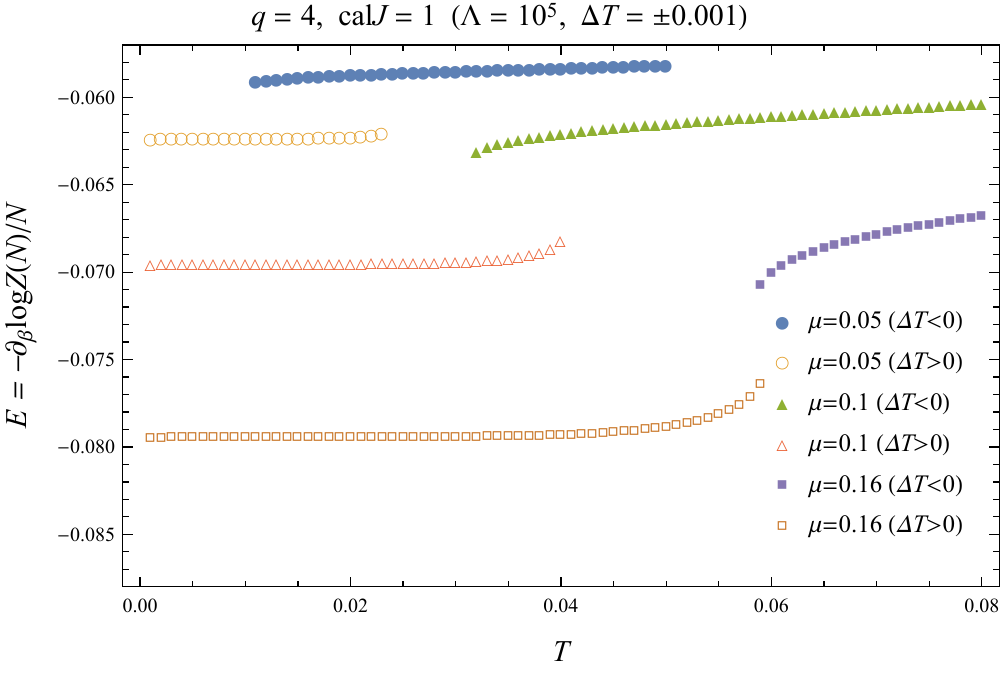}\\
\includegraphics[width=8cm]{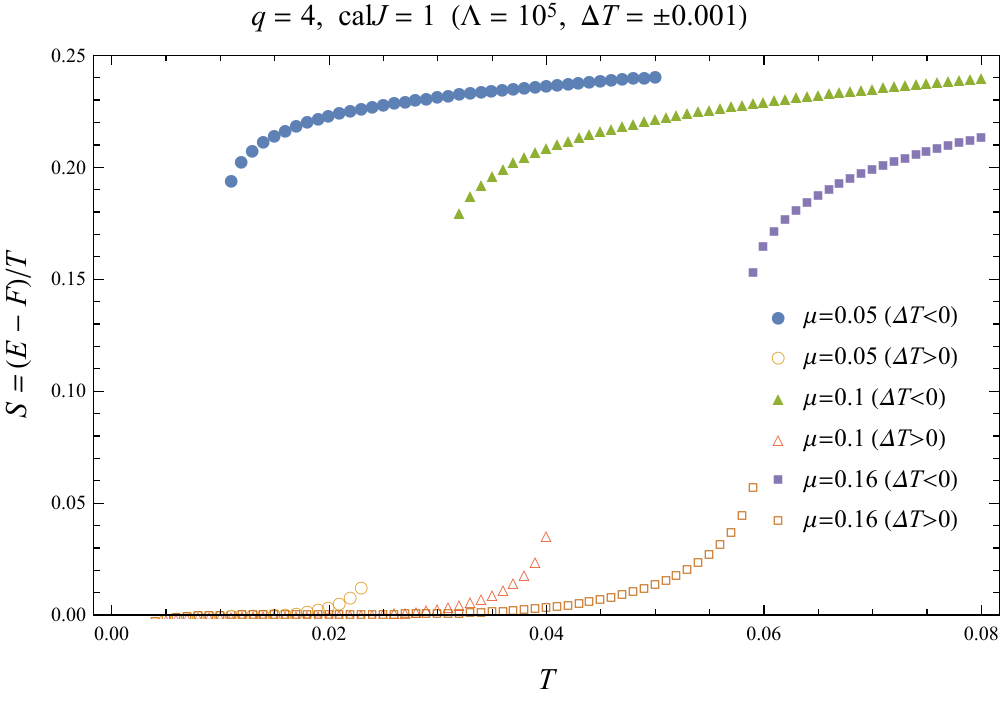}\quad\quad
\includegraphics[width=8cm]{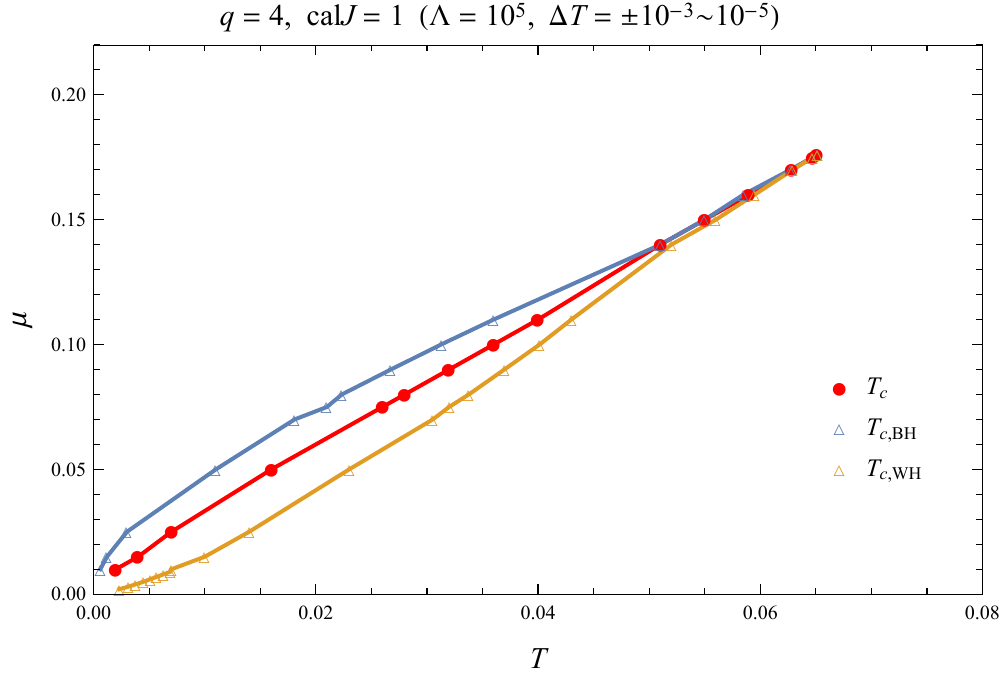}\\
\caption{
Top left/top right/bottom left: free energy/energy/entropy for the black hole solution and the wormhole solution; bottom right: phase diagram of the two coupled model.
In the phase diagram we have computed $T_{c,\text{BH}}$, $T_c$ and $T_{c,\text{WH}}$ for $\mu=0.07,0.08,0.09,0.1$ and $\mu\ge 0.16$ with $\Delta T=\pm 10^{-5}$, $T_{c,\text{BH}}$ for $\mu=0.01,0.015$ and $T_{c,\text{WH}}$ for $0.002\le \mu\le 0.009$ with $\Delta T=\pm 10^{-4}$, while all the other data points with $\Delta T=10^{-3}$.
}
\label{MQ_freeandTcTcBHTcWH}
\end{figure}

\begin{figure}
\includegraphics[width=8cm]{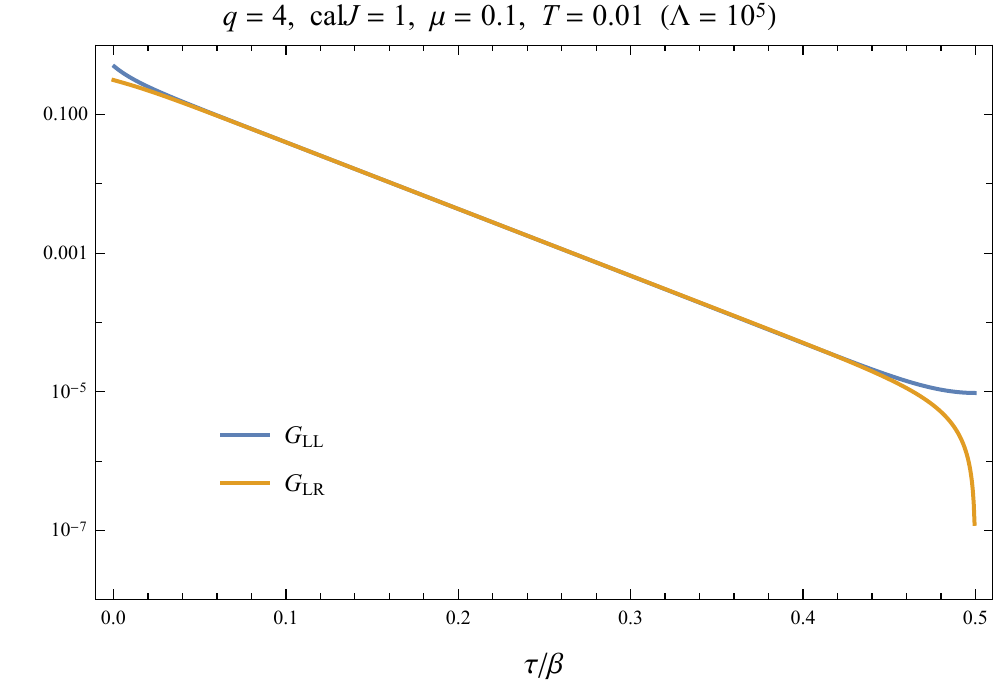}\quad\quad
\includegraphics[width=8cm]{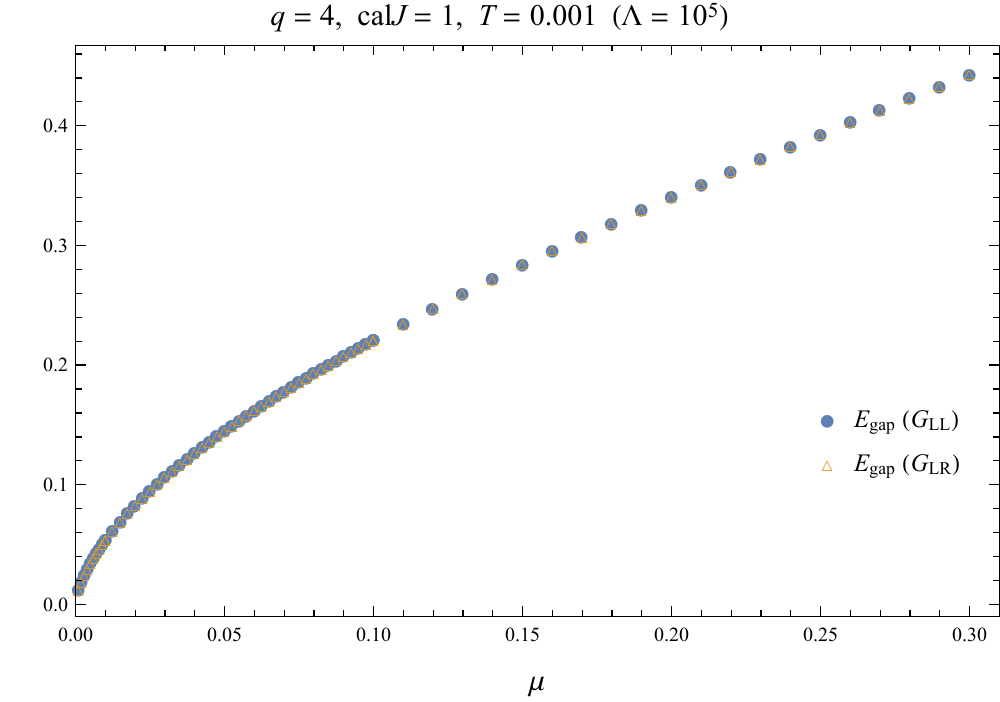}
\caption{
Left: Euclidean propagator for the wormhole phase around $\tau\sim 0$ which show exponential decay $G_{ab}(\tau)\sim e^{-E_\text{gap}\tau}$ with a common exponent $E_{\text{gap}}$ for $ab=LL,LR$.
Right: $E_\text{gap}$ obtained by fitting $G_{ab}(\tau)$ of the wormhole solutions at $T=0.001$.
}
\label{MQ_Egap}
\end{figure}

By solving the Schwinger-Dyson equations numerically by using the iteration method \cite{Maldacena:2016hyu}\footnote{
% By solving the Schwinger-Dyson equation numerically by using the iteration method \cite{Maldacena:2016hyu}\footnote{
For the numerics we have discretized $\tau$ as $\tau=\beta m/(2\Lambda)$ ($m=0,1,\cdots,2\Lambda-1$) with $\Lambda=10^5$.
% For the numerics we have discretized $\tau$ as $\tau=\frac{\beta m}{2\Lambda}$ ($m=0,1,\cdots,2\Lambda-1$) with $\Lambda=10^5$.
As the criterion for a configuration $G_{ab}(\tau),\Sigma_{ab}(\tau)$ to be a solution to the Schwinger-Dyson equations \eqref{EuclideanEoMoftwocoupledmodel2},\eqref{MQEoM1usedforrealtimeSD} we have adopted the following condition:
\begin{align}
&\quad \text{max}
\biggl\{
\biggl|{\widetilde G}_{LL}(\nu_n)+\frac{i\nu_n+{\widetilde\Sigma}_{LL}(\nu_n)}{(i\nu_n+{\widetilde\Sigma}_LL(\nu_n))^2+{\widetilde\Sigma}_{LR}(\nu_n)^2}\biggr|,
\biggl|{\widetilde G}_{LR}(\nu_n)-\frac{{\widetilde\Sigma}_{LR}(\nu_n)}{(i\nu_n+{\widetilde\Sigma}_LL(\nu_n))^2+{\widetilde\Sigma}_{LR}(\nu_n)^2}\biggr|
\biggr\}
_{n=-\Lambda}^{\Lambda-1}\nonumber \\
&\quad <2\times 10^{-9},
\label{convergencecriterion}
\end{align}
(${\widetilde G}_{ab}(\nu)=\int_0^\beta d\tau e^{i\nu\tau}G_{ab}(\tau)=(\beta/2\Lambda)\sum_{m=0}^{2\Lambda-1}e^{i\nu\beta m/(2\Lambda)}G_{ab}(\beta m/(2\Lambda))$) where $\nu_n=(2\pi/\beta)(n+1/2)$.
% (${\widetilde G}_{ab}(\nu)=\int_0^\beta d\tau e^{i\nu\tau}G_{ab}(\tau)=\frac{\beta}{2\Lambda}\sum_{m=0}^{2\Lambda-1}e^{\frac{i\nu\beta m}{2\Lambda}}G_{ab}(\frac{\beta m}{2\Lambda})$) where $\nu_n=\frac{2\pi}{\beta}(n+\frac{1}{2})$.
Note that this convergence criterion is more strict than the one adopted in \cite{Maldacena:2018lmt,Maldacena:2019ufo} (see eq(104) in \cite{Maldacena:2019ufo} which uses the average of the elements in \eqref{convergencecriterion} instead of the maximum.
}
we found\footnote{
Note that the numerical results of the Euclidean propagator (Fig.~\ref{MQ_GabEuclidean}), the phase diagram (Fig.~\ref{MQ_freeandTcTcBHTcWH}) and the energy gap $E_\text{gap}(\mu)$ (Fig.~\ref{MQ_Egap}) in this subsection as well as the real time propagator (Fig.~\ref{MQ_Grealtimeandspectralfunction}) and the first decay rate $\Gamma$ (Fig.~\ref{fittingresults}) in the next subsection were already obtained in the literatures \cite{Maldacena:2018lmt,Lantagne-Hurtubise:2019svg,Qi:2020ian,Maldacena:2019ufo,Plugge:2020wgc}.
Nevertheless, for completeness here we have repeated the same analyses in the current notation and displayed the results obtained by ourselves.
% Nevertheless, for completeness here we have repeated the same analyses in the current notation and displayed the results obtained by our own.
}
that when $\mu$ is smaller than $\mu_c\approx 0.177$ \cite{Garcia-Garcia:2019poj}, for each $\mu$ there are two distinctive solutions each of which varies continuously as the temperature is varied.
One of these two solutions exists only for $T>T_{c,\text{BH}}$ while the other exists only for $T<T_{c,\text{WH}}$ with some $T_{c,\text{BH}}(\mu),T_{c,\text{WH}}(\mu)$ which satisfies $T_{c,\text{BH}}<T_{c,\text{WH}}$.
For example for $\mu=0.1$ we have obtained $T_{c,\text{BH}}=0.032$, $T_{c,\text{WH}}=0.04$.
We call the solution exists at high temperature ``the black hole (BH) solution'' and the other ``the wormhole (WH) solution''.
See Fig.~\ref{MQ_GabEuclidean} for the profile of these two solutions.

The free energies evaluated at these two solution intersect at some $T_c(\mu)$ which satisfies $T_{c,\text{BH}}<T_c<T_{c,\text{WH}}$ hence the system exhibhts a first order phase transition at $T=T_c$.
% The free energies evaluated at these two solution intersect at some $T_c(\mu)$ which satisfies $T_{c,\text{BS}}<T_c<T_{c,\text{WH}}$ hence the system exhibhts a first order phase transition at $T=T_c$.
See Fig.~\ref{MQ_freeandTcTcBHTcWH} for the phase diagram together with the list of $T_{c,\text{BH}}$ and $T_{c,\text{WH}}$.
Note that the values of $T_{c,\text{BH}}$ and $T_{c,\text{WH}}$ we have obtained are respectively higher and lower compared with those displayed in \cite{Maldacena:2018lmt,Maldacena:2019ufo}.
These discrepancies are presumably because our criterion for a given configuration to be the solution \eqref{convergencecriterion} is more strict than that adopted in \cite{Maldacena:2019ufo}.
We have further evidence for the values of $T_{c,\text{BH}}$ and $T_{c,\text{WH}}$: (i) for several values of $\mu$ we ran the iterations with $\Delta T=\pm 0.0001$ as well as $\Delta T=\pm 0.001$, and obtained the same values of $(T_{c,\text{BH}},T_{c,\text{WH}})$; (ii) we have obtained the same values of $T_{c,\text{BH}},T_{c,\text{WH}}$ from the numerical study of the real time Schwinger-Dyson equation.
However, at present it is not clear which results are closer to the exact values of $T_{c,\text{BH}}$, $T_{c,\text{WH}}$.
% However, at present it is not clear which result is closer to the exact values of $T_{c,\text{BH}}$, $T_{c,\text{WH}}$.

\begin{figure}
\begin{center}
\includegraphics[width=9cm]{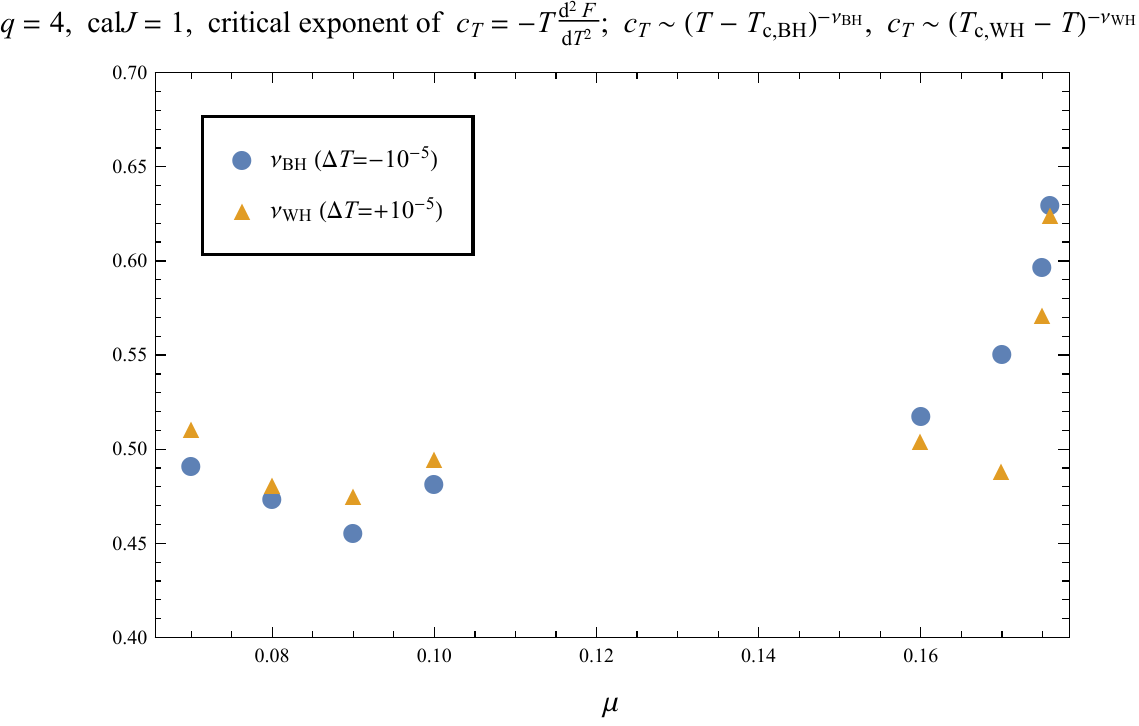}
\caption{
The critical exponents $\nu_\text{BH}$, $\nu_\text{WH}$ of the specific heat \eqref{criticalexponent_nu_by_cT} of the two coupled model with $q=4$, ${\cal J}=1$.
We have determined $\nu_\text{BH}$, $\nu_\text{WH}$ by fitting $(\partial E/\partial T)^{-1}$ near the discontinuity of $E(T)$ (Fig.~\ref{MQ_freeandTcTcBHTcWH}) by the ansatz $(\partial E/\partial T)^{-1}=c(T-T_{c,\text{BH}})^{\nu_\text{BH}}$ and $(\partial E/\partial T)^{-1}=c(T_{c,\text{WH}}-T)^{\nu_\text{WH}}$ with the fitting parameters $(c,\nu_\text{BH})$ and $(c,\nu_\text{WH})$.
}
\label{MQ_criticalexponentofcT}
\end{center}
\end{figure}

We observe that the slope of the energy $E(T)$ diverges as the temperature approaches $T_{c,\text{BH}}$ in the black hole phase or $T_{c,\text{WH}}$ in the wormhole phase, where we can define the critical exponents $\nu_\text{BH}$, $\nu_\text{WH}$ as
% We observe that the slope of the energy $E(T)$ diverges as the temperature approaches $T_{c,\text{BH}}$ in the black hole phase or $T_{c,\text{WH}}$ in the wormhole phase, where we can define the critical exponent $\nu_\text{BH}$, $\nu_\text{WH}$ as
\begin{align}
c_T=\frac{\partial E}{\partial T}\sim
\begin{cases}
(T-T_{c,\text{BH}})^{-\nu_\text{BH}}\quad (T\approx T_{c,\text{BH}},\,\text{black hole phase})\\
(T_{c,\text{WH}}-T)^{-\nu_\text{WH}}\quad (T\approx T_{c,\text{WH}},\,\text{wormhole phase})
\end{cases}.
\label{criticalexponent_nu_by_cT}
\end{align}
We have obtained $\nu_\text{BH},\nu_\text{WH}\approx 0.5$ when $\mu$ is not close to $\mu_c$, while $\nu_\text{BH},\nu_\text{WH}$ approaches $\approx 0.66$ as $\mu$ approaches $\mu_c$.
% We have obtained $\nu_\text{BH},\nu_\text{WH}\approx 0.5$ when $\mu$ is not close to $\mu_c$, while $\nu_\text{BH},\nu_\text{WH}$ approaches $\approx 0.66$ at $\mu$ approaches $\mu_c$.
See Fig.~\ref{MQ_criticalexponentofcT}.
These results are consistent with the claim that there are no phase transition in the micro canonical picture \cite{Maldacena:2018lmt}, where the black hole phase and the wormhole phase are smoothly connected by a canonically unstable intermediate phase which was called ``hot wormhole phase'' in \cite{Maldacena:2019ufo}.
Indeed, if $T(E)$ is infinitely differentiable with respect to $E$ at the canonical critical points $E_{c,\text{BH}}\equiv E(T_{c,\text{WH}})$ and $E_{c,\text{WH}}\equiv E(T_{c,\text{WH}})$, we have $T(E)=T_{c,\text{BH}}+(\cdots)(E-E_{c,\text{BH}})^m+\cdots$ and $T(E)=T_{c,\text{WH}}+(\cdots)(E_{c,\text{WH}}-E)^n+\cdots$ around these points, with $m,n$ being some integers greater than $1$.
% Indeed, if $T(E)$ is infinitely differentiable with respect to $E$ at the canonical critical points $E_{c,\text{BH}}\equiv E(T_{c,\text{WH}})$ and $E_{c,\text{WH}}\equiv E(T_{c,\text{WH}})$, we have $T(E)=T_{c,\text{BH}}+(\cdots)(E-E_{c,\text{BH}})^m+\cdots$, and $T(E)=T_{c,\text{WH}}+(\cdots)(E_{c,\text{WH}}-E)^n+\cdots$, respectively around these point, with $m,n$ being some integers greater than $1$.
Inverting these relations we find that the possible values of critical exponents are $1-1/\mathbb{N}_{>1}=1/2,2/3,\cdots$.
% Inverting these relations we find that the possible values of critical exponents are $1-\frac{1}{\mathbb{N}_{>1}}=\frac{1}{2},\frac{2}{3},\cdots$.
The critical exponents we found for $\mu<\mu_c$ and for $\mu\rightarrow \mu_c$ are close to $1/2$ and $2/3$ respectively.
% The critical exponents we found for $\mu<\mu_c$ and for $\mu\rightarrow \mu_c$ are close to $\frac{1}{2}$ and $\frac{2}{3}$ respectively.

As we can see from Fig.~\ref{MQ_Egap} (left), the wormhole solution exhibits exponential decay $G_{ab}(\tau)\sim e^{-E_\text{gap}\tau}$ which indicates that the system is gapped with the energy gap $E_\text{gap}$.
% As we can see from Fig.~\ref{MQ_Egap}, the wormhole solution exhibits exponential decay $G_{ab}(\tau)\sim e^{-E_\text{gap}\tau}$ which indicates that the system is gapped with the energy gap $E_\text{gap}$.
Indeed when the temperature is sufficiently low ($\beta$ is large) the two point functions can be expanded as
\begin{align}
\langle \psi^a_i(\tau)
\psi_i^b(0)
% \psi_i^b
\rangle_\beta
&=
\frac{1}{Z(\beta)}\text{Tr}
e^{\tau{\widehat H}}
{\widehat \psi}^a_i
e^{-\tau{\widehat H}}
{\widehat \psi}_i^b
e^{-\beta{\widehat H}}\nonumber \\
&=
\frac{1}{Z(\beta)}\sum_{m,n}
\langle E_m|{\widehat \psi}_i^a|E_n\rangle
\langle E_n|{\widehat \psi}_i^b|E_m\rangle
e^{-\beta E_m+\tau (E_m-E_n)}\nonumber \\
&\approx
\langle E_0|{\widehat \psi}_i^a|E_0\rangle
\langle E_0|{\widehat \psi}_i^b|E_0\rangle
+
\langle E_0|{\widehat \psi}_i^a|E_1\rangle
\langle E_1|{\widehat \psi}_i^b|E_0\rangle
e^{-(E_1-E_0)\tau}+\cdots,
\end{align}
where the fist term is zero since the Hamiltonian of the two coupled model preserves parity (fermion number) symmetry.
See Fig.~\ref{MQ_Egap} (right) for the values of $E_\text{gap}(\mu)$ which we have obtained by fitting the Euclidean propagators at $T=0.001$.
% See Fig.~\ref{MQ_Egap} for the $E_\text{gap}(\mu)$ which we have obtained by fitting the Euclidean propagators at $T=0.001$.

For $\mu>\mu_c$ there is only one solution with which both the BH/WH solutions in the subcritical regime are smoothly connected.

\subsubsection{Real time propagator $G_{ab}^>(t)$ and decay rates $\Gamma$}
\begin{figure}
\includegraphics[width=8cm]{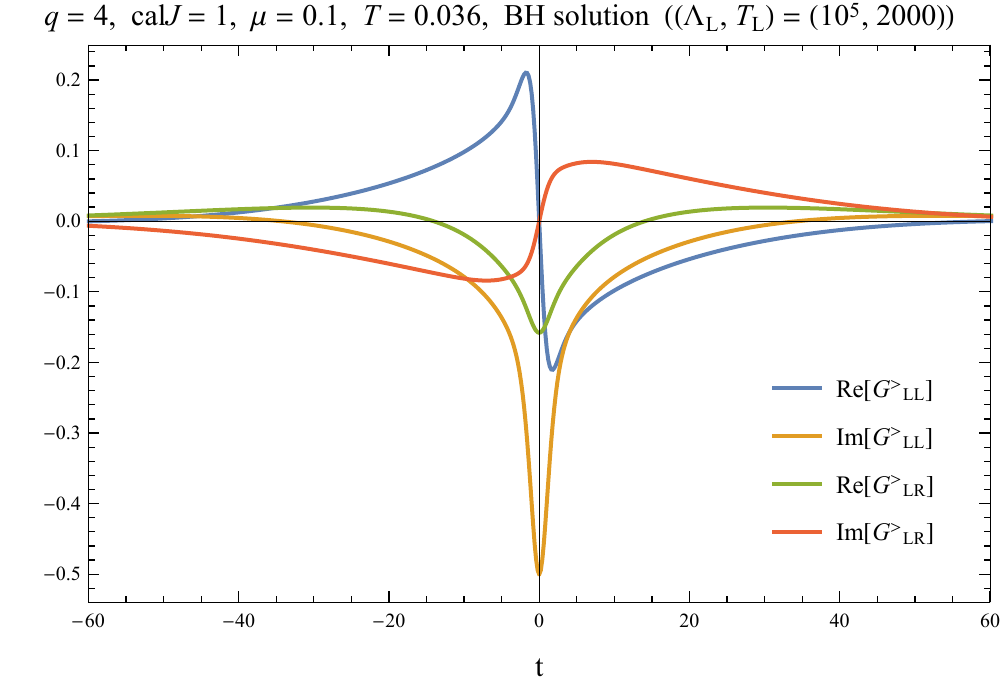}\quad\quad
\includegraphics[width=8cm]{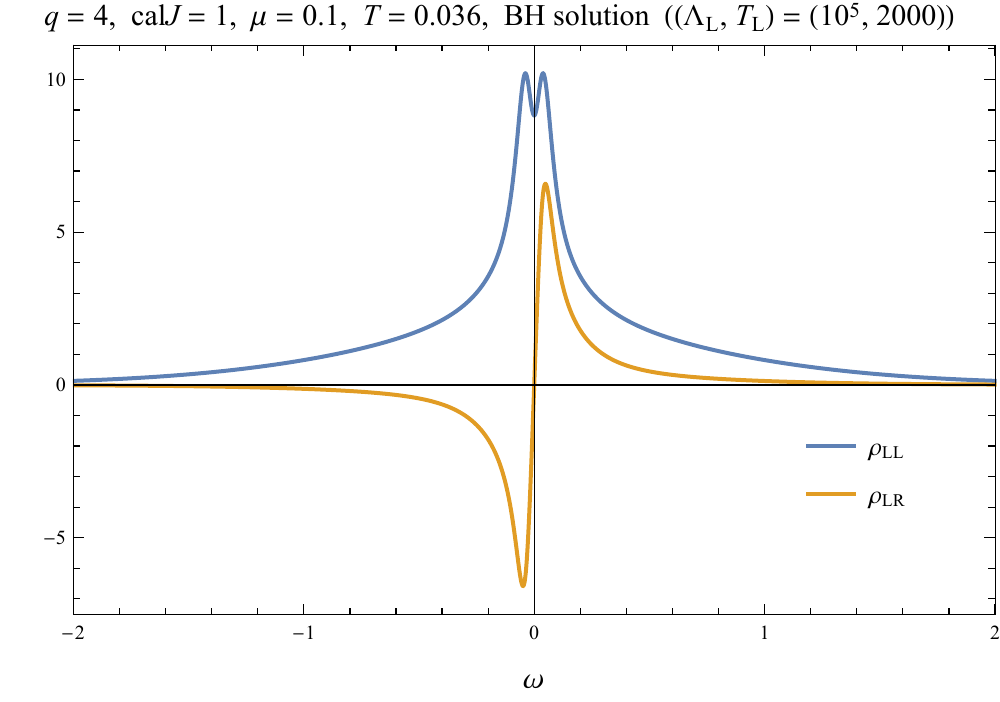}\\
\includegraphics[width=8cm]{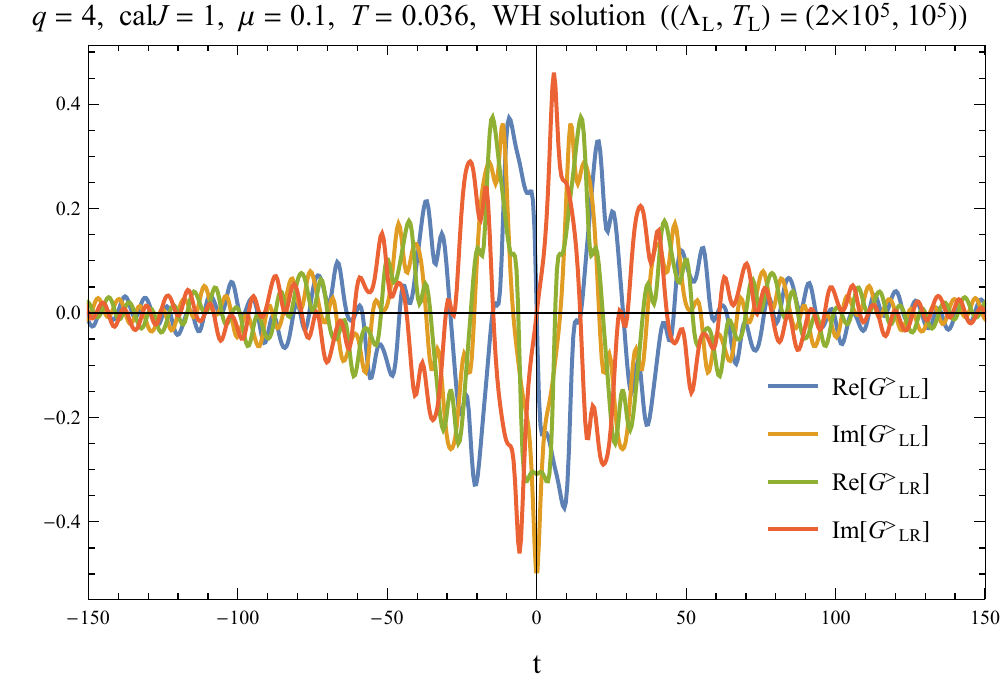}\quad\quad
\includegraphics[width=8cm]{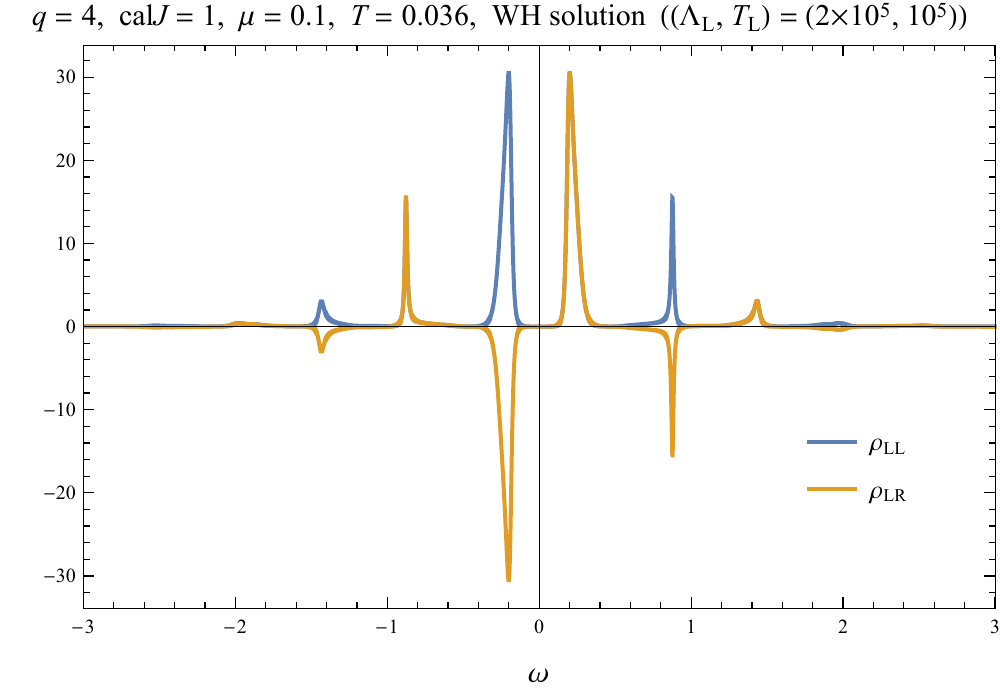}
\caption{
Top left/right: propagators/spectral functions $\rho_{LL}(\omega)=-2\text{Im}[{\widetilde G}^R_{LL}(\omega)]$, $\rho_{LR}(\omega)=-2\text{Re}[{\widetilde G}^R_{LR}(\omega)]$ for the black hope solution.
% Top left/right: propagators/spectral functions $\rho_{LL}=\text{Im}[{\widetilde G}^R_{LL}(\omega)]$, $\rho_{LR}=\text{Re}[{\widetilde G}^R_{LR}(\omega)]$ for the black hope solution.
Bottom left/right: propagators/spectral functions for the wormhole solution.
}
\label{MQ_Grealtimeandspectralfunction}
\end{figure}

% \subsubsection{Real time propagator $G_{ab}^R(t)$ and decay rates $\Gamma$}
\label{sec_result_realtime_MQ}
By solving the real time Schwinger-Dyson equations \eqref{MQrealtimeSDfinal1},\eqref{MQrealtimeSDfinal2} we found two distinctive solutions for each single point on the $\mu$-$T$ plane around the line $T=T_c(\mu)$ (see Fig.~\ref{MQ_Grealtimeandspectralfunction}).\footnote{
In the numerics for the real time formalism we have to introduce both the UV cutoff and the IR cutoff, since $t$ is not compactified like $\tau\sim \tau+\beta$.
We have chosen the UV/IR cutoff as $t=T_L/(2\Lambda_L)(m+1/2)$ ($m=-\Lambda_L,-\Lambda_L+1,\cdots,\Lambda_L-1$) with $(\Lambda_L,T_L)=(10^5,2000)$.
% We have chosen the UV/IR cutoff as $t=\frac{T_L}{2\Lambda_L}(m+\frac{1}{2})$ ($m=-\Lambda_L,-\Lambda_L+1,\cdots,\Lambda_L-1$) with $(\Lambda_L,T_L)=(10^5,2000)$.
As we mention later, in the wormhole phase we have also performed the numerics with $(\Lambda_L,T_L)=(2\times 10^5,10^5)$.
}
These two solutions correspond respectively to the black hole phase and the wormhole phase.
Indeed, by calculating the Euclidean propagator from the spectral function of each solution by \eqref{fixed200806}, we have obtained precisely the same configuration as those obtained by directly solving the Euclidean Schwinger-Dyson equations.
We have also found that the real time BH/WH solution stops to exist precisely at $T=T_{c,\text{BH}}$/$T=T_{c,\text{WH}}$ as we decrease/increase the temperature slowly, as we have mentioned in the previous subsection.
% We also found that the real time BH/WH solution stops to exist precisely at $T=T_{c,\text{BH}}$/$T=T_{c,\text{WH}}$ as we decrease/increase the temperature slowly, as we have mentioned in the previous subsection.
In Fig.~\ref{MQ_Grealtimeandspectralfunction} we have displayed the real time propagator $G_{ab}^>(t)$ together with the spectral functions $\rho_{LL}(\omega)=-2\text{Im}[{\widetilde G}_{LL}(\omega)]$, $\rho_{LR}(\omega)=-2\text{Re}[{\widetilde G}_{LR}(\omega)]$ of BH/WH phase for $\mu=0.1$, $T=0.036$.
% In Fig.~\ref{MQ_Grealtimeandspectralfunction} we have displayed the real time propagator $G_{ab}^>(t)$ together with the spectral functions $\rho_{LL}=\text{Im}[{\widetilde G}_{LL}(\omega)]$, $\rho_{LR}=\text{Re}[{\widetilde G}_{LR}(\omega)]$ of BH/WH phase for $\mu=0.1$, $T=0.036$.
Note that these two quantities $(G_{ab}^>(t),\rho_{ab}(\omega))$ are not independent with each other; given one of them one can construct the other through \eqref{definitionofRA},\eqref{MQrealtimeSDfinal2}.

\begin{figure}
\includegraphics[width=8cm]{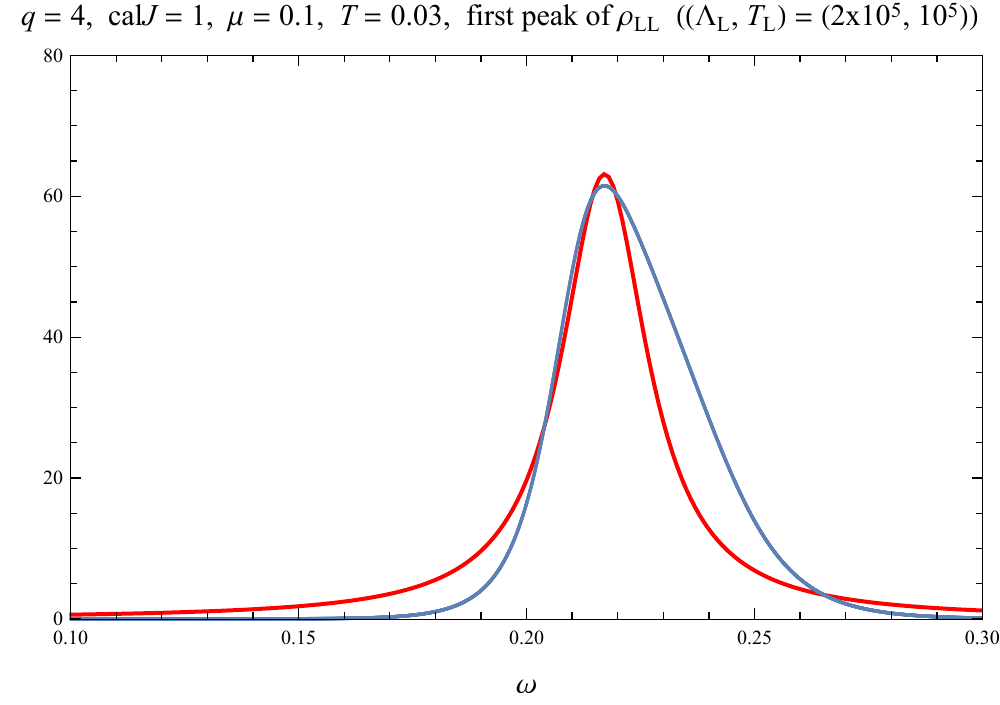}\quad
\includegraphics[width=8cm]{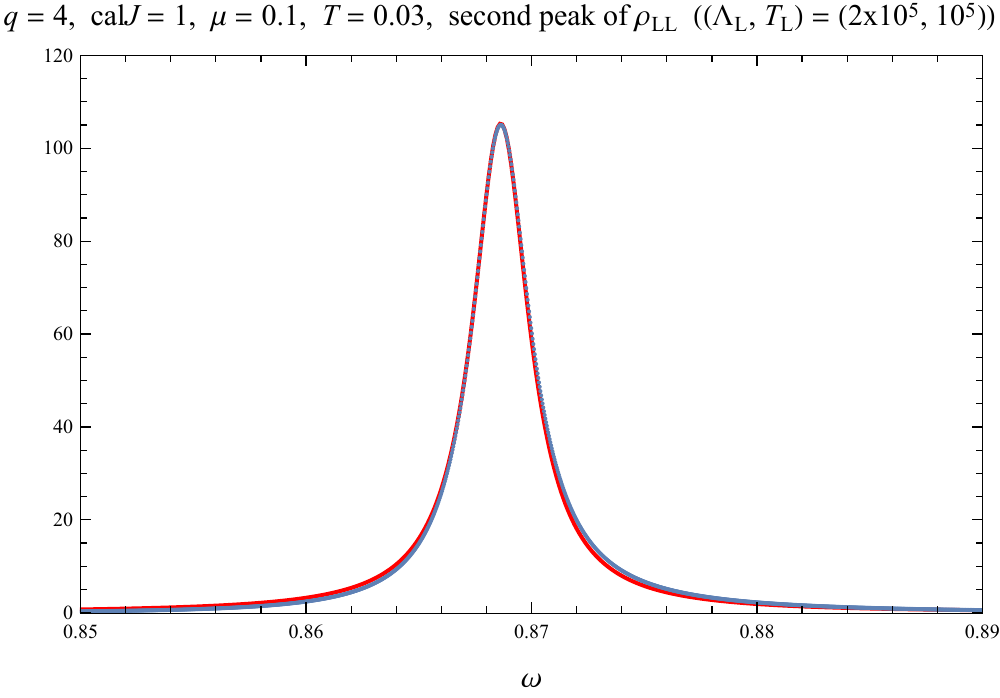}\\
\includegraphics[width=8cm]{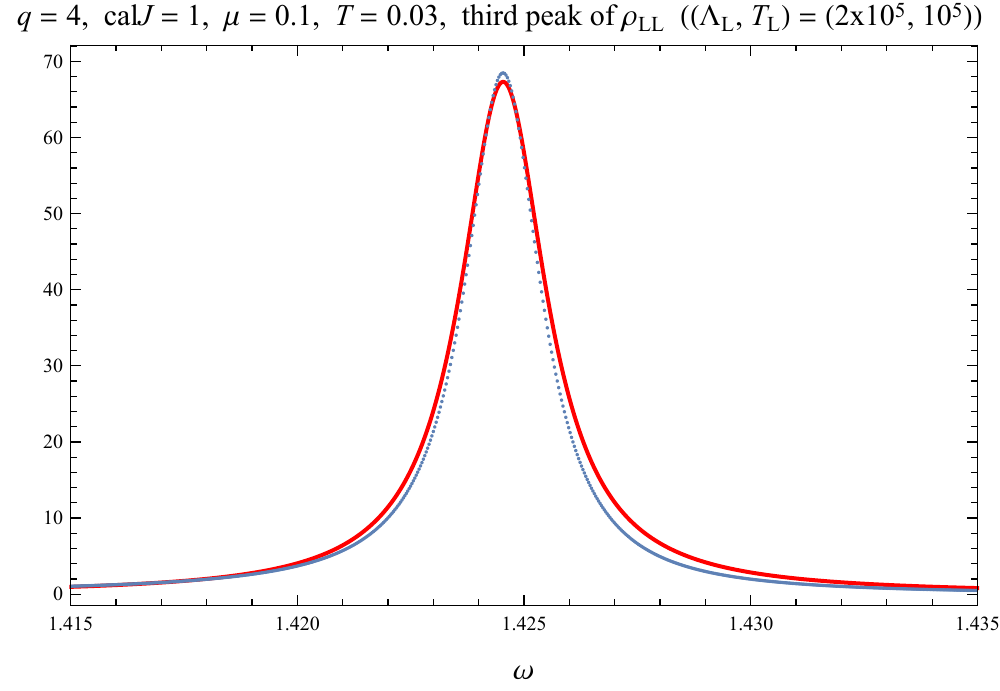}\quad\quad\quad\quad
\begin{tabular}{|c|c|c|c|c|c|c|}
\multicolumn{7}{c}{}\vspace{-6cm}\\
\hline
         &$\omega_0$&$A$       &$\Gamma$\\ \hline\hline
$LL$, 1st&$0.217$   &$2.28$    &$0.0115$\\ \hline
$LL$, 2nd&$0.869$   &$0.508$   &$0.00153$\\ \hline
$LL$, 3rd&$1.42$    &$0.243$   &$0.00115$\\ \hline\hline
$LR$, 1st&$0.217$   &$2.28$    &$0.0115$\\ \hline
$LR$, 2nd&$0.869$   &$0.508$   &$0.00153$\\ \hline 
$LR$, 3rd&$1.42$    &$0.242$   &$0.00114$\\ \hline
\end{tabular}
\caption{
Top left/top right/bottom left: first/second/third peak of the spectral function $\rho_{LL}(\omega)$ for $q=4$, ${\cal J}=1$, $\mu=0.1$, $T=0.03$ with $(\Lambda_L,T_L)=(2\times 10^5,10^5)$.
Red lines: fitting curve $\frac{A}{\pi}\frac{\Gamma}{(\omega-\omega_0)^2+\Gamma^2}$.
Here we have determined $A,\omega_0$ separately as the integration of $\rho_{LL}$ around the peak and the local maximim, and used only $\Gamma$ as the fitting parameter.
Bottom right: results of fitting for the first three peaks of $\rho_{LL}(\omega)$, $\rho_{LR}(\omega)$.
}
\label{fittingresults}
\end{figure}

Here are additional remarks on the wormhole solution.
As shown in Fig.~\ref{MQ_Grealtimeandspectralfunction}, the spectral functions $\rho_{ab}(\omega)$ of the wormhole solution split into sharp peaks.
We find that the position of the peaks are same for $\rho_{LL}(\omega)$ and $\rho_{LR}(\omega)$ and, in particular, the position of the first peak is in good agreement with $E_\text{gap}$ obtained by fitting the Euclidean propagator (see Fig.~\ref{MQ_Egap}).
Indeed, if $\rho_{ab}(\omega)$ were given as $\rho_{LL}(\omega)=A_{LL}(\delta(\omega-E_\text{gap})+\delta(\omega+E_\text{gap}))$ and $\rho_{LR}(\omega)=A_{LR}(\delta(\omega-E_\text{gap})-\delta(\omega+E_\text{gap}))$, then from \eqref{fixed200806} we obtain
\begin{align}
G_{LL}(\tau)&=\frac{A_{LL}}{\pi}\Bigl(
\frac{e^{-E_\text{gap}\tau}}{1+e^{-\beta E_\text{gap}}}
+\frac{e^{E_\text{gap}\tau}}{1+e^{\beta E_\text{gap}}}
\Bigr)
\approx \frac{A_{LL}}{\pi}e^{-E_\text{gap}\tau},\nonumber \\
G_{LR}(\tau)&=-\frac{iA_{LR}}{\pi}\Bigl(
\frac{e^{-E_\text{gap}\tau}}{1+e^{-\beta E_\text{gap}}}
-\frac{e^{E_\text{gap}\tau}}{1+e^{\beta E_\text{gap}}}
\Bigr)
\approx -\frac{iA_{LR}}{\pi}e^{-E_\text{gap}\tau}.
\end{align}
The situation is not completely same with the actual result of $\rho_{ab}$ where there are infinitely many other peaks and each peak is of finite width.
Each peak can be fit well with $A\delta_\Gamma(\omega-\omega_0)$ with $\delta_\Gamma(\omega)=-\frac{1}{\pi}\text{Im}[\frac{1}{\omega-\omega_0+i\Gamma}]$ (see Fig.~\ref{fittingresults}), which corresponds to a particle of finite lifetime:
% Each peak can be fit well with $A\delta_\Gamma(\omega-\omega_0)$ with $\delta_\Gamma(\omega)=\frac{1}{\pi}\text{Im}[\frac{1}{\omega-\omega_0+i\Gamma}]$ (see Fig.~\ref{fittingresults}), which corresponds to a particle of finite lifetime:
\begin{align}
&\rho_{LL}(\omega)=
A_{LL}(\delta_\Gamma(\omega-E_\text{gap})+\delta_\Gamma(\omega+E_\text{gap}))\nonumber \\
&\quad \Rightarrow\quad G^>_{LL}=-\frac{iA_{LL}}{\pi}\Bigl(
\frac{e^{-iE_\text{gap}t}}{1+e^{-\beta E_\text{gap}}}
+\frac{e^{iE_\text{gap}t}}{1+e^{\beta E_\text{gap}}}
\Bigr)
e^{-\Gamma|t|},\nonumber \\
&\rho_{LR}(\omega)=
A_{LR}(\delta_\Gamma(\omega-E_\text{gap})-\delta_\Gamma(\omega+E_\text{gap}))\nonumber \\
&\quad \Rightarrow\quad G^>_{LR}=-\frac{A_{LR}}{\pi}\Bigl(
\frac{e^{-iE_\text{gap}t}}{1+e^{-\beta E_\text{gap}}}
-\frac{e^{iE_\text{gap}t}}{1+e^{\beta E_\text{gap}}}
\Bigr)
e^{-\Gamma|t|}.
\end{align}
The decay width of each peak decreases as the temperature decreases.
In order the finite IR cutoff $|t|<T_L/2$ to be a good approximation to the reality $t\in(-\infty,\infty)$, $T_L$ has to be sufficiently larger than the inverse of the decay rates so that $G_{ab}^R(t),G_{ab}^>(t)\approx 0$ at the IR cutoff and the effect of compactification $t\sim t+t_L$ is negligible.
% In order the finite IR cutoff $|t|<\frac{T_L}{2}$ to be a good approximation to the reality $t\in(-\infty,\infty)$, $T_L$ has to be sufficiently larger than the inverse of the decay rates so that $G_{ab}^R(t),G_{ab}^>(t)\approx 0$ at the IR cutoff and the effect of compactification $t\sim t+t_L$ is negligible.
For example, for $\mu=0.1$, if we choose $\Lambda_L,T_L$ as $\Lambda_L=10^5$, $T_L=2000$ we could solve the Schwinger-Dyson equation only for $T\ge 0.03$.
In general when we increase $T_L$ we also have to increase the number of the discrete points $\Lambda_L$ at the same rate to keep the UV resolution, which makes the numerics at low temperature difficult.
However, we found that the weight of the peak $A$ is smaller for the higher peaks.
% On the other hand, we found that the weight of the peak $A$ is smaller for the higher peaks.
In particular, for $\mu=0.1$, $T=0.03$ the total weight of the first three peaks of $\rho_{LL}$ is $6.04$,
which is $96.1\%$ of the total weight
$\int d\omega\rho_{LL}(\omega)=2\pi i(G_{LL}^R(+0)-(G_{LL}^R(-0))^*)=2\pi$;
$\rho_{LL}$ is well approximated by the contributions of only first three peaks.
We found this is the case also for other values of $\mu,T$ as long as the temperature is low enough so that the peaks are well separated.
This fact implies that the sufficient value of $\Lambda_L$ relative to $T_L$ is such that $\omega_\text{max}=(\pi/T_L)(\Lambda_L-1/2)$ is larger than the position of the third peak.
This required value is much smaller than $\omega_\text{max}=157$ for $(\Lambda_L,T_L)=(10^5,2000)$, hence we can improve the numerics at low temperature by just increasing $T_L$ with $\Lambda_L$ kept the same.
Indeed, by choosing $(\Lambda_L,T_L)=(2\times 10^5,10^5)$, for $\mu=0.1$ we achieved to reach down to $T=0.019$.

\begin{figure}
\includegraphics[width=8cm]{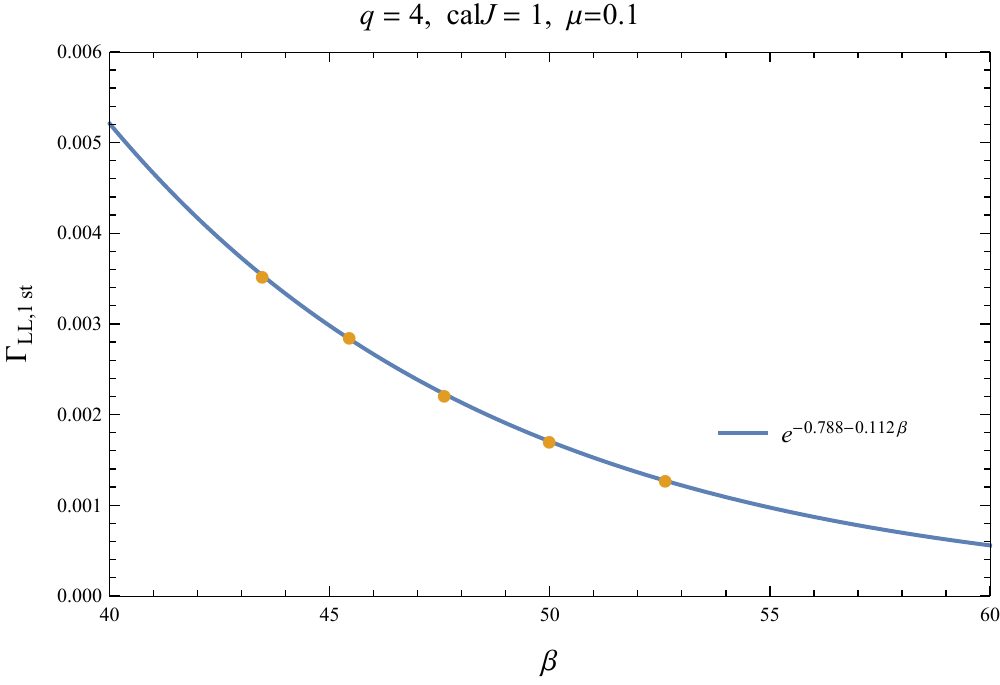}\quad
\includegraphics[width=8cm]{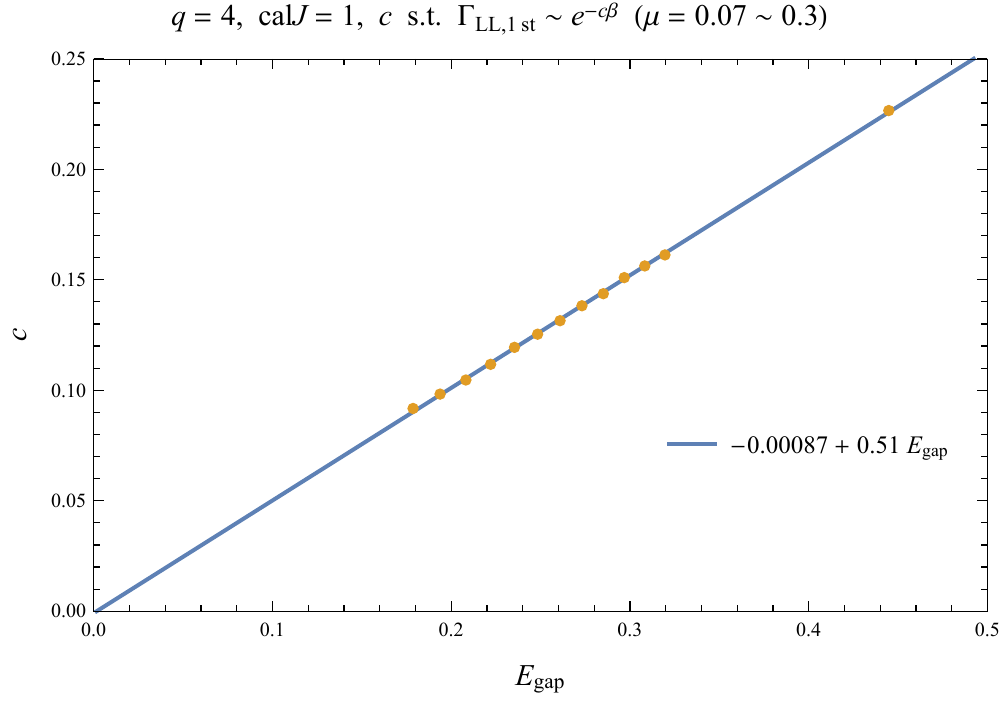}
\caption{
Left: Fitting of decay width of the first peak of $\rho_{LL}$ $\Gamma_{LL,\text{1st}}$ for $\mu=0.1$ with $e^{c_1-c_2\beta}$.
Right: Comparison of the fitting coefficient $c_2$ with $E_\text{gap}$.
}
\label{fittingofGammaLL1st}
\end{figure}

We have displayed in Fig.~\ref{fittingresults} the fitting results for the first three peaks of the wormhole solution at $\mu=0.1$, $T=0.03$.
The results we have obtained are consistent with those displayed in \cite{Qi:2020ian,Plugge:2020wgc}.
We have also found that the decay rate of the first peak $\Gamma_{\text{1st}}$ obeys the following relation with $E_\text{gap}$
\begin{align}
\Gamma_\text{1st}\sim e^{-\frac{1}{2}E_\text{gap}\beta}
\label{Gamma1st_QiZhangformula}
\end{align}
up to some overall constant which is independent of $T$, as argued in \cite{Qi:2020ian}.
% up to some overall constant which is possibly depends on $\mu$ but independent of $T$, as argued in \cite{Qi:2020ian}.
See Fig.~\ref{fittingofGammaLL1st}.
Interestingly, we have found that the chaos exponent $\lambda_L$ also obeys the same formula in the wormhole phase, as we display in the next subsection.
% Interestingly, we found that the chaos exponent $\lambda_L$ also obeys the same formula in the wormhole phase, as we display in the next subsection.

\subsubsection{Chaos exponent}
\begin{figure}
\begin{center}
\includegraphics[width=8cm]{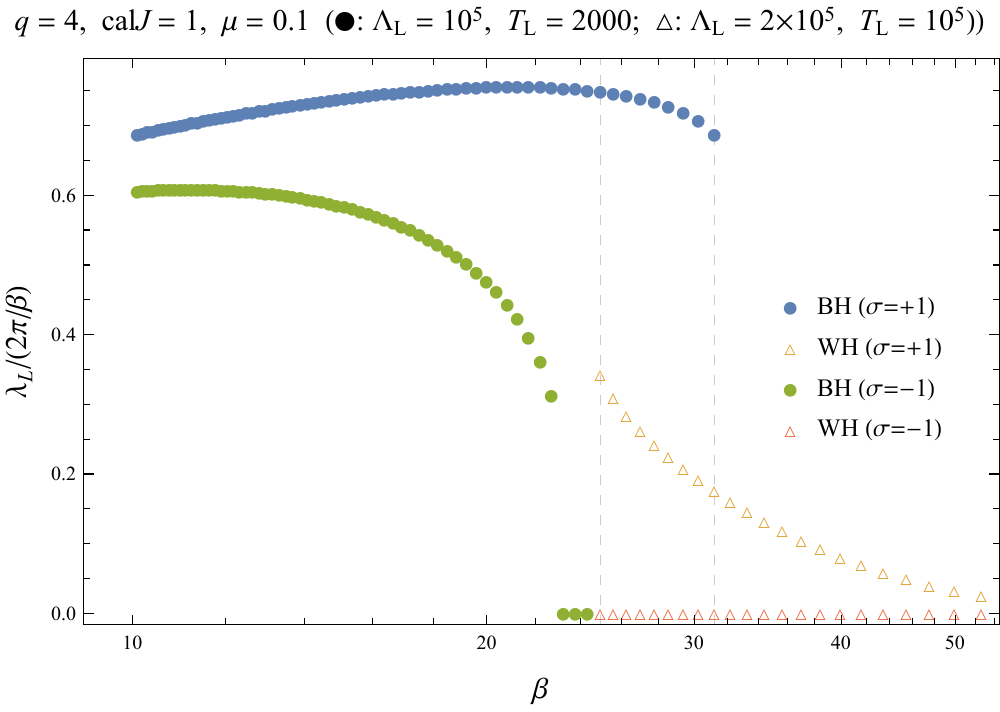}
\end{center}
\caption{Chaos exponent of the two coupled model with $q=4$, ${\cal J}=1$, $\mu=0.1$ computed separately for $L\leftrightarrow R$ even/odd sector ($\sigma=\pm 1$).}
\label{fig_MQLyapunovsplitbysymmetry}
\end{figure}

We can compute the chaos exponent of the two coupled model by solving the ladder equation \eqref{ladderfinalMQ} with the ladder kernel evaluated on the real time propagators obtained in the previous section.
As we have seen in \eqref{ladderMQfinalsimplified}, the ladder equation decomposes into the two sectors which are even/odd under the $L\leftrightarrow R$ flipping \eqref{flipsymmetryofMQladdereq}, hence we can compute the chaos exponent for each sector separately.
We have observed that the chaos exponent of the even ($\sigma=+1$) sector is always larger than that of the odd ($\sigma=-1$) sector (see Fig.~\ref{fig_MQLyapunovsplitbysymmetry}), hence below we focus on the even sector.

\begin{figure}
\includegraphics[width=8cm]{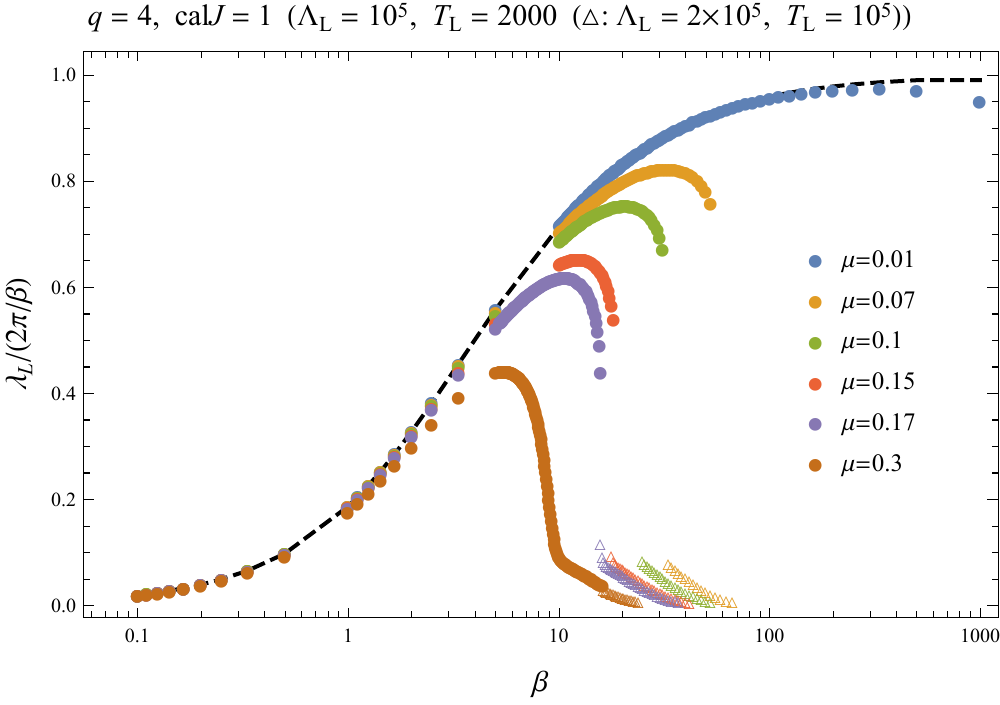}\quad\quad
\includegraphics[width=8cm]{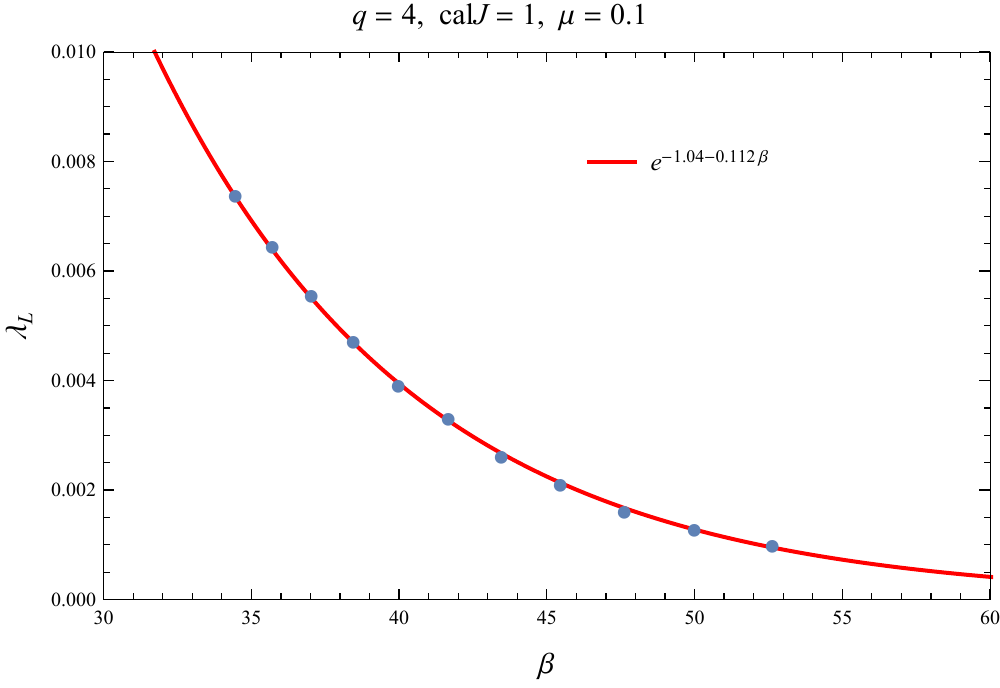}\\
\includegraphics[width=8cm]{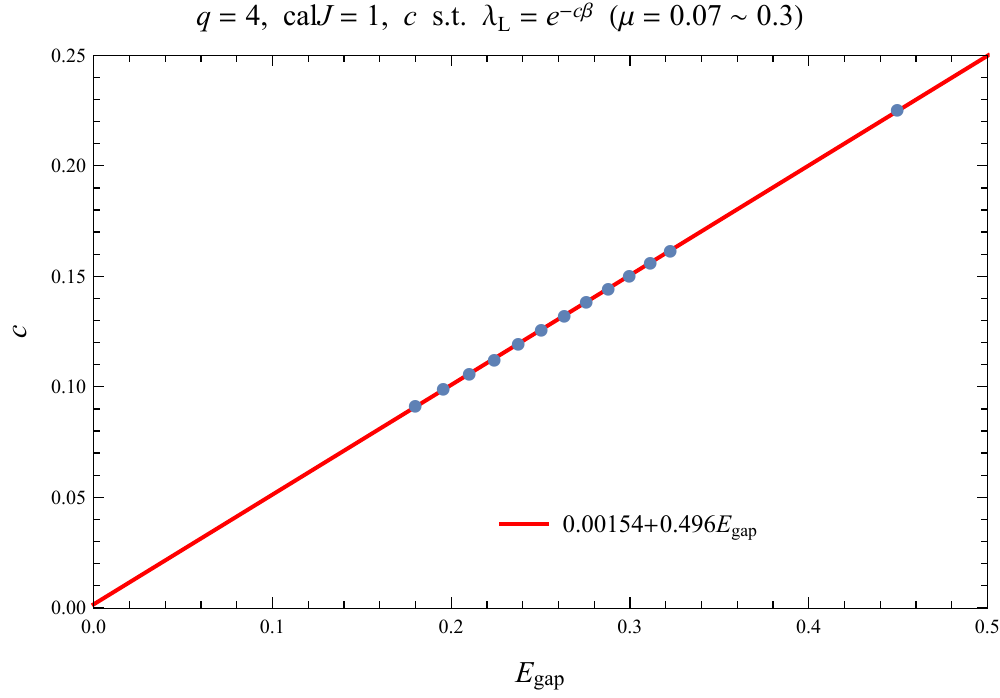}
\caption{
Top left: chaos exponent for the black hole solution and the wormhole solution, where the dashed black line is the chaos exponent of the pure SYK model $\mu=0$.
Top right: Fitting of the chaos exponent for $\mu=0.1$ with $e^{c_1-c_2\beta}$.
Bottom: Comparison of the fitting coefficient $c_2$ with $E_\text{gap}$.
}
\label{Lyapunovfigure}
\end{figure}

In Fig.~\ref{Lyapunovfigure} we have displayed the chaos exponent of the even sector for various $\mu$ in the two phases.
% In Fig.~\ref{Lyapunovfigure} we have displayed the chaos exponent of the two coupled model obtained by the procedure explained in section \ref{sec_chaosexponent_MQ}.
It is remarkable that the chaos exponent is small but non-zero even in the wormhole phase.
This is indeed consistent with the fact that the decay rate is small but non-zero in the same phase, as we have seen in the previous subsection; the system thermalize, which is another indication for the system to be quantum chaotic.
Furthermore, we have found that the chaos exponent obeys completely the same formula \eqref{Gamma1st_QiZhangformula} as the decay rate of the first peak when the temperature is low enough
\begin{align}
\lambda_L\sim e^{-\frac{1}{2}\beta E_\text{gap}},
\label{LambdaL_QiZhangformula}
\end{align}
up to an overall factor which is independent of $T$.
See Fig.~\ref{Lyapunovfigure}.
We have found this formula is also satisfied for $\mu>\mu_c$ where there are no phase transition, if the temperature is sufficiently low, as was the case also for $\Gamma_\text{1st}$.
% We found this formula is also satisfied for $\mu>\mu_c$ where there are no phase transition, if the temperature is sufficiently low, as was the case also for $\Gamma_\text{1st}$.

\begin{figure}
\includegraphics[width=8cm]{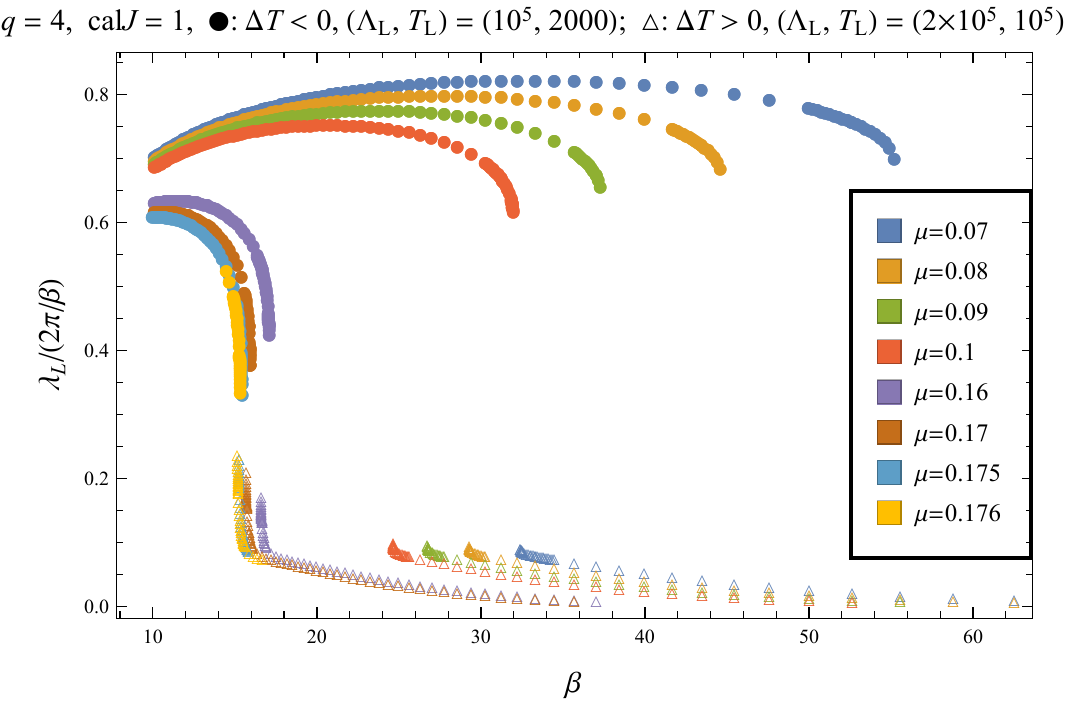}\quad\quad
\includegraphics[width=8cm]{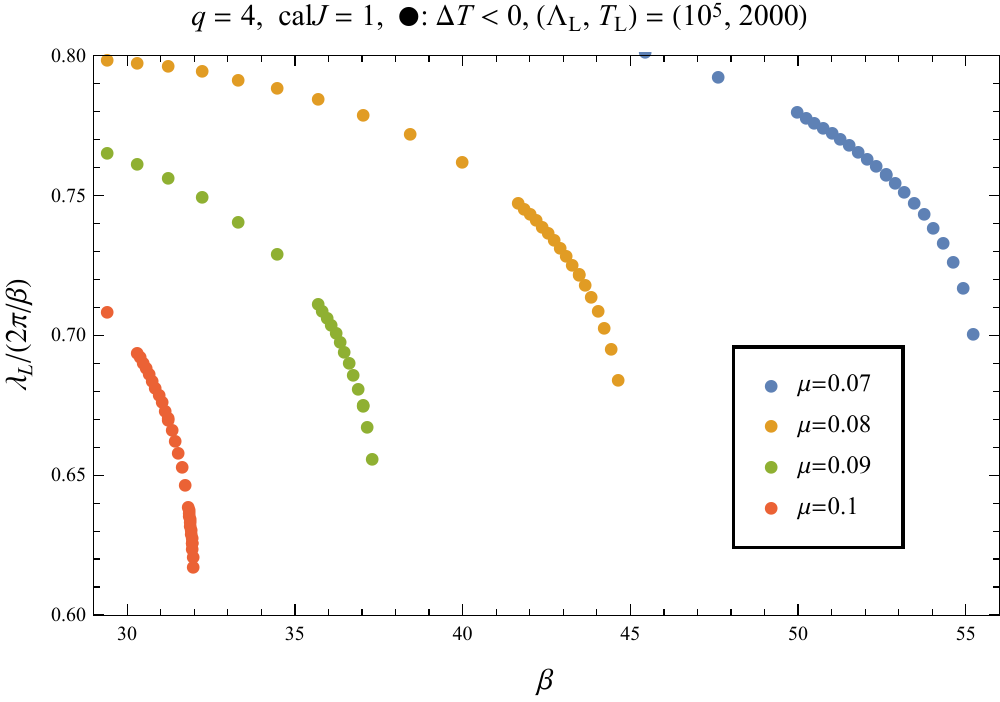}\\
\includegraphics[width=8cm]{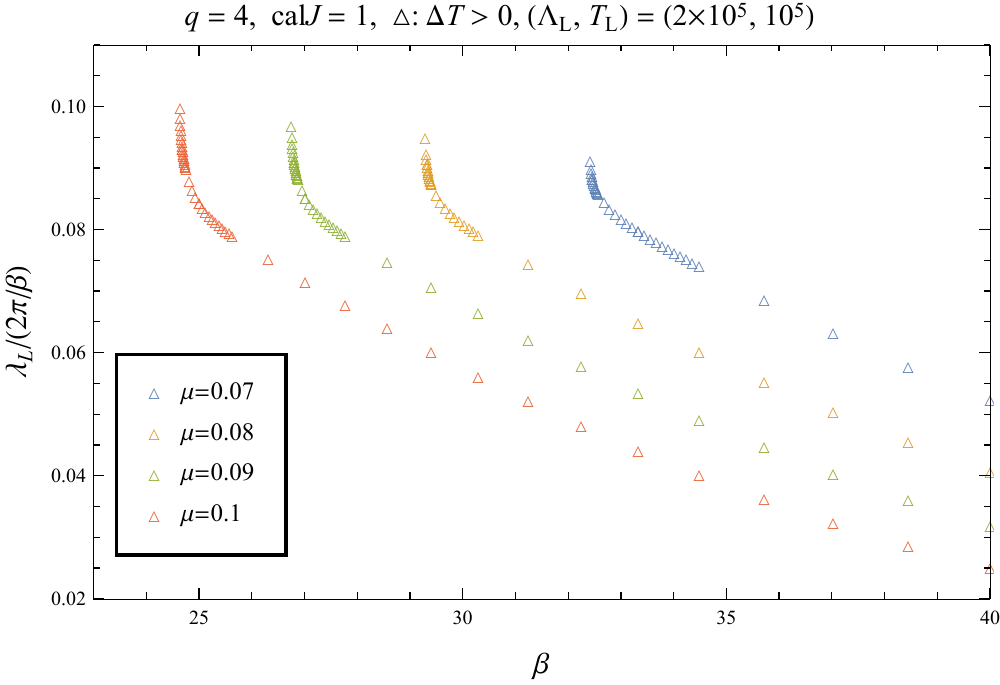}\quad\quad
\includegraphics[width=8cm]{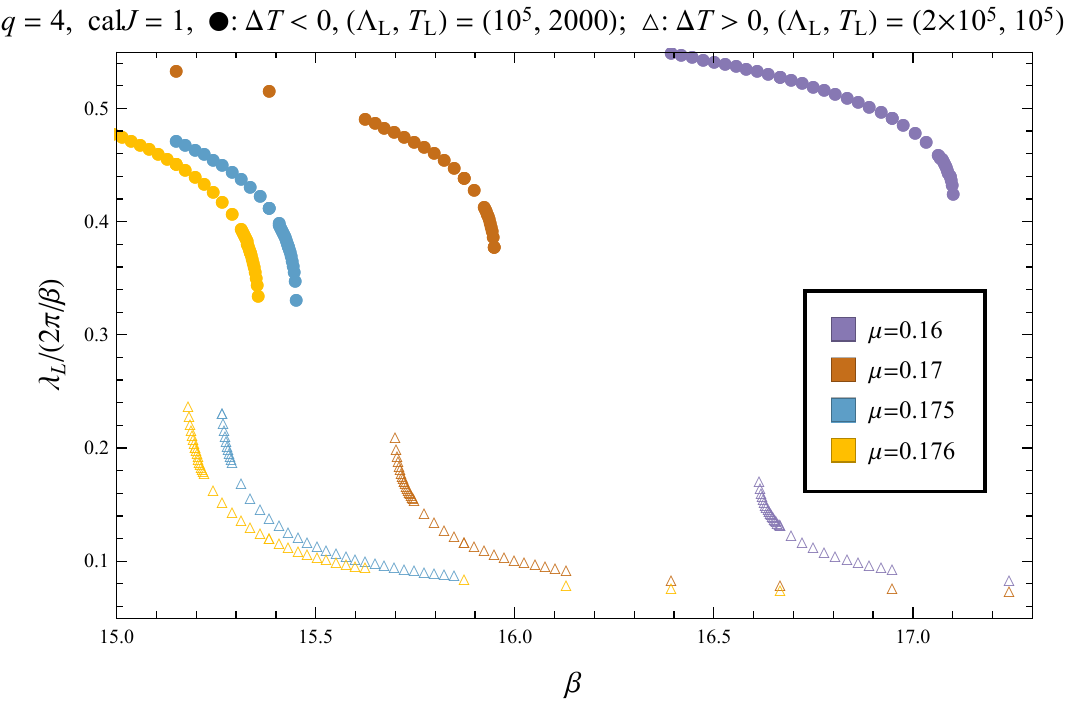}
\caption{
The chaos exponent of the two coupled model near $T=T_{c,\text{BH}}$ and $T=T_{c,\text{WH}}$ ($q=4$, ${\cal J}=1$).
The data point in the black hole phase are generated with $(\Lambda_L,T_L)=(10^5,2000)$ and $\Delta T=-10^{-3},-10^{-4},-10^{-5}$, while the data points in the wormhole phase are generated with $(\Lambda_L,T_L)=(2\times 10^5,10^5)$ and $\Delta T=10^3,10^4,10^5$.
}
\label{fig_MQLyapunovnearTcBHandTcWH}
\end{figure}

\begin{figure}
\begin{center}
\includegraphics[width=10cm]{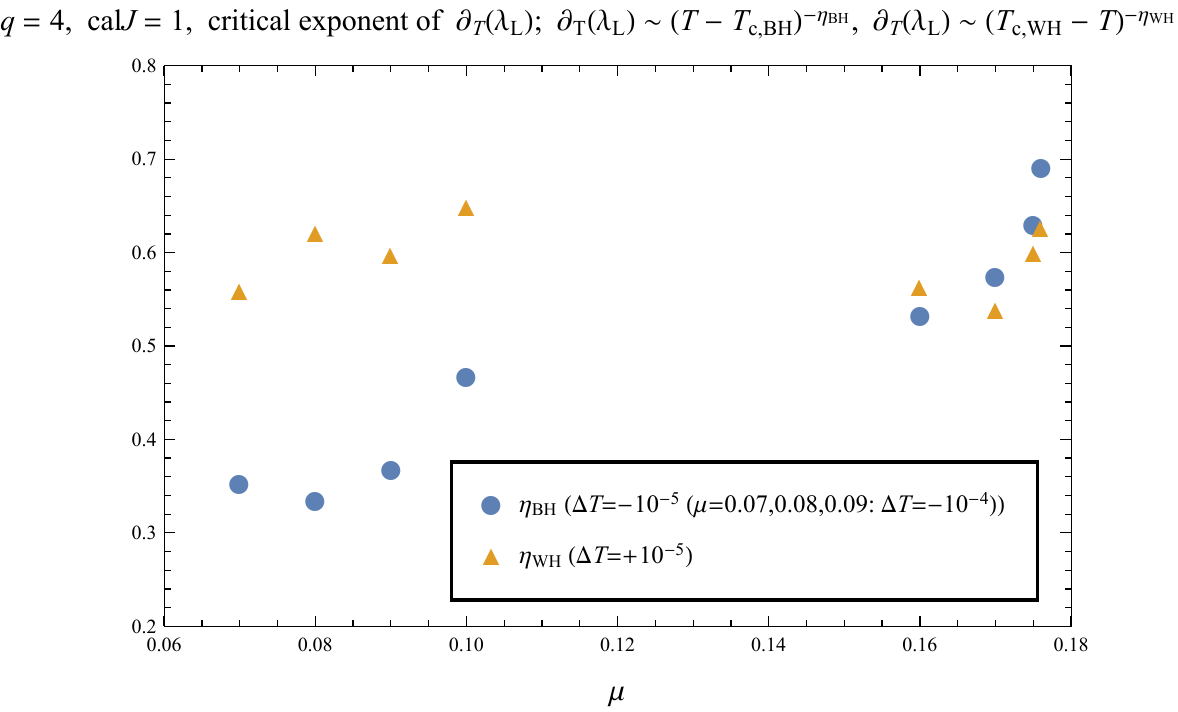}
\caption{
The critical exponent $\eta_{\text{BH}}$ and $\eta_{\text{WH}}$ of the chaos exponent \eqref{criticalexponent_eta_by_LambdaL} of the two coupled model with $q=4$, ${\cal J}=1$.
We have determined $\eta_\text{BH}$, $\eta_\text{WH}$ by fitting $(\partial_T\lambda_L)^{-1}$ by the ansatz $(\partial_T\lambda_L)^{-1}=c(T-T_{c,\text{BH}})^{\eta_\text{BH}}$ and $(\partial_T\lambda_L)^{-1}=c(T_{c,\text{WH}}-T)^{\eta_\text{WH}}$ with the fitting parameters $(c,\eta_\text{BH})$ and $(c,\eta_\text{WH})$.
}
\label{fig_criticalexponent}
\end{center}
\end{figure}

We also observe that the slope of the chaos exponent of the black hole phase $\partial \lambda_L/\partial T$ diverges as the temperature approaches $T_{c,\text{BH}}$ (Fig.~\ref{fig_MQLyapunovnearTcBHandTcWH}).
% We also observe that the slope of the chaos exponent of the black hole phase $\frac{\partial \lambda_L}{\partial T}$ diverges as the temperature approaches $T_{c,\text{BH}}$ (Fig.~\ref{fig_MQLyapunovnearTcBHandTcWH}).
Similarly, the slope also seems to diverge in the wormhole phase at $T=T_{c,\text{WH}}$.
From the detailed analysis close to $T=T_{c,\text{BH}}$ and $T=T_{c,\text{WH}}$ we have identified the critical exponent as
\begin{align}
\frac{\partial \lambda_L}{\partial T}\sim\begin{cases}
&(T-T_{c,\text{BH}})^{\eta_{\text{BH}}}\quad (T\approx T_{c,\text{BH}},\, \text{black hole phase})\\
&(T_{c,\text{WH}}-T)^{\eta_{\text{WH}}}\quad (T\approx T_{c,\text{WH}},\, \text{wormhole phase})
\end{cases}
\label{criticalexponent_eta_by_LambdaL}
\end{align}
with $\eta_{\text{BH}}$ and $\eta_\text{WH}$ displayed in Fig.~\ref{fig_criticalexponent}.
In particular, as $\mu$ approaches $\mu_c=0.177$ the two critical exponents almost coincide around $\eta\approx 2/3$.
% In particular, as $\mu$ approaches $\mu_c=0.177$ the two critical exponents almost coincide around $\eta=0.66\approx \frac{2}{3}$.
This agrees with the behavior of the critical exponent $\nu_\text{BH},\nu_\text{WH}$ defined by the specific heat \eqref{criticalexponent_nu_by_cT}, and is consistent with the fact that for $\mu\ge \mu_c$ the phase transition disappears and the two phases are smoothly connected.
On the other hand, for $\mu\le 0.1$ the critical exponents deviate significantly from those of the specific heat $\nu_\text{BH},\nu_\text{WH}\approx 1/2$, except $\eta_{\text{BH}}(\mu=0.1)$ and $\eta_{\text{WH}}(\mu=0.07)$.
% On the other hand, for $\mu\le 0.1$ the critical exponents deviate significantly from those of the specific heat $\nu_\text{BH},\nu_\text{WH}\approx \frac{1}{2}$, except $\eta_{\text{BH}}(\mu=0.1)$ and $\eta_{\text{WH}}(\mu=0.07)$.

\subsubsection{Chaos exponent in quasi-particle approximation}
In the wormhole phase at $T\ll T_\text{WH}$, we can reduce the ladder equation \eqref{MQladderFKF}, which is originally a set of integral equations, to a simple differential equation.
This enables us to evaluate the chaos exponent in this regime analytically in terms of $E_\text{gap}$ and the decay rate of the first peak $\Gamma$.

The calculation goes as follows.
When the temperature is sufficiently low, the spectral function is dominated by the first peak and its mirror image
\begin{align}
\rho_{LL}(\omega)\approx \pi(\delta_\Gamma(\omega-E_\text{gap})+\delta_\Gamma(\omega+E_\text{gap})),\quad
\rho_{LR}(\omega)\approx \pi(\delta_\Gamma(\omega-E_\text{gap})-\delta_\Gamma(\omega+E_\text{gap})),
\end{align}
from which we obtain, via \eqref{MQrealtimeSDfinal2},
\begin{align}
G_{LL}^R(t)&\approx -\frac{i}{2}\theta(t)(e^{-iE_\text{gap}t}+e^{iE_\text{gap}t})e^{-\Gamma t},\quad
G_{LL}\Bigl(\frac{\beta}{2}+it\Bigr)\approx e^{-\frac{\beta E_\text{gap}}{2}}\cos(E_\text{gap}t)e^{-\Gamma|t|},\nonumber \\
G_{LR}^R(t)&\approx -\frac{1}{2}\theta(t)(e^{-iE_\text{gap}t}-e^{iE_\text{gap}t})e^{-\Gamma t},\quad
G_{LR}\Bigl(\frac{\beta}{2}+it\Bigr)\approx -e^{-\frac{\beta E_\text{gap}}{2}}\sin(E_\text{gap}t)e^{-\Gamma|t|}.
\end{align}
By substituting these $G_{LL}^R$, $G_{LR}^R$ into the ladder equation \eqref{MQladderFKF} we obtain, under the assumption ${\cal F}_{RL}=-{\cal F}_{LR}$, ${\cal F}_{RR}={\cal F}_{LL}$ (here we suppress the last two indices of ${\cal F}_{abcd}$ to which the ladder kernel does not act, and we denote ${\cal F}_{abcd}$ simply as ${\cal F}_{ab}$), the following equations
\begin{align}
&{\cal F}_{LL}(t_1,t_2)\pm i{\cal F}_{LR}(t_1,t_2)\nonumber \\
&=\frac{{\cal J}^2\cdot 2^{q-1}(q-1)}{q}\int dtdt'\theta(t_1-t)\theta(t_2-t')
e^{(\mp iE_\text{gap}-\Gamma)(t_1-t)}
e^{(\pm iE_\text{gap}-\Gamma)(t_2-t')}\nonumber \\
&\quad \Bigl[G_{LL}\Bigl(\frac{\beta}{2}+i(t-t')\Bigr)^{q-2}{\cal F}_{LL}(t,t')\pm i(-1)^{\frac{q}{2}}G_{LR}\Bigl(\frac{\beta}{2}+i(t-t')\Bigr)^{q-2}{\cal F}_{LR}(t,t')\Bigr].
\label{quasiparticleapprox3}
\end{align}
Now we differentiate both sides of this equation by $\partial_{t_1}+\Gamma\pm iE_\text{gap}$ and $\partial_{t_2}+\Gamma\mp iE_\text{gap}$.
Since the exponential factors in \eqref{quasiparticleapprox3} are eliminated by these differential operators, from the right-hand side of \eqref{quasiparticleapprox3} only gain $\partial_{t_1}\partial_{t_2}\theta(t_1-t)\theta(t_2-t')=\delta(t_1-t)\delta(t_2-t')$, which cancel the integrations and we obtain a differential equation
\begin{align}
&(\partial_{t_1}+\Gamma\pm iE_\text{gap})
(\partial_{t_2}+\Gamma\mp iE_\text{gap})
({\cal F}_{LL}(t_1,t_2)\pm i{\cal F}_{LR}(t_1,t_2))
=\frac{{\cal J}^2\cdot 2^{q-1}(q-1)}{q}\nonumber \\
&\quad \Bigl[
G_{LL}\Bigl(\frac{\beta}{2}+i(t_1-t_2)\Bigr)^{q-2}{\cal F}_{LL}(t_1,t_2)
\pm i(-1)^{\frac{q}{2}}G_{LR}\Bigl(\frac{\beta}{2}+i(t_1-t_2)\Bigr)^{q-2}{\cal F}_{LR}(t_1,t_2)
\Bigr].
\label{quasiparticleapprox4}
\end{align}
If we assume the $t_1+t_2$ dependence of ${\cal F}_{ab}(t_1,t_2)$ as ${\cal F}_{ab}=e^{\lambda_L(t_1+t_2)/2}f_{ab}(t_1-t_2)$ and also assume $f_{ab}\in\mathbb{R}$, the ladder equation \eqref{quasiparticleapprox4} becomes ($t\equiv t_1-t_2$)
\begin{align}
&\Bigl[-\partial_t^2+E_\text{gap}^2+\Bigl(\frac{\lambda_L}{2}+\Gamma\Bigr)^2\Bigr]f_{LL}(t)+2E_\text{gap}\partial_tf_{LR}(t)\nonumber \\
&\quad =\frac{{\cal J}^2\cdot 2^{q-1}(q-1)}{q}G_{LL}\Bigl(\frac{\beta}{2}+it\Bigr)^{q-2}f_{LL}(t),\nonumber \\
&\Bigl[-\partial_t^2+E_\text{gap}^2+\Bigl(\frac{\lambda_L}{2}+\Gamma\Bigr)^2\Bigr]f_{LR}(t)-2E_\text{gap}\partial_tf_{LL}(t)\nonumber \\
&\quad =\frac{(-1)^{\frac{q}{2}}{\cal J}^2\cdot 2^{q-1}(q-1)}{q}G_{LR}\Bigl(\frac{\beta}{2}+it\Bigr)^{q-2}f_{LR}(t),
\end{align}
For $q\in 4\mathbb{N}$ these equations simplify drastically with the following additional ansatz\footnote{
For $q\in 4\mathbb{N}+2$ we could not find a way to simplify the differential equation where a non-trivial solution still exists.
% \textcolor{nosaka2}{
% $q=6$, ${\cal F}_{RR}=-{\cal F}_{LL},{\cal F}_{RL}=+{\cal F}_{LR}$もやってみたけど駄目だったので, 以下消す:
% We suspect that some of the assumptions we have made above: ${\cal F}_{RL}=-{\cal F}_{LR}$, ${\cal F}_{RR}={\cal F}_{LL}$, $f_{ab}(t)\in\mathbb{R}$, which are motivated from the numerically obtained eigenvectors of the ladder kernel for $q=4$, might not be true for $q\in\mathbb{N}+2$.
% }
}
\begin{align}
f_{LL}(t)=\cos(E_\text{gap}t)g(t),\quad
f_{LR}(t)=-\sin(E_\text{gap}t)g(t),
\end{align}
as
\begin{align}
\Bigl[-\partial_t^2+\Bigl(\frac{\lambda_L}{2}+\Gamma\Bigr)^2\Bigr]g(t)&=\frac{{\cal J}^2\cdot 2^{q-1}(q-1)}{q}G_{LL}\Bigr(\frac{\beta}{2}+it\Bigr)^{q-2}g(t),\nonumber \\
\Bigl[-\partial_t^2+\Bigl(\frac{\lambda_L}{2}+\Gamma\Bigr)^2\Bigr]g(t)&=\frac{{\cal J}^2\cdot 2^{q-1}(q-1)}{q}G_{LR}\Bigr(\frac{\beta}{2}+it\Bigr)^{q-2}g(t).
\end{align}
By assuming that $g(t)$ varies slowly compared to the scale $E_\text{gap}^{-1}$, we can replace $\cos^{q-2}E_\text{gap}t$ in $G_{LL}(t)^{q-2}$ and $\sin^{q-2}E_\text{gap}t$ in $G_{LR}(t)^{q-2}$ with their average over the period as
\begin{align}
\cos^{q-2}E_\text{gap}t,\quad\sin^{q-2}E_\text{gap}t\,\longrightarrow\, \frac{(q-2)!}{2^{q-2}((\frac{q}{2}-1)!)^2},
\end{align}
hence we obtain
\begin{align}
\Bigl[-\partial_t^2+\Bigl(\frac{\lambda_L}{2}+\Gamma\Bigr)^2-\frac{2{\cal J}^2(q-1)!}{q((\frac{q}{2}-1)!)^2} e^{-(\frac{q}{2}-1)\beta E_\text{gap}} e^{-(q-2)\Gamma|t|}\Bigr]g(t)=0.
\end{align}
If we rescale $t$ as $t'=(q-2)\Gamma t$ and use the expression for $\Gamma$ under the quasi-particle approximation $\Gamma\approx \sqrt{2{\cal J}^2(q-2)!/(((q/2)!)^2)} e^{-(q/2-1)\beta E_\text{gap}/2}$ \cite{Qi:2020ian}, we finally obtain
\begin{align}
\Bigl[-\frac{\partial^2}{\partial t'^2}+\frac{1}{(q-2)^2}\Bigl(\frac{\lambda_L}{2\Gamma}+1\Bigr)^2-\frac{q(q-1)}{(q-2)!}e^{-|t'|}\Bigr]g(t')=0.
\label{quasiparticleapproxfinaldiffeq}
\end{align}
It is not difficult to solve the differential equation \eqref{quasiparticleapproxfinaldiffeq}; the solution for $t'>0$ and $t'<0$ are separately given by Bessel function $J[n,z]$ with $n=2(\lambda_L/(2\Gamma)+1)/(q-2)$ and $z=2\sqrt{q(q-1)/(q-2)^2}e^{-|t'|/2}]$, and the value of $\lambda_L/\Gamma$ is determined by requiring a smooth connection of $g(t')$ at $t'=0$, as
\begin{align}
\frac{\lambda_L}{\Gamma}=(q-2)n-2,\quad\quad
\frac{\partial J[n,z]}{\partial z}\biggr|_{z=2\sqrt{\frac{q(q-1)}{(q-2)^2}}}=0.
\label{simplerelationbetweenLyapunovandGamma1st}
\end{align}
For $q=4$ this gives $\lambda_L/\Gamma\approx 2.706$.
Actually it is not easy to reproduce this value (as well as the overall factor $\sqrt{2{\cal J}^2(q-2)!/(((q/2)!)^2)}$ of $\Gamma$) precisely from the numerical analysis.
However, the remarkable point of this conclusion is rather that when the temperature is sufficiently low the ratio $\lambda_L/\Gamma$ is completely independent of the temperature and the other parameters of the two coupled model ${\cal J},\mu$.

\subsection{Single sided model}
\begin{figure}
\includegraphics[width=8cm]{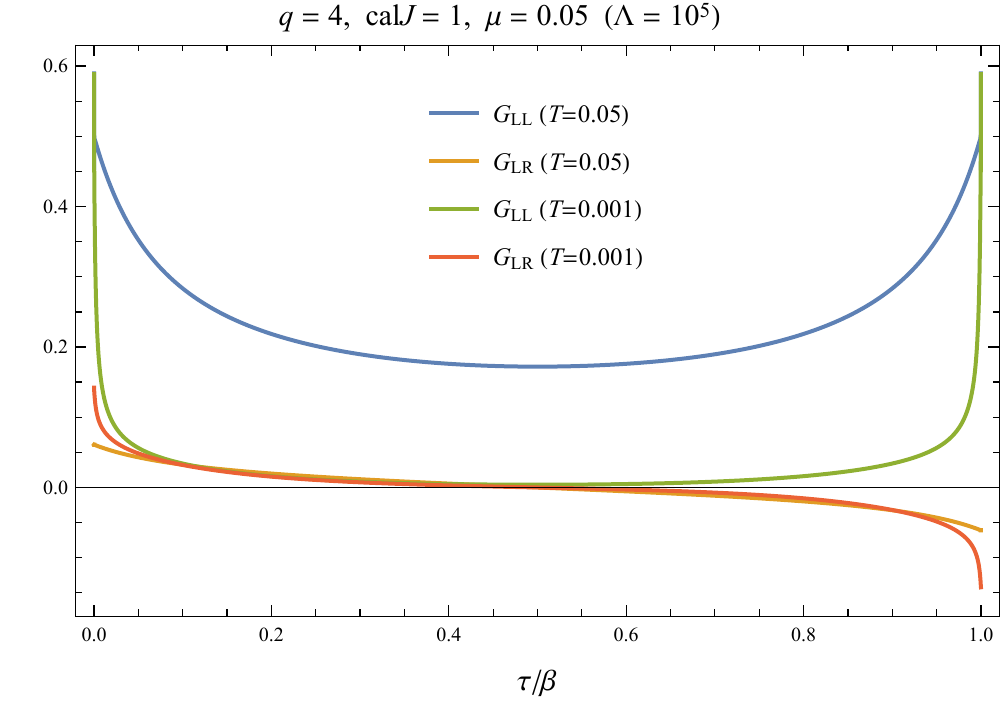}\quad\quad
\includegraphics[width=8cm]{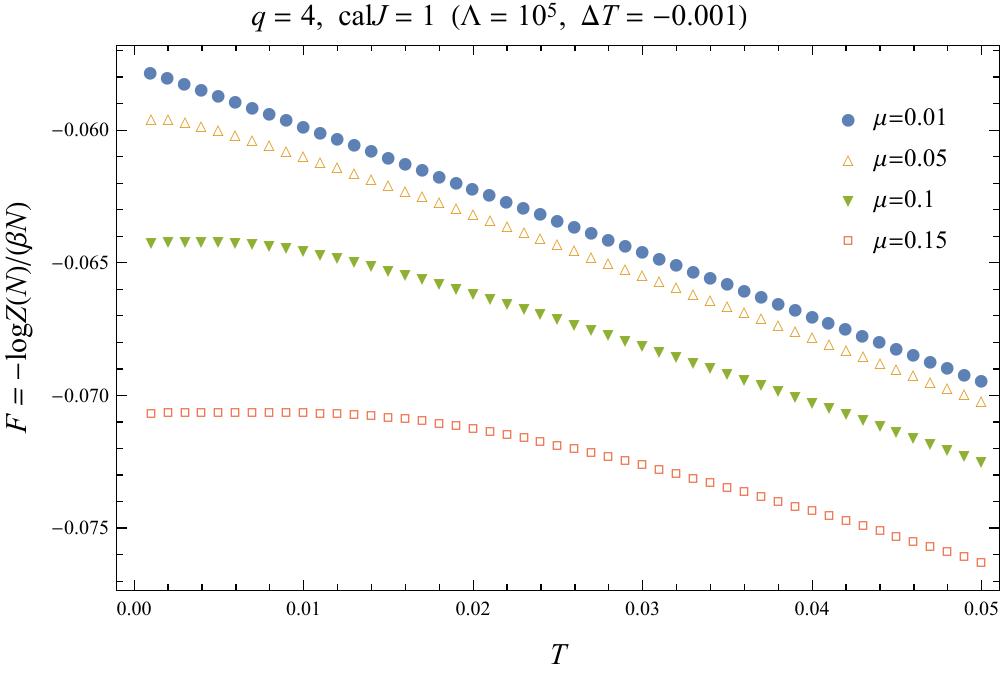}\\
\includegraphics[width=8cm]{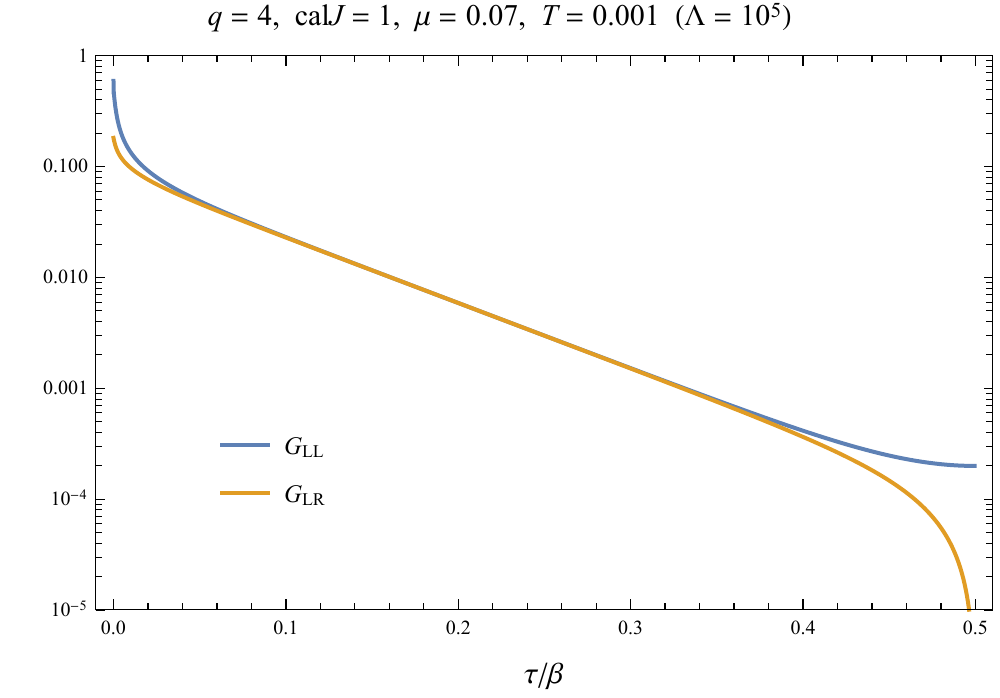}\quad\quad
\includegraphics[width=8cm]{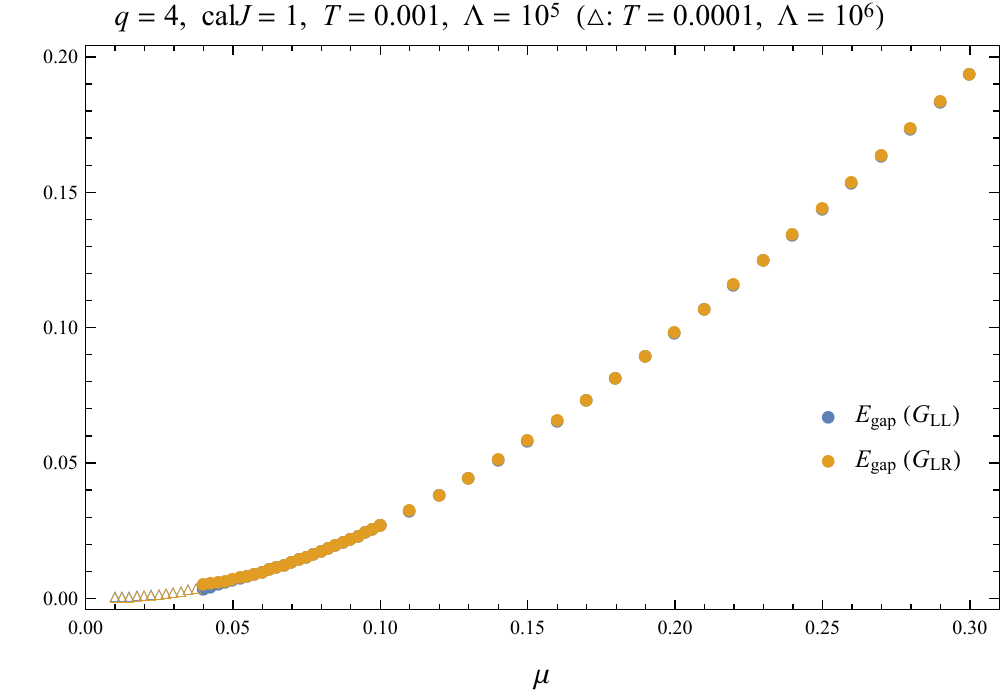}
\caption{
Top left: Euclidean propagator $G_{ab}(\tau)$ of the single sided model with $\mu=0.05$, $T=0.001,0.05$.
Top right: Free energy.
Bottom left: Euclidean propagator at low temperature $\mu=0.05$, $T=0.001$ where the exponential decay is significant.
% Bottom left: Euclidean propagator at low temperature $\mu=0.05$, $T=0.001$ where exponential decay is significant.
Bottom right: $E_\text{gap}$ of the single sided model obtained by fitting $G_{ab}(\tau)$ at $T=0.001$ ($\mu\ge 0.04$) and at $T=0.0001$ ($\mu\le 0.03$).
For $\mu\le 0.03$ we have set $\Lambda=10^6$.
}
\label{fig_EuclideanGabandfreeenergyandEgapofsinglesidedmodel}
\end{figure}

In Fig.~\ref{fig_EuclideanGabandfreeenergyandEgapofsinglesidedmodel} we have displayed the Euclidean propagators and the free energy of the single sided model for $\mu=0.1$.
% In Fig.~\ref{fig_EuclideanGabandfreeenergyandEgapofsinglesidedmodel} display the Euclidean propagators and the free energy of the single sided model for $\mu=0.1$.
This model does not exhibit a phase transition regardless of the value of $\mu$ \cite{Nosaka:2019tcx}.
When the temperature is sufficiently low, however, the Euclidean propagators exhibits exponential decay, which indicates that the system is gapped.
We find that $E_\text{gap}$ of the single sided model is smaller than that of the two coupled model at same value of $\mu$ and that it behaves as $E_\text{gap}\sim \mu^2$ at small $\mu$ \cite{Nosaka:2019tcx}, which is in contrast to the two coupled model where $E_\text{gap}\sim \mu^{2/3}$ \cite{Maldacena:2018lmt}.

\begin{figure}[H]
\includegraphics[width=8cm]{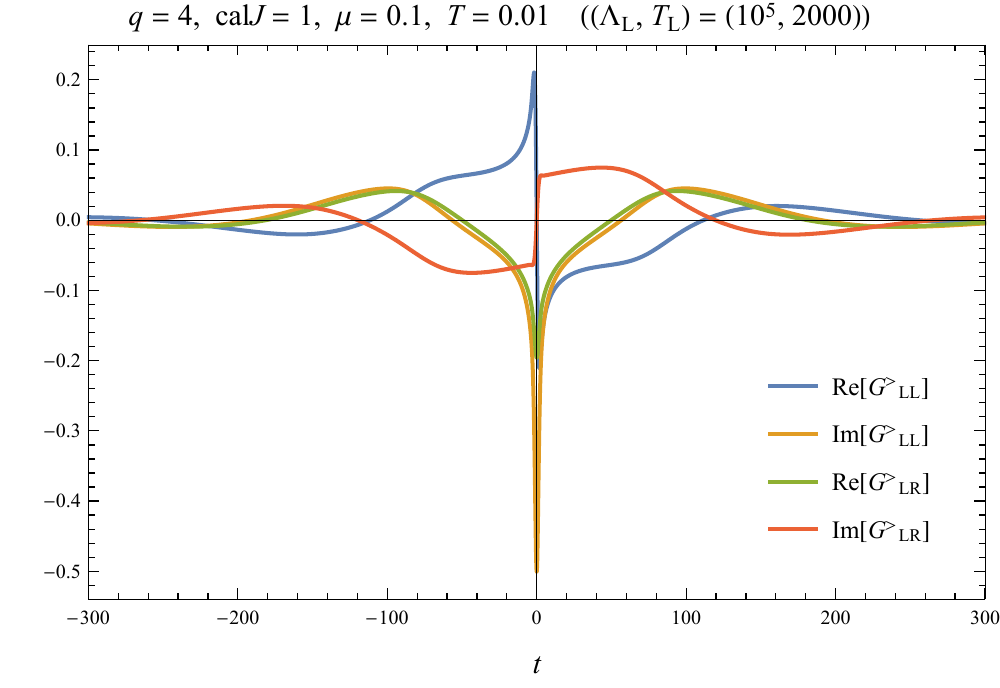}\quad\quad
\includegraphics[width=8cm]{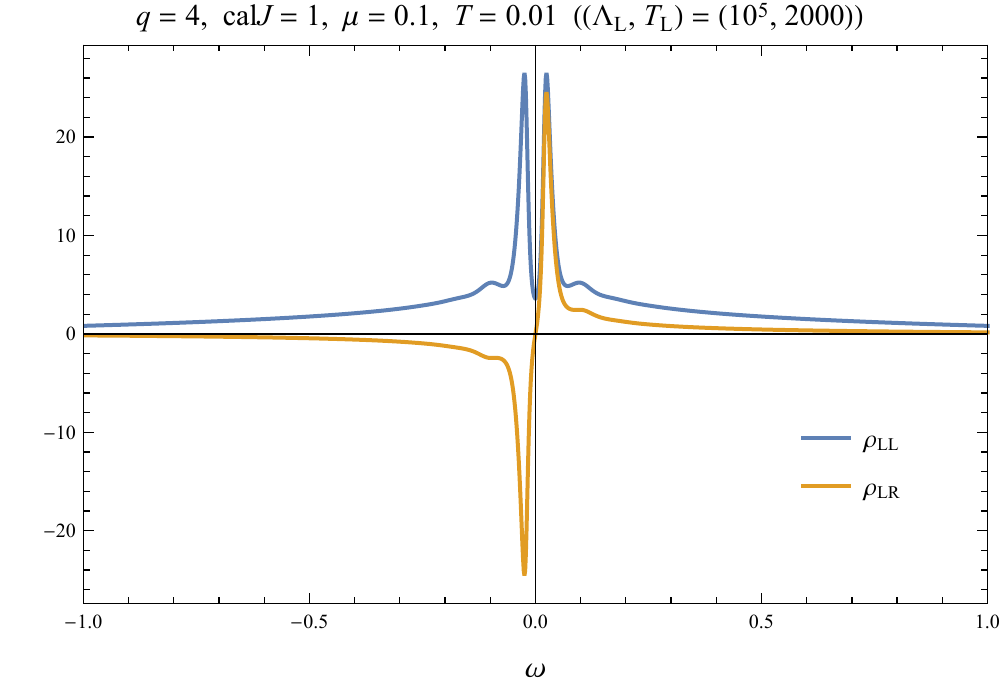}\\
\includegraphics[width=8cm]{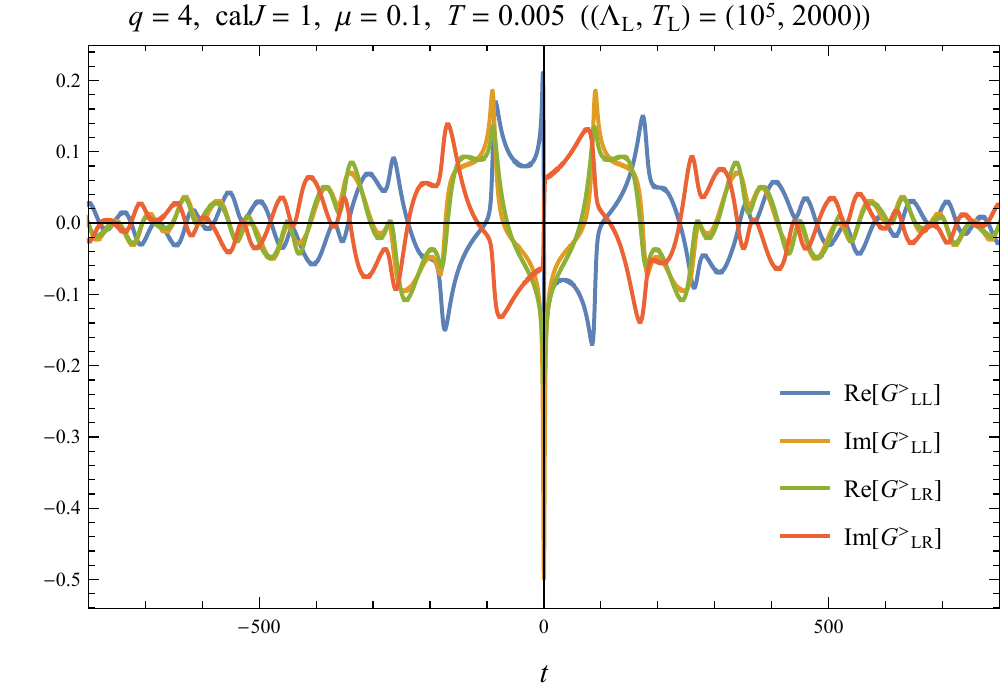}\quad\quad
\includegraphics[width=8cm]{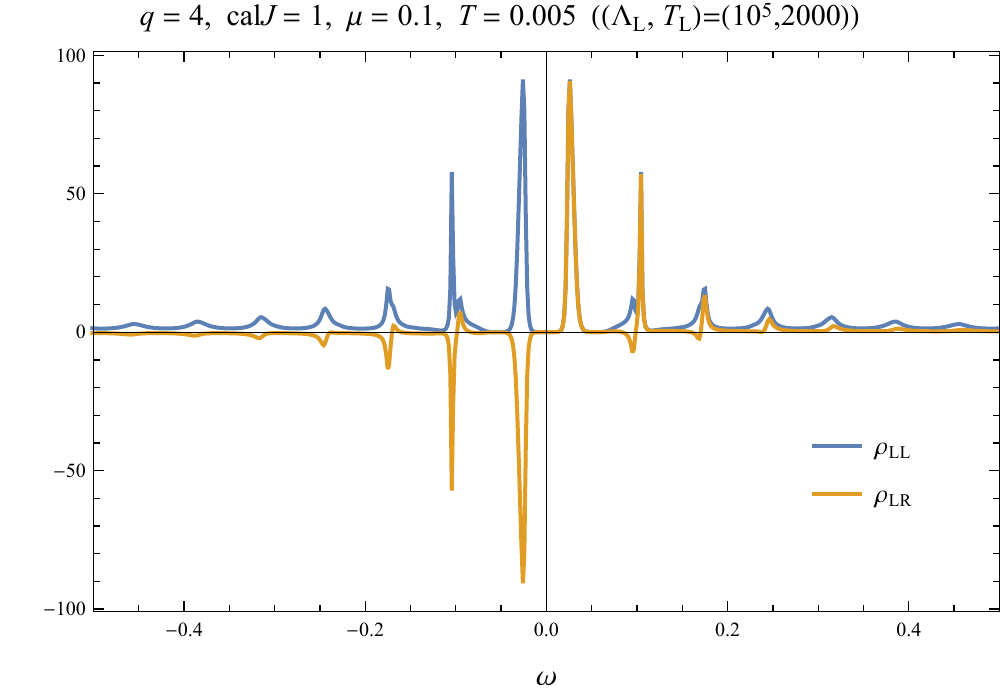}\\
\includegraphics[width=8cm]{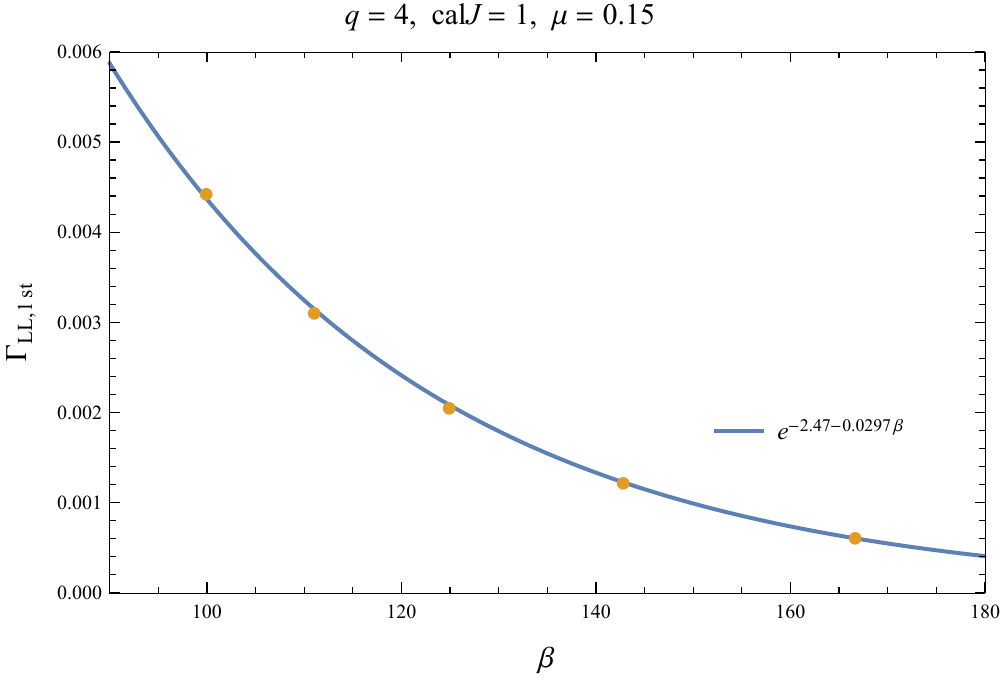}\quad\quad
\includegraphics[width=8cm]{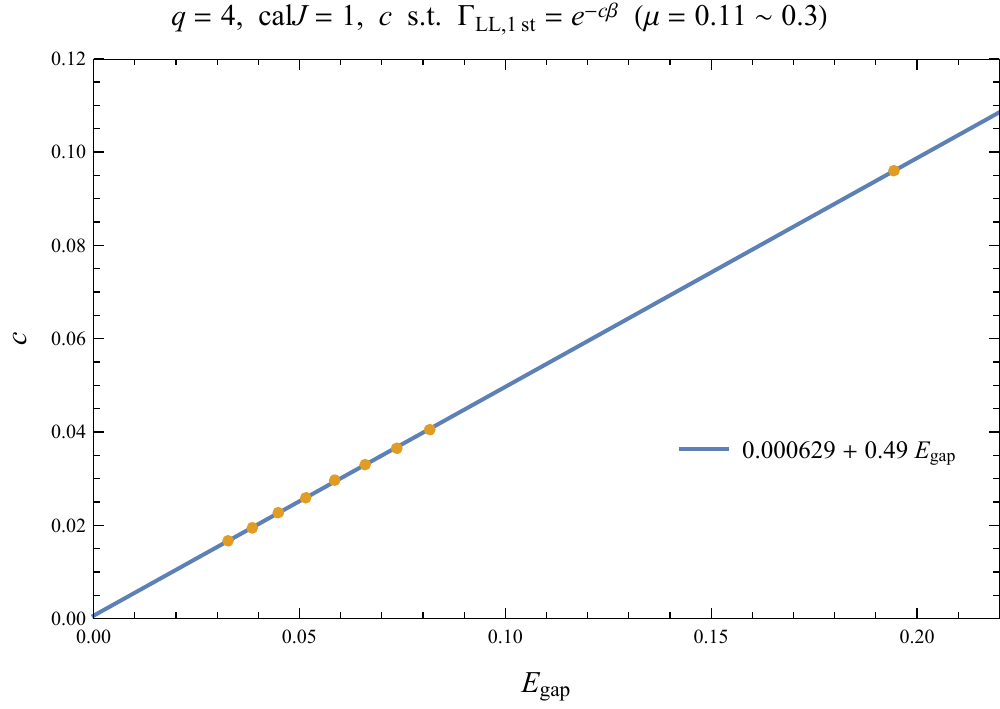}
\caption{
Top left/top right: propagator and spectral function of the single sided model at $\mu=0.1$, $T=0.01$ which are qualitatively same as those in the black hole phase of the two coupled model.
Middle left/middle right: propagator and spectral funcntion at $\mu=0.1$, $T=0.01$ where the spectral functions split into well separated peaks as in the wormhole phase of the two coupled model.
Bottom left: fitting of the decay width of the first peak with $e^{c_1-c_2\beta}$, at sufficiently low temperature where the first peak is well separated from the second peak and the mirror of the first peak at $\omega<0$.
% Bottom left: fitting of the decay width of the first peak at sufficiently low temperature where the first peak is well separated from the second peak and the mirror of the first peak at $\omega<0$ with $e^{c_1-c_2\beta}$.
Bottom right: Comparison of $c_2$ with $E_\text{gap}$ of the single sided model.
}
\label{GabandrhoabandGamma_KMmodel}
\end{figure}

The real time propagators also behave similarly to those in the two coupled model both at high temperature and at low temperature.
In particular when the temperature is sufficiently low the spectral functions split into sharp peaks, which corresponds to the fact that the system is gapped. 
The height of the first peak is lower than that for the two coupled model.
This is not because the weight $A$ is smaller but rather because the decay width $\Gamma$ is larger.
Indeed we have found the decay width of the first peak for the single sided model again obeys the formula \eqref{Gamma1st_QiZhangformula} when the temperature is sufficiently low
% Indeed we found the decay width of the first peak for single sided model again obeys the formula \eqref{Gamma1st_QiZhangformula} when the temperature is sufficiently low
\begin{align}
\Gamma_{\text{1st}}\sim e^{-\frac{1}{2}\beta E_\text{gap}}.
\end{align}
Here $E_\text{gap}$ is the energy gap of the single sided model.
See Fig.~\ref{GabandrhoabandGamma_KMmodel}.

\begin{figure}
\includegraphics[width=8cm]{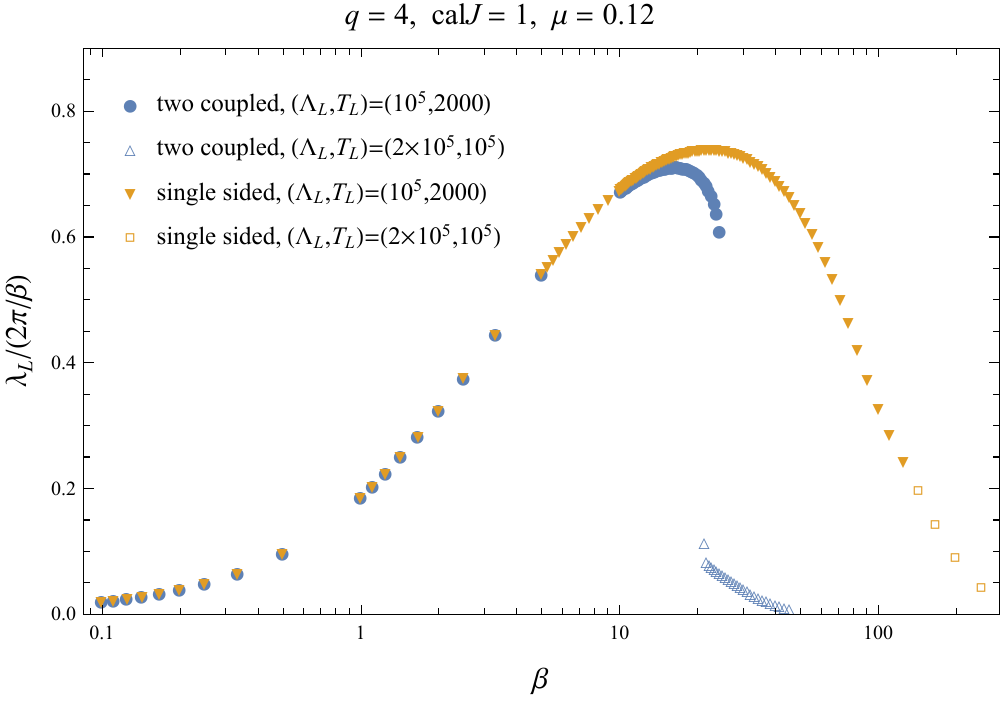}\quad\quad
\includegraphics[width=8cm]{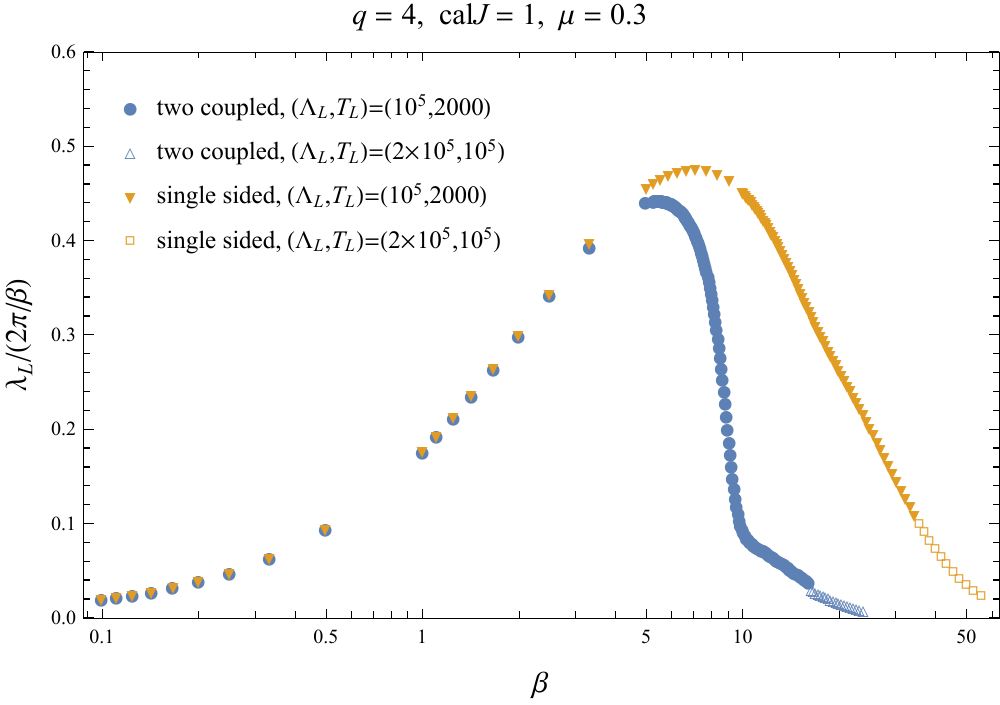}
\caption{Comparison of the chaos exponent between the two copuled model and the single sided model.
}
\label{fig_LambdaLMQvsKM}
\end{figure}

\begin{figure}
\includegraphics[width=8cm]{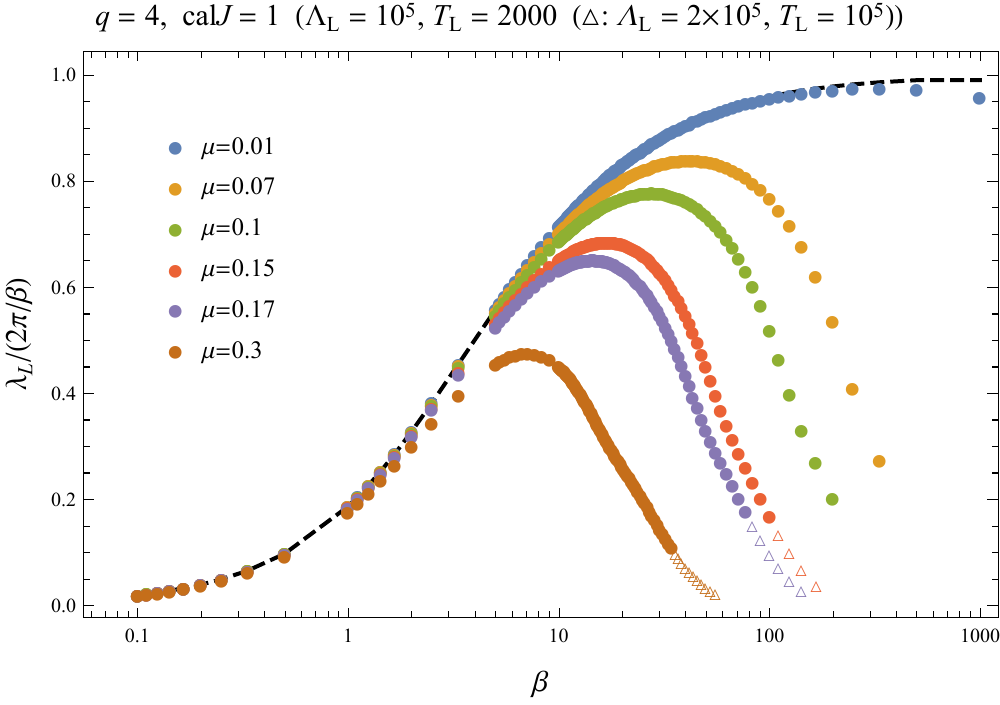}\quad\quad
\includegraphics[width=8cm]{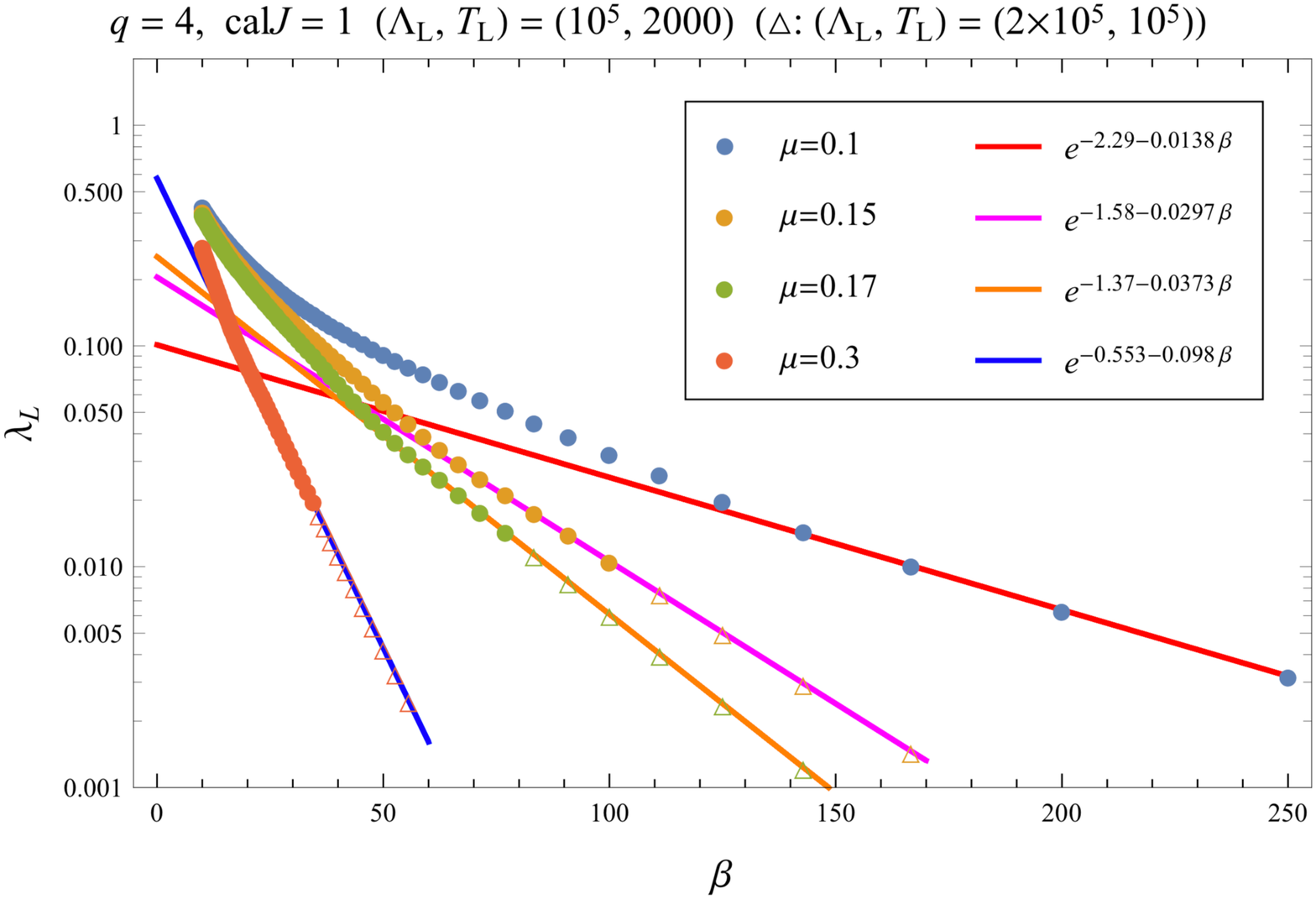}\\
\includegraphics[width=8cm]{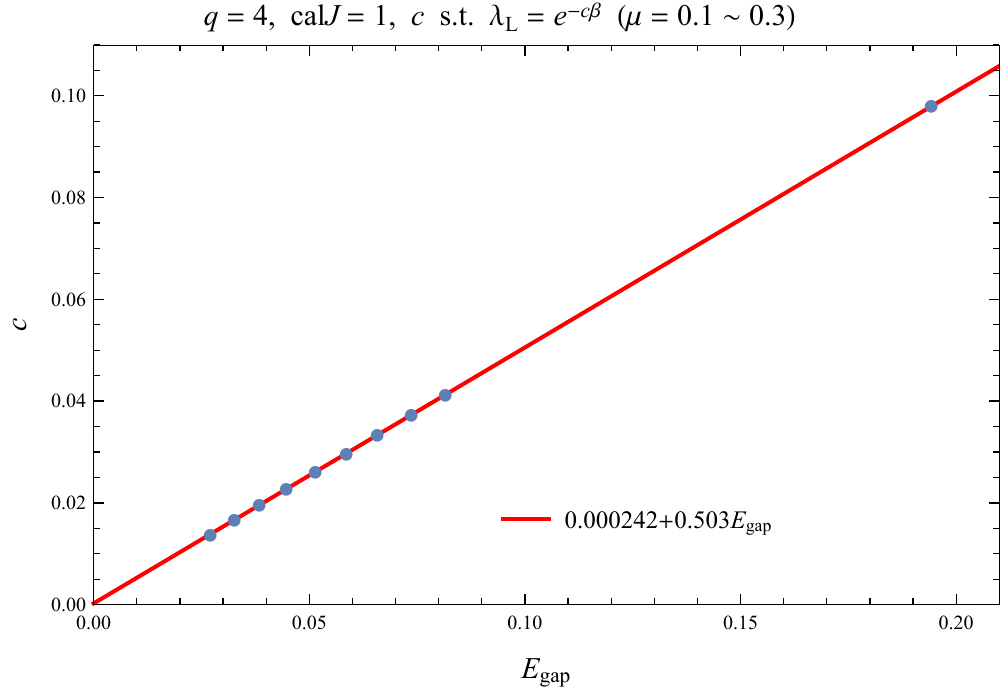}
\caption{
Top left: Chaos exponent $\lambda_L$ of the single sided model.
Top right: Comparison of $\lambda_L$ with the exponential decay $e^{c_1-c_2\beta}$ obtained by fitting last three to six data points with the largest values of $\beta$ for each $\mu$.
Bottom: Comparison of the fitting coefficient $c_2$ with $E_\text{gap}$.
}
\label{fig_KMLyapunov}
\end{figure}

Lastly we display the chaos exponent $\lambda_L$ in Fig.~\ref{fig_KMLyapunov}.
The overall behavior of the chaos exponent is qualitatively same as that of the two coupled model except the absence of the phase transition.
We also observe that $\lambda_L$ for the single sided model is always greater than that of the two coupled model at the same values of $(\mu,T)$ as displayed in Fig.~\ref{fig_LambdaLMQvsKM}.
At low temperature we again found that $\lambda_L$ obeys \eqref{LambdaL_QiZhangformula}
\begin{align}
\lambda_L\sim e^{-\frac{1}{2}\beta E_\text{gap}}.
\end{align}
As $E_\text{gap}$ for the single sided model is smaller than that of the two coupled model, this explains the fact that $\lambda_L$ for the single sided model is greater than that of the two coupled model.
For the same dominance persisting at higher temperature we do not have such clear explanation, but we argue a possible interpretation of it in section \ref{sec_Discussion}.
% The same dominance persists also at higher temperature, which may be understood more clearly if we interpret the results for the single sided model as that of the two coupled model with $J^L_{ijk\ell}$ and $J^R_{ijk\ell}$ uncorrelated.
% Although the two models are different at finite $N$, they are equivalent in the $G\Sigma$ formalism at the level of on-shell.
% We elavolate this point further in section \ref{sec_Discussion}.

\section{Discussion}
\label{sec_Discussion}

In this paper we have studied the chaos exponent of the Maldacena-Qi model \cite{Maldacena:2018lmt} in detail.
The analysis of the level statistics at finite $N$ \cite{Garcia-Garcia:2019poj} suggests that there is a quantum chaos transition below $\mu = \mu_c=0.177$ where the Hawking-Page like transition in the Maldacena-Qi model disappears.
This motivate us to study the chaos exponents, which can be analyzed in the large $N$ limit using the $G,\Sigma$ formalism.
Since the Hawking-Page like transition originates from the exchange of the dominance of two different saddles, one may think that the coincidence of the thermal phase transition and a chaos transition is not so surprising.
However, it is still non trivial how both phases are characterized from the view of quantum chaos.
% However, it is still non trivial that how both phases are characterized from the view of quantum chaos.
We have found that when the system goes to the wormhole phase from the black hole phase, the chaos exponent jumps to extremely small values, which is consistent with the expectation in \cite{Garcia-Garcia:2019poj}.

Another motivation of our analysis is to study the chaos exponent in the gapped phase, which we expect to be an integrable phase.
Surprisingly, however, it was found \cite{Qi:2020ian} that the two point function shows an exponential decay, which indicates that the system is still chaotic even in this regime.
Indeed, we have found that that the chaos exponent is small but non-zero also in the wormhole phase.
Moreover, we have found a quantitative relation \eqref{simplerelationbetweenLyapunovandGamma1st} between the chaos exponent and the decay rate which was found to behave as $\Gamma_\text{1st}\approx {\cal J}\sqrt{(q-2)!/(2((q/2)!)^2)}e^{-(q/2-1)E_\text{gap}\beta/2}$ \cite{Qi:2020ian}.
% Moreover, we have found a quantitative relation \eqref{simplerelationbetweenLyapunovandGamma1st} between the chaos exponent and the decay rate which was found to behave as $\Gamma_\text{1st}\approx {\cal J}\sqrt{\frac{(q-2)!}{2((\frac{q}{2})!)^2}}e^{-(q/2-1)E_\text{gap}\beta/2}$ \cite{Qi:2020ian}.
Note that these formulas imply that both the decay rate and the chaos exponent vanishes non-perturbatively in the large $q$ limit with $q\mu$ and $qT$ kept fixed\footnote{
Although the prefactor $\sqrt{(q-2)!/(2((q/2)!)^2)}$ grows exponentially in $q$ as $\sim 2^q$, the exponential decay of $e^{-(q/2-1)\beta E_\text{gap}/2}$ is even faster due to the rescaling of $T$.
% Although the prefactor $\sqrt{\frac{(q-2)!}{2((\frac{q}{2})!)^2}}$ grows exponentially in $q$ as $\sim 2^q$, the exponential decay of $e^{-(q/2-1)\beta E_\text{gap}/2}$ is even faster due to the rescaling of $T$.
}, which is consistent with the fact that we did not observe these chaotic properties in the direct large $q$ analysis \cite{Maldacena:2018lmt}.
Also note that such a simple relation would not hold in general.
For example, in a general conformal field theory the two point function is completely determined by the conformal dimension of the two operators, while to calculate the four point function, which encodes the chaos exponent, we also have to know the OPE coefficients.
% For example, in a conformal field theory the two point function is completely determined by the conformal dimension of the two operators, while to calculate the four point function, which encodes the chaos exponent, we also have to know the OPE coefficients.
It would be interesting to understand how the simple relation \eqref{simplerelationbetweenLyapunovandGamma1st} between the chaos exponent and the decay rate, if it exists, will be generalized in other chaotic systems.

\begin{figure}
\begin{center}
\includegraphics[width=8cm]{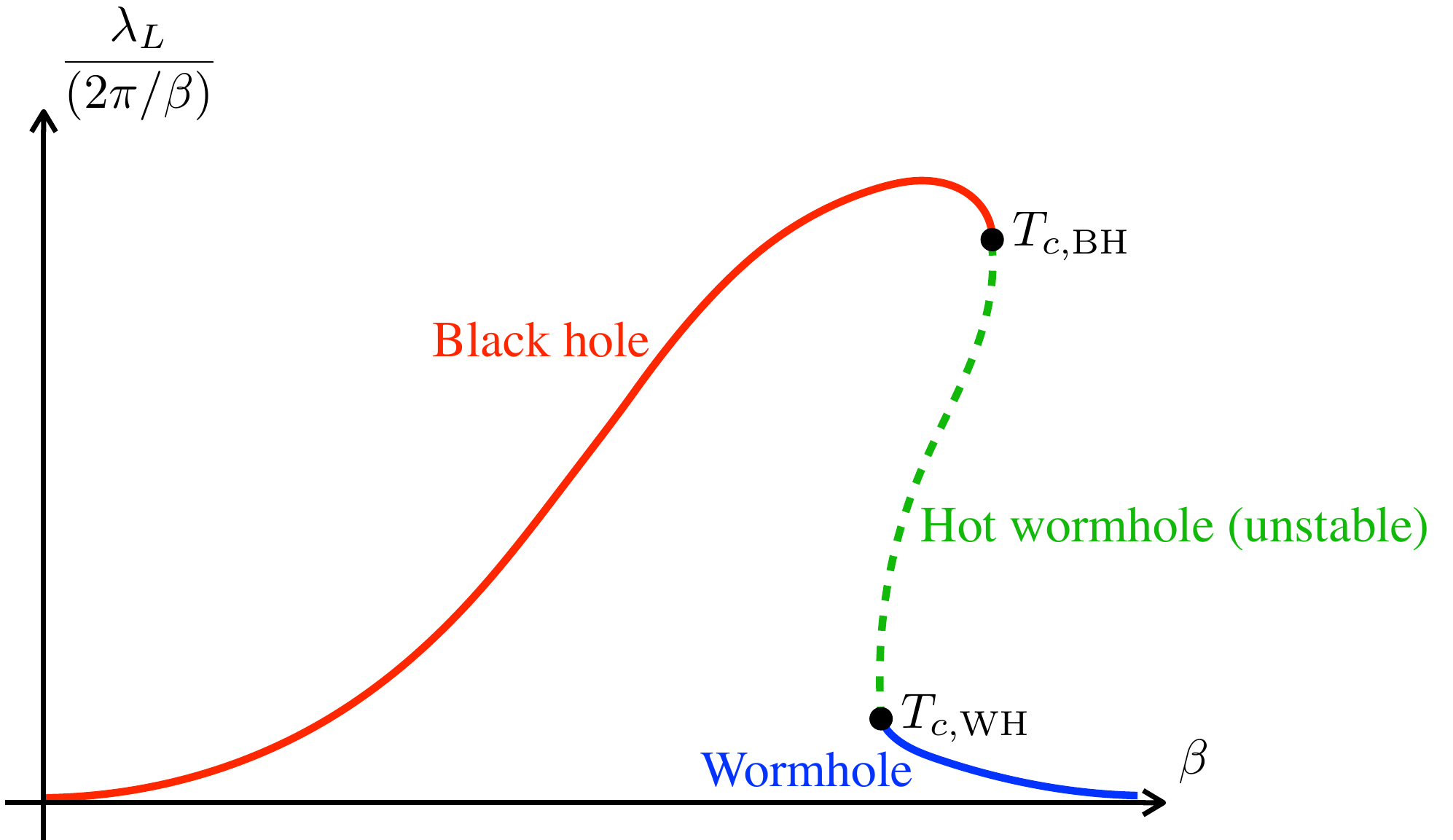}
\caption{
Schematic picture for the expected behavior of the chaos exponent in the unstable hot wormhole phase for $\mu<\mu_c$.
}
\label{sketch}
\end{center}
\end{figure}

We have also found that the slope of the chaos exponent $\partial_T\lambda_L(T)$ diverges at the end of the two phases $T=T_{c,\text{BH}}$, $T=T_{c,\text{WH}}$.
% We have also found that the slope of the chaos exponent $\partial_T\lambda_L(T)$ diverges at the end of the two phases $T=T_{c,\text{BH}}$, $T=T_{c,\text{WH}}$.
These divergent behaviors resemble that of the energy $E(T)$ and the entropy $S(T)$, rather than of the free energy $F$ whose slope is finite (almost constant) in each phase even near $T_{c,\text{BH}},T_{c,\text{WH}}$.
% These divergent behaviors resemble that of the energy $E$ and the entropy $S$, rather than of the free energy $F$ whose slope is finite (almost constant) in each phases even near $T_{c,\text{BH}},T_{c,\text{WH}}$.
In \cite{Maldacena:2018lmt} it was claimed that for $T_{c,\text{BH}}<T<T_{c,\text{WH}}$ there exists another canonically unstable phase throught which the energy varies completely smoothly in the all parameter regime \cite{Maldacena:2019ufo}.
% In \cite{Maldacena:2018lmt} it was claimed that for $T_{c,\text{BH}}<T<T_{c,\text{WH}}$ there exist another canonically unstable phase throught which the energy varies completely smoothly in the all parameter regime \cite{Maldacena:2019ufo}.
Although we could not reach the unstable phase in the current analysis, we expect that the chaos exponent shows a similar behavior as the energy, as sketched in Fig.~\ref{sketch}.
This would be confirmed by solving the Kadanoff-Baym equation of the two coupled system coupled to a cool bath and evaluating the chaos exponent by using the propagators at each time $t_1+t_2$ before it reaches the equilibrium with $T_\text{bath}$ \cite{Maldacena:2019ufo,Almheiri:2019jqq,Numasawatoappear}.
% This would be confirmed by solving the Kadanoff-Baym equation of the two coupled system coupled to a cool bath and evaluating the chaos exponent by using the propagators at each time $t_1+t_2$ before it reaches the equilibrium with $T_\text{bath}$ \cite{Numasawatoappear}.

We have also considered a model with single SYK with a mass deformation \eqref{Hsinglesided} \cite{Kourkoulou:2017zaj}.
While the single sided model is qualitatively same as the two coupled model in the limit of $\mu\rightarrow 0$ and $\mu\rightarrow\infty$, in contrast to the two coupled model, this model does not exhibit a phase transition.
% While the single sided model is qualitatively same with the two coupled model in the limit of $\mu\rightarrow 0$ and $\mu\rightarrow\infty$, in contrast to the two coupled model, this model does not exhibit a phase transition.
% We have found that the chaos exponent of the single sided model is always greater than that of the two coupled model in the whole parameter regime.
Correspondingly, the chaos exponent we have obtained varies smoothly at all $(\mu,T)$.
We have also found that the chaos exponent obeys the same exponential formula $\lambda_L\sim e^{-(q/2-1)\beta E_\text{gap}/2}$ \eqref{lowtemperatureformula} as the decay rate of the first peak \cite{Qi:2020ian}.
As displayed in appendix \ref{sec_KMq6q8}, for the single sided model we have reached the low temperature regime also for $q=6,8$, where we have confirmed the formula \eqref{lowtemperatureformula} holds also for $q=6,8$.

We have further found that the chaos exponent of the single sided model is always greater than that of the two coupled model in the whole parameter regime.
Though in this paper we have regarded the two models in independent ways, we can treat the two models as two different parameter points of a unifed model, where the direct comparison would be more reasonable.
We can consider a generalization of the two coupled model with the correlation between the random couplings of the two sides being incomplete $\langle J_{i_1i_2\cdots i_q}^L J_{i_1i_2\cdots i_q}^R \rangle<\langle (J_{i_1i_2\cdots i_q}^L)^2\rangle= \langle (J_{i_1i_2\cdots i_q}^R)^2\rangle$, where $\langle J_{i_1i_2\cdots i_q}^L J_{i_1i_2\cdots i_q}^R \rangle/\langle (J_{i_1i_2\cdots i_q}^L)^2\rangle$ is a new tunable parameter of the theory.
% We can consider a generalization of the two coupled model with the random couplings of the two sides are not completely correlated $\langle J_{i_1i_2\cdots i_q}^L J_{i_1i_2\cdots i_q}^R \rangle<\langle (J_{i_1i_2\cdots i_q}^L)^2\rangle= \langle (J_{i_1i_2\cdots i_q}^R)^2\rangle$, where $\langle J_{i_1i_2\cdots i_q}^L J_{i_1i_2\cdots i_q}^R \rangle/\langle (J_{i_1i_2\cdots i_q}^L)^2\rangle$ is a new tunable parameter of the theory.
As we have commented in \cite{Nosaka:2019tcx}, the single sided model \eqref{Hsinglesided} is equivanlent to this model with $\langle J^L_{i_1i_2\cdots i_q}J^R_{i_1i_2\cdots i_q}\rangle/\langle(J^L_{i_1i_2\cdots i_q})^2\rangle=0$ at the level of the large $N$ $G\Sigma$ formalism.\footnote{
% As we have commented in \cite{Nosaka:2019tcx}, the single sided model \eqref{Hsinglesided} is equivanlent at the level of the large $N$ $G\Sigma$ formalism.\footnote{
Precisely speaking, the $G\Sigma$ effective action and its first variation are identical for the two models after imposing the ansatz $G_{LL}=G_{RR}$, while the second variation of the effective action is not the same even after the substitution of the solution to the equations of motion with $G_{LL}=G_{RR}$.
One can show, however, that this discrepancy does not affect the leading chaos exponent \cite{Nosakatoappear}.
}
Hence this model unifies the two coupled model and the single sided model, and our observation can be rephrased that the model is less chaotic when $J^L_{i_1i_2\cdots i_q}$ and $J^R_{i_1i_2\cdots i_q}$ are more correlated.
It would be interesting to study the chaotic property of this unifying model and see whether the chaos exponent monotonically decreases with respect to $0<\langle J_{i_1i_2\cdots i_q}^L J_{i_1i_2\cdots i_q}^R \rangle/\langle (J_{i_1i_2\cdots i_q}^L)^2\rangle<1$ \cite{Nosakatoappear}.

\section*{Acknowledgement}
The numerical analyses in this paper were performed on sushiki server in Yukawa Institute Compute Facility and on Ulysses cluster v2 in SISSA.
T.~Nosaka is also grateful to the online conference ``4th INTERNATIONAL CONFERENCE on HOLOGRAPHY, STRING THEORY and DISCRETE APPROACH in HANOI, VIETNAM'' where he presented the preliminary results of this work.
% T.~Nosaka is also grateful to the online conference ``4th INTERNATIONAL CONFERENCE on HOLOGRAPHY, STRING THEORY and DISCRETE APPROACH in HANOI, VIETNAM'' where he presented the preliminary results on this work.

\appendix

\section{Numerical results for $q=6,8$}
\label{sec_q6q8}
\subsection{two coupled model}
\begin{figure}
\includegraphics[width=8cm]{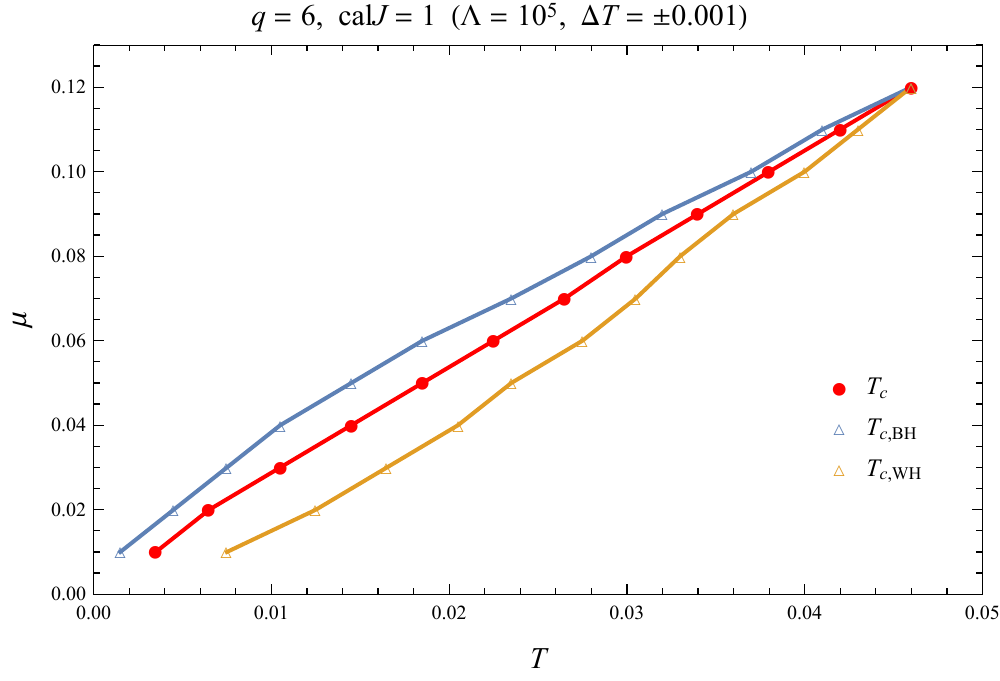}\quad\quad
\includegraphics[width=8cm]{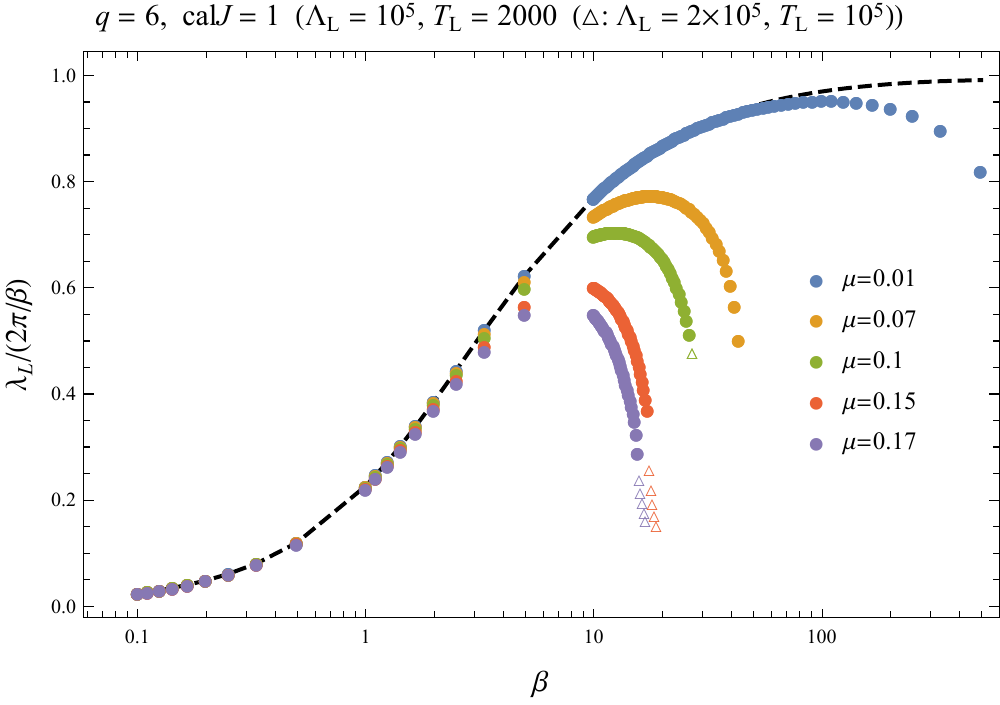}
\caption{
Phase diagram (left) and the chaos exponent (right) of the two coupled model with $q=6$, ${\cal J}=1$.
}
\label{fig_MQq6}
\end{figure}

\begin{figure}
\includegraphics[width=8cm]{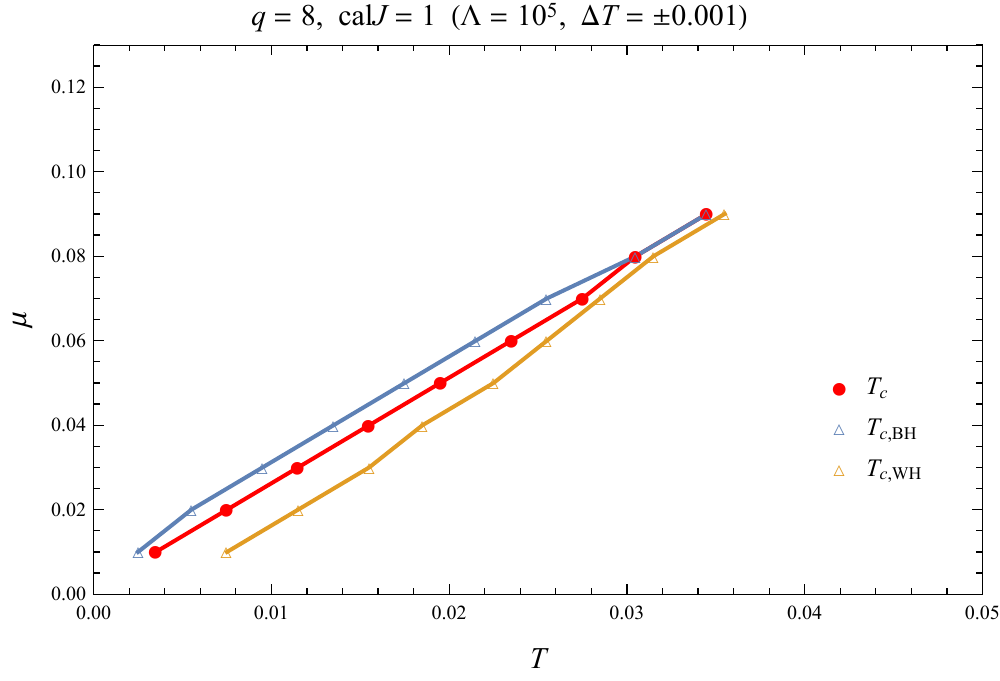}\quad\quad
\includegraphics[width=8cm]{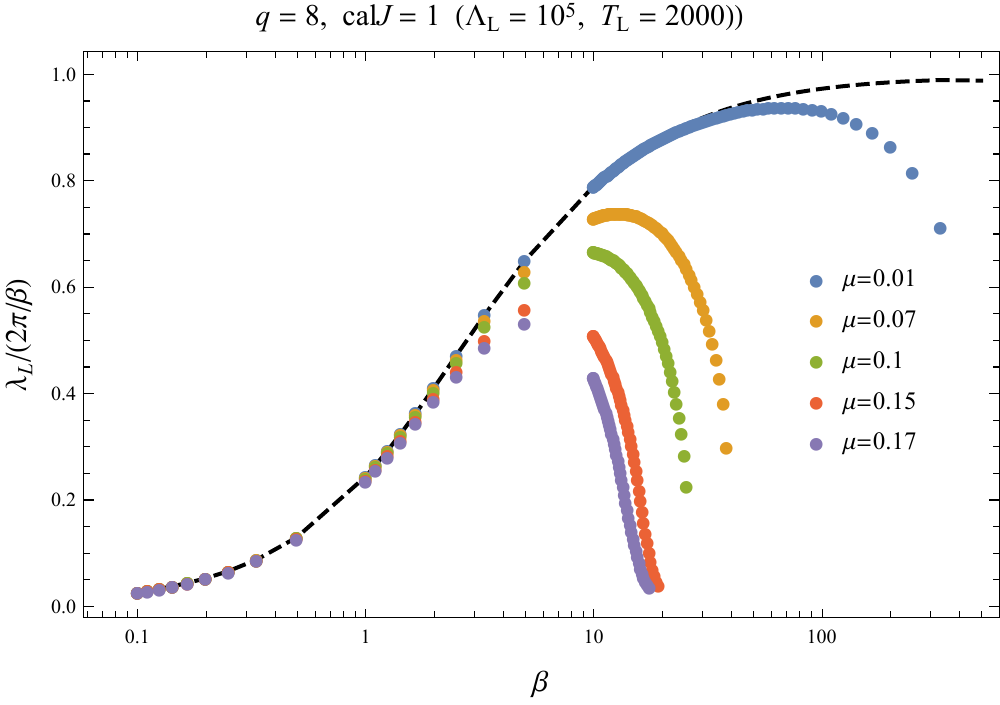}
\caption{
Phase diagram (left) and the chaos exponent (right) of the two coupled model with $q=8$, ${\cal J}=1$.
}
\label{fig_MQq8}
\end{figure}

Below we display the results for the phase diagram obtained by solving the Euclidean Schwinger-Dyson equations, and the chaos exponent obtained by solving the real time Schwinger-Dyson equations.
See Fig.~\ref{fig_MQq6},\ref{fig_MQq8}.
In contrast to the $q=4$ case in the real time analysis we could not reach the convergence in the wormhole regime even with the method of taking $\Lambda/T_L$ small explained in the end of section \ref{sec_result_realtime_MQ}.

\subsection{single sided model}
\label{sec_KMq6q8}
\begin{figure}
\includegraphics[width=8cm]{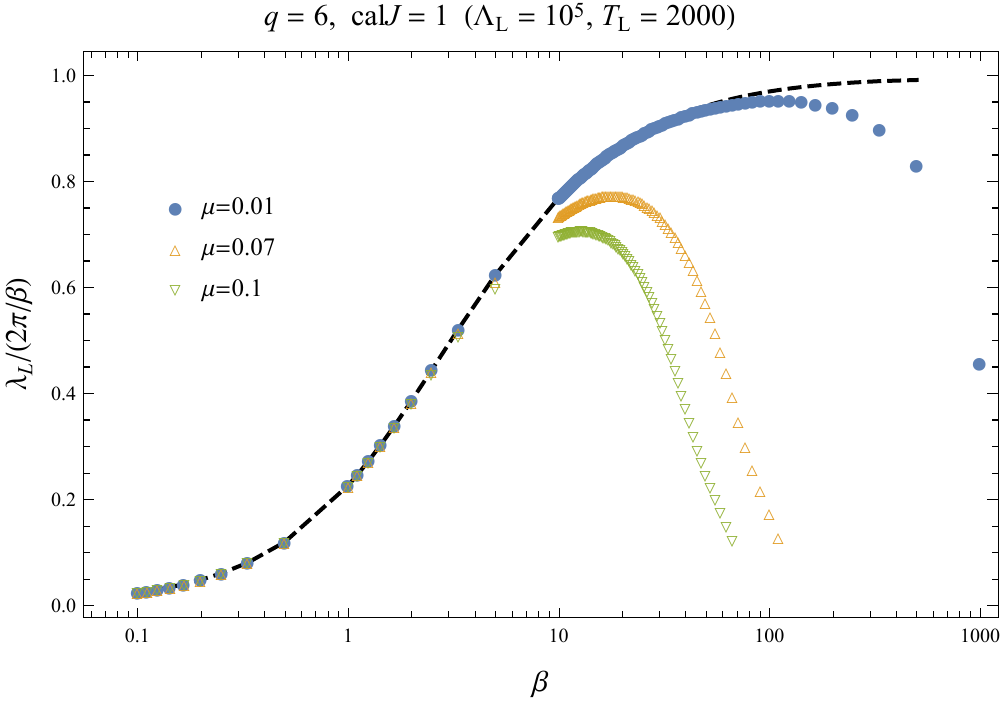}\quad\quad
\includegraphics[width=8cm]{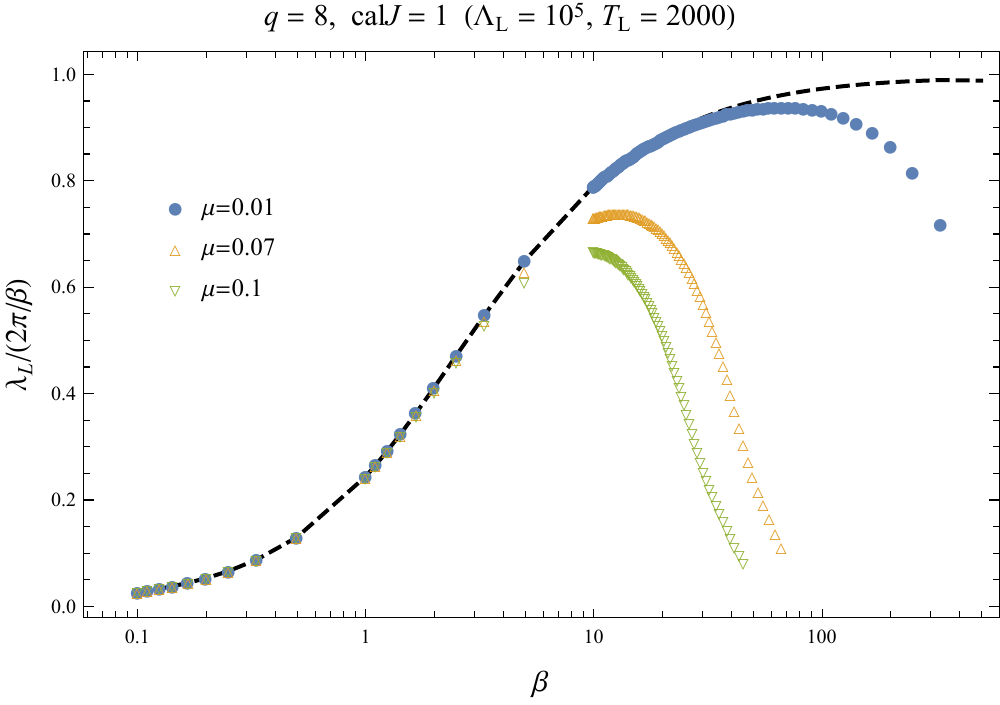}\quad\quad
\includegraphics[width=8cm]{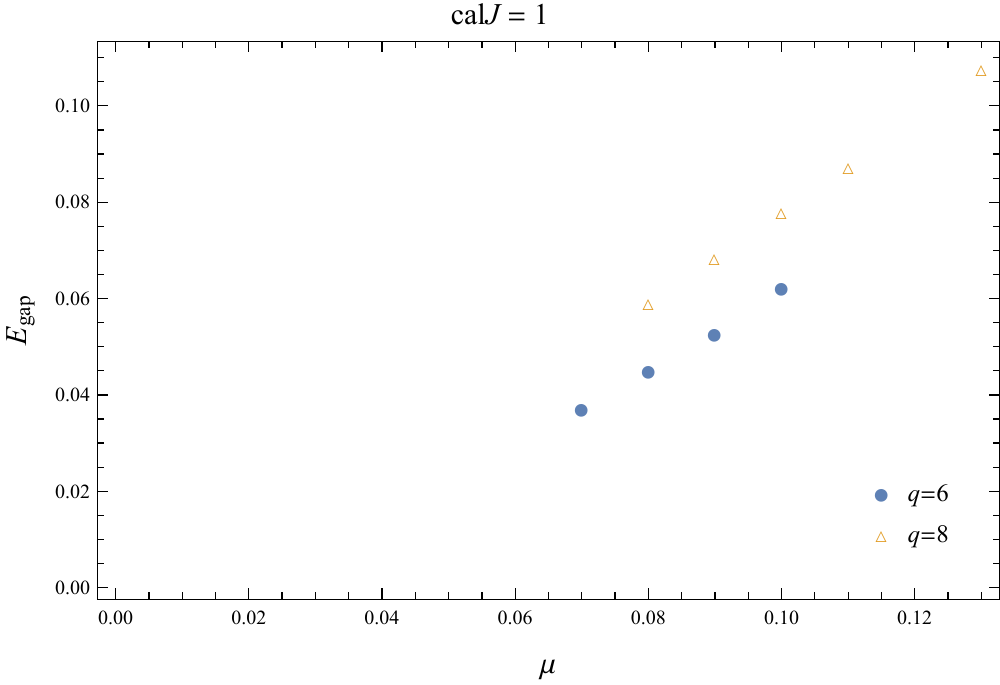}\quad\quad
\includegraphics[width=8cm]{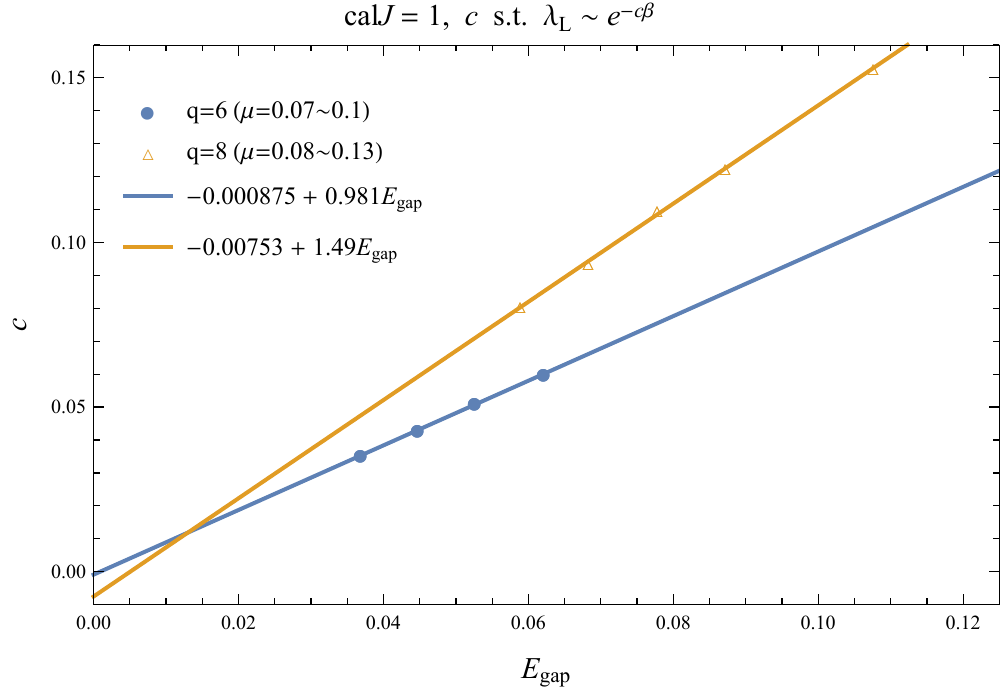}
\caption{
Top left/right: The chaos exponent of the single sided model with $q=6$, ${\cal J}=1$ and $q=8$, ${\cal J}=1$.
Bottom left/right: The energy gap $E_\text{gap}$ and its comparison with the decay exponent $c$ of the chaos exponent $\lambda_L\sim e^{-c\beta}$ at low temperature regime.
% Bottom left/right: The energy gap $E_\text{gap}$ and its comparison with the decay exponent of the chaos exponent at low temperature regime.
}
\label{fig_KMq6q8}
\end{figure}

Below we display the results for the chaos exponent of the single sided model with $q=6,8$.
See Fig.~\ref{fig_KMq6q8}.
As in the case of $q=4$, there are no phase transition.
At low temperature we found that the chaos exponent obeys the following formula
\begin{align}
\lambda_L\sim e^{-\frac{q/2-1}{2}\beta E_\text{gap}}.
\end{align}
Interestingly, this behavior is completely same as that of the decay rate of the first peak \cite{Qi:2020ian}.

\bibliography{KMNNbunken1_200917Nosaka.bib}
\end{document}